\documentclass[preprint,12pt,3p]{elsarticle}

\usepackage{graphicx}
\usepackage[hidelinks]{hyperref}
\usepackage{amsmath} 
\usepackage{amssymb}
\usepackage{cleveref} 
\usepackage{xcolor}
\usepackage{graphicx} 
\usepackage{array}
\usepackage{makecell}
\usepackage{adjustbox}
\usepackage{subcaption} 
\usepackage{natbib}
\biboptions{sort&compress}

\newcommand{\mihi}[1]{\textcolor{black}{#1}}

\newcolumntype{C}[1]{>{\centering\arraybackslash}m{#1}}

\begin{document}
	\begin{frontmatter}
		\title{An Efficient Bayesian Framework for Inverse Problems via Optimization and Inversion: Surrogate Modeling, Parameter Inference, and Uncertainty Quantification}

        \address[UNIPV]{Department of Civil Engineering and Architecture - University of Pavia, Via Ferrata 3, 27100, Pavia, Italy}
        \address[TUDresden]{Chair of Computational and Experimental Solid Mechanics, TU Dresden, Helmholtzstr 10,
        	01069, Dresden, Germany}
        \address[DCMS]{Dresden Center for Computational Materials Science (DCMS), TU Dresden, Dresden, Germany}
        \address[IMATI]{Istituto di Matematica Applicata e Tecnologie Informatiche “E. Magenes” - Consiglio Nazionale delle Ricerche, Via Ferrata, 5/A, 27100, Pavia, Italy}
        
		\author[UNIPV]{Mihaela Chiappetta} 
		\ead{mihaela.chiappetta@unipv.it}
		\author[UNIPV]{Massimo Carraturo}
		\ead{massimo.carraturo@unipv.it}
		\author[TUDresden]{Alexander Ra\ss loff}
		\ead{alexander.rassloff@tu-dresden.de}
		\author[TUDresden,DCMS]{Markus K\"astner}
	    \ead{markus.kaestner@tu-dresden.de}
	    \author[UNIPV,IMATI]{Ferdinando Auricchio}
	    \ead{ferdinando.auricchio@unipv.it}

		\begin{abstract}
			The present paper proposes a Bayesian framework for inverse problems that seamlessly integrates optimization and inversion to enable rapid surrogate modeling, accurate parameter inference, and rigorous uncertainty quantification. 
			Bayesian optimization is employed to adaptively construct accurate Gaussian process surrogate models using a minimal number of high-fidelity model evaluations, strategically focusing sampling in regions of high predictive uncertainty. 
			The trained surrogate model is then leveraged within a Bayesian inversion scheme to infer optimal parameter values by combining prior knowledge with observed quantities of interest, resulting in posterior distributions that rigorously characterize epistemic uncertainty. 
			The framework is theoretically grounded, computationally efficient, and particularly suited for engineering applications in which high-fidelity models — whether arising from numerical simulations or physical experiments — are computationally expensive, analytically intractable, or difficult to replicate, and data availability is limited.
			Furthermore, the combined use of Bayesian optimization and inversion outperforms their separate application, highlighting the synergistic benefits of unifying the two approaches.
				The performance of the proposed Bayesian framework is demonstrated on a suite of one- and two-dimensional analytical benchmarks, including the Mixed Gaussian-Periodic, Lévy, Griewank, Forrester, and Rosenbrock functions, which provide a controlled setting to assess surrogate modeling accuracy, parameter inference robustness, and uncertainty quantification.
				The results demonstrate the framework’s effectiveness in efficiently solving inverse problems while providing informative uncertainty quantification and supporting reliable engineering decision-making at reduced computational cost.
		\end{abstract}

		\end{frontmatter}
		
		\section{Introduction}
		\label{sec:introduction}
		\noindent
		Inverse problems \cite{tarantola2005inverse, groetsch1993inverse, aster2018parameter, neto2012introduction, stuart2010inverse, kirsch2011introduction} play a fundamental role across a wide range of engineering and scientific disciplines, providing a rigorous mathematical framework for inferring unknown or uncertain model parameters from observed data. 
		In contrast to forward problems, which predict system responses — i.e., quantities of interest — based on known input parameters, inverse problems aim to identify the parameters that best reproduce observed output quantities of interest \cite{tarantola2005inverse, kirsch2011introduction}. 
		Such problems frequently arise in applications including structural health monitoring, material characterization, geophysics, and fluid dynamics \cite{liu2024data, bureerat2018inverse, wang2022inverse, zunger2018inverse, zhdanov2015inverse, liu2017development, allen2022inverse, molesky2018inverse}, where direct measurement of governing parameters is often infeasible or prohibitively expensive. 
		However, inverse problems are typically ill-posed, meaning that small perturbations in the input parameters can lead to large variations in the inferred parameters and, consequently, in the predicted quantities of interest. 
		Therefore, robust mathematical and computational strategies are required to ensure reliable and interpretable solutions \cite{sullivan2015introduction, ghanem2017handbook}.
		
		Uncertainty quantification \cite{soize2017uncertainty, abdar2021review, sullivan2015introduction, ghanem2017handbook, richardson2012uncertainty, zhang2020basic, ahu2017uncertainty} is a fundamental component in inverse problems, providing the theoretical and computational tools needed to assess the reliability of inferred parameters by accounting for various sources of uncertainty, including measurement noise, model approximations, limited data availability, and incomplete prior knowledge.
		
		A key challenge in inverse problems is the high computational cost of evaluating high-fidelity models, or the high cost of performing physical experiments. High-fidelity models are designed to accurately capture the physics of complex systems, but they are often computationally intensive or lack a closed-form analytical representation. Physical experiments, on the other hand, can be expensive, time-consuming, or infeasible under certain conditions.
		To address the challenges outlined above, surrogate models are commonly employed to approximate high-fidelity behaviour at a significantly reduced computational cost \cite{sudret2017surrogate}. 
		Surrogate models may be deterministic — such as polynomial approximations, splines, or neural networks \cite{werner1984polynomial, alpert1991fast, mckinley1998cubic, knott1999interpolating, najm2009uncertainty, wang2003artificial} — or stochastic, with Gaussian Processes (GPs) being among the most widely used and effective approaches \cite{sudret2017surrogate, mackay1998introduction, williams1995gaussian, wilson2011gaussian, mosegaard2002probabilistic}.
		
		Stochastic surrogate models offer a distinct advantage in inverse problems, as they not only approximate high-fidelity models but also quantify the predictive uncertainty associated with model approximation. 
		Within this context, Bayesian methods \cite{carlin2008bayesian, ramoni1999bayesian} are particularly well-suited, offering a probabilistic framework for integrating prior knowledge — expressed as parameter distributions — with observed quantities of interest, and producing posterior distributions that rigorously characterize epistemic uncertainty in the inferred parameters and quantities of interest \cite{stuart2010inverse, idier2013bayesian, dashti2013bayesian, cotter2009bayesian, fitzpatrick1991bayesian}.
		
		Among Bayesian techniques, two are especially relevant in the context of inverse problems: Bayesian Optimization (BO) and Bayesian Inversion (BI). 
		BO \cite{frazier2018bayesian, garnett2023bayesian, frazier2018tutorial2, wang2023recent} is a global optimization strategy that leverages surrogate models to efficiently explore the parameter space, balancing exploration and exploitation via uncertainty-aware sampling. 
		BI \cite{stuart2010inverse, adler2018deep, calvetti2018inverse, ambikasaran2013fast, cotter2009bayesian, cotter2010approximation}, by contrast, formulates the inverse problem in a probabilistic setting, yielding posterior distributions that support robust, uncertainty-aware parameter inference.
		
		An increasing number of studies successfully apply Bayesian techniques to inverse problems. 
		Kennedy et al. \cite{kennedy2001bayesian} introduce a foundational Bayesian calibration framework for complex mathematical models applied to nuclear radiation data and a simulated nuclear accident. 
		Higdon et al. \cite{higdon2008computer} extend this methodology to thermal problems. 
		Kugler et al. \cite{kugler2022fast} propose scalable BO techniques for high-dimensional inverse problems in planetary remote sensing. 
		Alexanderian et al. \cite{alexanderian2024optimal} employ BI for optimal experimental design in PDE-constrained systems. 
		Raßloff et al. \cite{rassloff2025inverse} develop a BO-based inverse design strategy tailored to small-data regimes for the characterization of spinodoid architected materials. 
		Zheng et al. \cite{zheng2020inverse} propose and validate a Gauss–Bayesian approach for inverse problems in additive manufacturing. 
		Lartaud et al. \cite{lartaud2024sequential} introduce a sequential Bayesian design strategy for surrogate-based inverse design modeling. 
		Chiappetta et al. \cite{chiappetta2023sparse, chiappetta2024data} apply BI to reduce uncertainty in residual strain predictions for the additive manufacturing of Inconel 625 cantilever beam.
		
		Building on these complementary Bayesian techniques, we propose a Bayesian framework for inverse problems that integrates BO and BI in a computationally efficient manner. 
		BO accelerates the adaptive construction of accurate GP surrogate models using a limited number of high-fidelity model evaluations, whereas BI enables robust probabilistic inference of model parameters based on prior knowledge and observed quantities of interest. 
		In contrast to previous approaches such as \cite{alexanderian2024optimal, kennedy2001bayesian}, our BI implementation relies on a least-squares formulation, offering a computationally efficient and theoretically robust inference procedure \cite{stuart2010inverse, bjorck1990least, miller2006method}. 
		The integration of BO and BI enhances computational efficiency, systematically quantifies uncertainty, and improves parameter inference without compromising predictive accuracy, highlighting the synergistic benefit of unifying the two approaches. 
		This makes the proposed framework particularly suited to scenarios characterized by limited data availability and high-fidelity computational costs, offering a practical trade-off between efficiency, accuracy, and uncertainty-awareness.
		
		\mihi{
			The capabilities of the proposed framework are demonstrated through a suite of well-established one- and two-dimensional analytical benchmarks — including the Mixed Gaussian–Periodic, Lévy, Griewank, Forrester, and Rosenbrock functions — which allow for a controlled and interpretable assessment of nonlinearity, multimodality, and non-convexity in inverse problem settings.
		}
		
		\mihi{
			The numerical results confirm that, within these low-dimensional settings, the proposed methodology enables efficient surrogate modeling, accurate parameter inference, and comprehensive uncertainty quantification.
		}
		
		The remainder of the manuscript is organized as follows. 
		Section~\ref{sec:bayesian_method} presents the theoretical background on BO and BI. 
		Section~\ref{sec:inverse-design} describes the proposed Bayesian framework for inverse problems. 
		Section~\ref{analytic_benchmark} discusses numerical results and analytic benchmark validation. 
		Finally, Section~\ref{sec:conclusion} summarizes the key findings and outlines future research directions.
		
		\section{Theoretical Background on Bayesian Methods}
		\label{sec:bayesian_method}
		\noindent
		The present section outlines the theoretical background underlying the two Bayesian techniques, BO and BI~\cite{frazier2018bayesian, garnett2023bayesian, frazier2018tutorial2, wang2023recent, stuart2010inverse, adler2018deep, calvetti2018inverse, ambikasaran2013fast, cotter2009bayesian, cotter2010approximation}, integrated into the proposed  framework.
		
		Section~\ref{subsec:bayesian_optimization} introduces BO, emphasizing its reliance on GP surrogate models~\cite{sudret2017surrogate, mackay1998introduction, williams1995gaussian, wilson2011gaussian, mosegaard2002probabilistic} and uncertainty-aware acquisition strategies for efficient exploration of the parameter space~\cite{gan2021acquisition, archetti2019acquisition, wilson2018maximizing}.
		Section~\ref{subsec:bayesian_inversion} presents BI, which formulates the inverse problem in a probabilistic framework and enables parameter inference through the computation of posterior distributions.
		
			\subsection{Bayesian Optimization}
			\label{subsec:bayesian_optimization}
			\noindent
			BO~\cite{frazier2018bayesian, garnett2023bayesian, frazier2018tutorial2, wang2023recent} is a sample-efficient strategy for the global optimization of objective functions that are defined implicitly through computational models or physical experiments. 
			In inverse problems, the objective function evaluations often requires high-fidelity numerical simulations or costly experimental procedures, making BO particularly suitable for contexts where only a limited number of objective function evaluations can be performed.
		
			The BO framework relies on a probabilistic surrogate model, typically a GP~\cite{sudret2017surrogate, mackay1998introduction, williams1995gaussian, wilson2011gaussian, mosegaard2002probabilistic}, to approximate the objective function \( f: \mathbb{R}^d \rightarrow \mathbb{R} \). 
			A training dataset \( \mathcal{D}_n = \{ (\mathbf{x}_i, y_i) \}_{i=1}^n \), where \( \mathbf{x}_i \in \mathbb{R}^d \) are the input parameters and \( y_i = f(\mathbf{x}_i) + \epsilon_i \) are the corresponding function evaluations (i.e., the quantities of interest) with additive noise \( \epsilon_i \sim \mathcal{N}(0, \sigma_n^2) \), is used to fit the surrogate model.
			A GP defines a prior distribution over functions of the form:
			\begin{equation}
				\label{eq:gp}
				f(\mathbf{x}) \sim f_{\mathcal{GP}}(\mathbf{x}) = \mathcal{GP}(\mu(\mathbf{x}), k(\mathbf{x}, \mathbf{x}'))
			\end{equation}
			where \( \mu(\mathbf{x}) \) is the mean function, typically assumed to be zero without loss of generality, and \( k(\mathbf{x}, \mathbf{x}') \) is the covariance kernel that encodes spatial correlation over the input parameters domain.
		
			Common choices for the kernel include the Radial Basis Function (RBF) kernel~\cite{buhmann2000radial}:
			\begin{equation}
				\label{eq:rbf}
				k_{\text{RBF}}(\mathbf{x}, \mathbf{x}') = \sigma^2 \exp\left( -\frac{\| \mathbf{x} - \mathbf{x}' \|^2}{2\ell^2} \right)
			\end{equation}
			where \( \ell > 0 \) is the length-scale parameter and \( \sigma^2 \) is the signal variance. 
			Another widely used kernel is the Matérn kernel~\cite{melkumyan2011multi}:
			\begin{equation}
				\label{eq:matern}
				k_{\text{Matérn}}(\mathbf{x}, \mathbf{x}') = \sigma^2 \frac{2^{1-\nu}}{\Gamma(\nu)} \left( \frac{\sqrt{2\nu} \| \mathbf{x} - \mathbf{x}' \|}{\ell} \right)^\nu K_\nu \left( \frac{\sqrt{2\nu} \| \mathbf{x} - \mathbf{x}' \|}{\ell} \right)
			\end{equation}
			where \( \nu > 0 \) controls the smoothness of the realizations and \( K_\nu \) is the modified Bessel function of the second kind. 
			The choice between RBF and Matérn kernels typically depends on the expected smoothness of the underlying function.
			The Matérn kernel with $\nu = 5/2$ is frequently adopted, as it offers a good compromise between smoothness and flexibility.
		
			Once the GP surrogate model is trained, it is queried through an acquisition function \( \alpha(\mathbf x): \mathbb{R}^d \rightarrow \mathbb{R} \), which is used to identify the next input parameters \( \mathbf{x}_{\text{next}} \) to improve the surrogate model approximation. 
			The acquisition function formalizes the trade-off between exploration of uncertain regions and exploitation of areas likely to contain the optimum. 
			Two widely adopted acquisition functions are Expected Improvement (EI) and Upper Confidence Bound (UCB)~\cite{gan2021acquisition, archetti2019acquisition, wilson2018maximizing}.
		
			The EI acquisition function is defined as:
			\begin{equation}
				\label{eq:ei}
				\alpha_{\text{EI}}(\mathbf{x}) = \mathbb{E}\left[ \max(f_{\mathcal{GP}}(\mathbf{x}) - y^*, 0) \right]
			\end{equation}
			where \( y^* \) denotes the best observed quantity of interest. 
			This criterion favours sampling points expected to improve upon the best observation, although it may underexplore regions of high uncertainty.
		
			Alternatively, the UCB acquisition function explicitly incorporates predictive uncertainty:
			\begin{equation}
				\label{eq:ucb}
				\alpha_{\text{UCB}}(\mathbf{x}) = \mu(\mathbf{x}) + \kappa \sigma(\mathbf{x})
			\end{equation}
			where \( \mu(\mathbf{x}) \) and \( \sigma(\mathbf{x}) \) denote the GP posterior predictive mean and standard deviation, conditioned on the training dataset, and \( \kappa > 0 \) is a user-defined parameter that governs the balance between exploration and exploitation.
			In particular, the choice between EI and UCB depends on the specific inverse problem scenario: EI is preferred for improvement search, while UCB promotes a more balanced exploration of the parameter space.
		
			At each iteration, the next input parameter are selected as:
			\begin{equation}
				\label{eq:next}
				\mathbf{x}_{\text{next}} = \arg\max_{\mathbf{x}} \alpha(\mathbf{x})
			\end{equation}
			and the objective function is evaluated at this location. 
			The dataset is updated, and the GP surrogate model is retrained. 
			The process continues iteratively until a convergence criterion is satisfied, such as reaching a maximum number of evaluations, achieving stabilization of the acquisition function, or achieving a desired surrogate model accuracy.
		
			\subsection{Bayesian Inversion}
			\label{subsec:bayesian_inversion}
			\noindent
			BI~\cite{stuart2010inverse, adler2018deep, calvetti2018inverse, ambikasaran2013fast, cotter2009bayesian, cotter2010approximation} offers a rigorous probabilistic framework for solving inverse problems by leveraging Bayes’ theorem~\cite{bernardo2009bayesian, koch1990bayes, swinburne2004bayes, damien2013bayesian, ghosh2007introduction}. 
			The approach is particularly effective in ill-posed settings, where classical deterministic methods may yield unstable or non-unique solutions due to data noise, model inaccuracies, or limited observations~\cite{ghanem2017handbook}.
			
			Let \( \bar{y} \in \mathbb{R}^m \) denote the observed quantity of interest, and let \( \mathbf{x} \in \mathbb{R}^d \) be the parameters to be inferred. 
			Bayes’ theorem defines the posterior distribution over \( \mathbf{x} \), conditioned on the data \( \bar{y} \), as:
			\begin{equation}
				\label{eq:posterior_bayes2}
				p(\mathbf{x} \mid \bar{y}) = \frac{p(\bar{y} \mid \mathbf{x}) \, p(\mathbf{x})}{p(\bar{y})} \propto p(\bar{y} \mid \mathbf{x}) \, p(\mathbf{x})
			\end{equation}
			where \( p(\bar{y} \mid \mathbf{x}) \) is the likelihood function, \( p(\mathbf{x}) \) is the prior distribution encoding prior knowledge about the parameters, and \( p(\bar{y}) \) is the marginal likelihood, also known as  evidence. 
			Since the evidence does not depend on \( \mathbf{x} \) and is typically intractable~\cite{lotfi2022bayesian}, BI is typically performed on the unnormalized posterior distribution.
			
			Assuming additive Gaussian noise, the observed quantities of interest are expressed as:
			\begin{equation}
				\label{eq:synthetic_data}
				\bar{y} = f(\mathbf{x}_{\text{true}}) + \varepsilon, \quad \varepsilon \sim \mathcal{N}(0, \sigma^2 I_m)
			\end{equation}
			where \( f(\cdot) \) denotes the high-fidelity model, \( \mathbf{x}_{\text{true}} \) are the unknown true parameters value, \( \varepsilon \) is zero-mean Gaussian noise with known variance \( \sigma^2 \), and \( I_m \in \mathbb{R}^{m \times m} \) denotes the identity matrix of dimension \( m \).
			Under this assumption, the likelihood function takes the form:
			\begin{equation}
				\label{eq:likelihood_bi1}
				p(\bar{y} \mid \mathbf{x}) = \frac{1}{(2\pi\sigma^2)^{m/2}} \exp\left( -\frac{1}{2\sigma^2} \left\| \bar{y} - f(\mathbf{x}) \right\|^2 \right)
			\end{equation}
			
			The prior distribution \( p(\mathbf{x}) \) serves as a regularizing term, incorporating prior assumptions on \( \mathbf{x} \) before the quantity of interest is observed. 
			A common and analytically convenient choice is the multivariate Gaussian prior distribution:
			\begin{equation}
				\label{eq:prior_bi1}
				p(\mathbf{x}) = \frac{1}{(2\pi)^{d/2} |\Gamma|^{1/2}} \exp\left( -\frac{1}{2} (\mathbf{x} - \mu_{\mathbf{x}})^\top \Gamma^{-1} (\mathbf{x} - \mu_{\mathbf{x}}) \right)
			\end{equation}
			where \( \mu_{\mathbf{x}} \in \mathbb{R}^d \) and \( \Gamma \in \mathbb{R}^{d \times d} \) are the prior mean and covariance matrix, respectively.
			
			A widely adopted point estimator in BI is the Maximum A Posteriori (MAP) estimate~\cite{gauvain1994maximum, bassett2019maximum}, defined as:
			\begin{equation}
				\label{eq:map_estimation1}
				\mathbf{x}_{\text{MAP}} = \arg\max_{\mathbf{x}} \, p(\mathbf{x} \mid \bar{y})
			\end{equation}
			
			This is equivalent to minimizing the negative log-posterior distribution:
			\begin{equation}
				\label{eq:map_nll10}
				\mathbf{x}_{\text{MAP}} = \arg\min_{\mathbf{x}} \left[ -\log p(\mathbf{x} \mid \bar{y}) \right]
			\end{equation}
			
			Under the Gaussian likelihood and prior distribution assumptions, this leads to the following optimization problem:
			\begin{equation}
				\label{eq:map_log_likelihood}
				\mathbf{x}_{\text{MAP}} = \arg\min_{\mathbf{x}} \left[ \frac{1}{2\sigma^2} \left\| \bar{y} - f(\mathbf{x}) \right\|^2 + \frac{1}{2} (\mathbf{x} - \mu_{\mathbf{x}})^\top \Gamma^{-1} (\mathbf{x} - \mu_{\mathbf{x}}) \right]
			\end{equation}
			
			This formulation explicitly reveals the balance between fidelity to the observed quantities of interest and adherence to prior knowledge, which is central to the application of BI in inverse problems. 
			
		\section{Bayesian Framework for Inverse Problems via Optimization and Inversion}
		\label{sec:inverse-design}
		\noindent
		Building upon the probabilistic foundations established in Section~\ref{sec:bayesian_method}, the present section introduces the proposed Bayesian framework for inverse problems that synergistically integrates BO and BI~\cite{frazier2018bayesian, garnett2023bayesian, frazier2018tutorial2, wang2023recent, stuart2010inverse, adler2018deep, calvetti2018inverse, ambikasaran2013fast, cotter2009bayesian, cotter2010approximation}. 
		
\begin{figure}[htbp]
			\centering
			\includegraphics[width=1\textwidth]{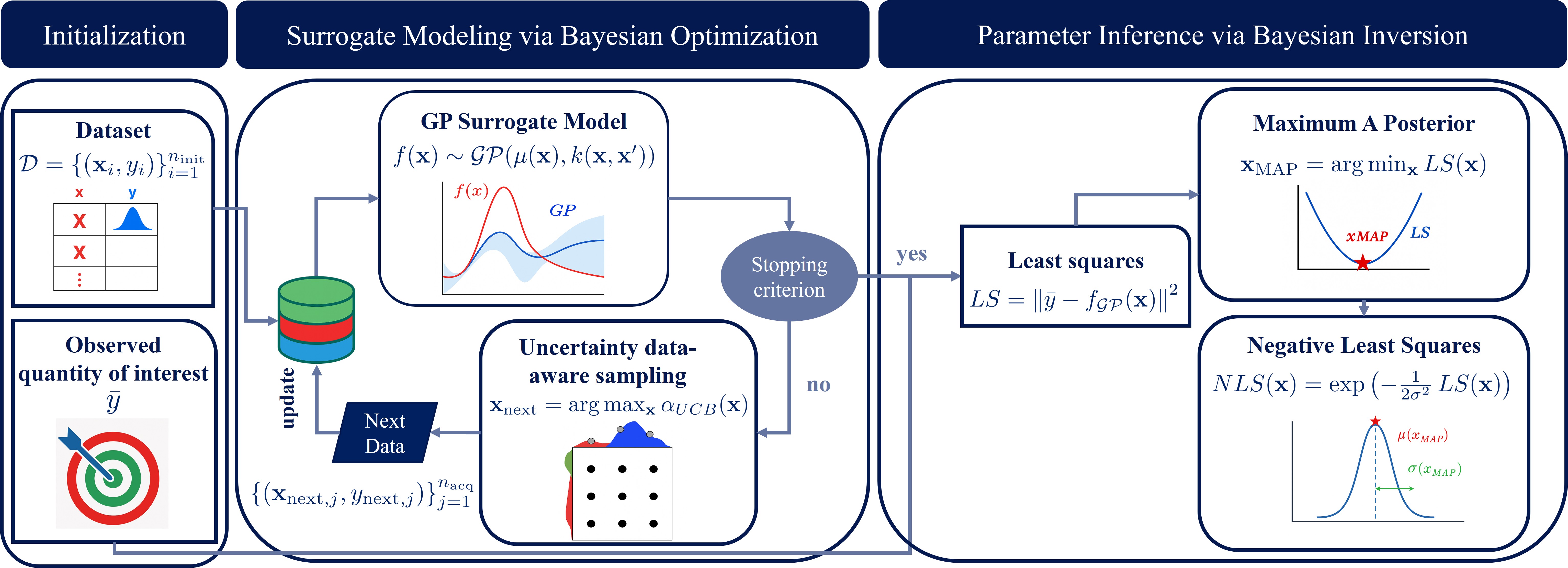}
			\caption{Schematic representation of the proposed Bayesian framework for inverse problems, integrating BO and BI. 
				The process begins with the initialization stage (left), where an initial training dataset and an observed quantity of interest are defined. 
				A GP surrogate model is then iteratively refined via BO using uncertainty-aware sampling guided by the UCB acquisition strategy (center). 
				Once convergence is achieved and given an observed quantity of interest, the surrogate model is employed in a BI procedure to perform parameter inference through LS (Least Squares) minimization, yielding the MAP estimate and the posterior distribution via the NLS (Negative Least Squares) formulation (right).}
			\label{fig:graphical}
		\end{figure}

		The overarching goal is to identify the input parameters that most plausibly generate a given observed quantity of interest, while rigorously characterizing the epistemic uncertainty associated with the inferred parameters through BI. 
		To alleviate the computational burden of repeated high-fidelity model evaluations, the framework leverages BO to adaptively construct a GP surrogate model~\cite{sudret2017surrogate, mackay1998introduction, williams1995gaussian, wilson2011gaussian, mosegaard2002probabilistic}.
		A schematic overview of the proposed framework is provided in Figure~\ref{fig:graphical}. 

		The remainder of the present section is organized as follows. 
		Section~\ref{subsec:bobo} details the surrogate model construction through BO. Section~\ref{subsec:bobi} describes the parameter inference procedure through BI based on the surrogate model constructed in Section~\ref{subsec:bobo}. Finally, Section~\ref{subsec:efficient} illustrates the computational efficiency of the proposed Bayesian framework.
		
			\subsection{Surrogate Modeling via Bayesian Optimization}
			\label{subsec:bobo}
			\noindent
			As illustrated in Figure~\ref{fig:graphical}, the first stage of the proposed framework consists in constructing a surrogate model that approximates the high-fidelity mapping from input parameters to quantities of interest. 
			The process begins from an initial training dataset:
			\begin{equation}
				\label{eq:dataset}
				D = \{(\mathbf{x}_i, y_i)\}_{i=1}^{n_{\text{init}}}
			\end{equation}
			where \( \mathbf{x}_i \in \mathbb{R}^{d} \) are the input parameters, \( y_i \in \mathbb{R}^m \) are the corresponding quantities of interest, and \( n_{\text{init}} \) is the initial number of samples. 
			Based on \( D \), a GP surrogate model \( f_{\mathcal{GP}}(\mathbf{x}) \) is trained to approximate the high-fidelity model \( f(\mathbf{x}) \), according to \Cref{eq:gp}.
			
			To iteratively refine the surrogate model, the UCB acquisition function (\Cref{eq:ucb}) is used to guide the selection of new input parameters. 
			In the present framework, no prior knowledge of the observed quantity of interest is assumed during the surrogate model construction phase; such information is instead incorporated during the parameter inference step, as detailed in Section~\ref{subsec:bobi}. 
			To reflect this assumption, the exploration parameter \( \kappa \) (see \Cref{eq:ucb}) is set to a sufficiently large value, effectively promoting a purely exploratory sampling strategy that prioritizes regions of high predictive uncertainty. 
			\mihi{
				This choice enables the construction of a globally accurate surrogate model over the parameter space considered in the present study, which can subsequently support multiple inverse inference tasks within the same setting without requiring retraining.
			}
			
			At each iteration, a batch of \( n_{\text{acq}} \) acquired new number of samples \( \{\mathbf{x}_{\text{next},j}\}_{j=1}^{n_{\text{acq}}} \) is selected by maximizing the acquisition function (\Cref{eq:next}). 
			This strategy directs sampling toward regions of high predictive uncertainty, thereby improving surrogate model accuracy.
			
			The dataset is updated as:
			\begin{equation}
				\label{eq:dataset_update}
				D \leftarrow D \cup \{(\mathbf{x}_{\text{next}, j}, y_{\text{next}, j})\}_{j=1}^{n_{\text{acq}}}
			\end{equation}
			where \( y_{\text{next}, j} = f(\mathbf{x}_{\text{next}, j}) \) are new evaluations of the high-fidelity model. 
			The GP surrogate model is retrained on the updated dataset, progressively enhancing the surrogate model fidelity and reducing predictive uncertainty.
			
			The iterative surrogate model refinement procedure continues until a convergence criterion is satisfied.
			In the present work, convergence is assessed based on the Mean Squared Error (MSE) between high-fidelity model outputs and surrogate model predictions on an independent validation set of \( n_{\text{val}} \) samples:
			\begin{equation}
				\label{eq:mse}
				\text{MSE} = \frac{1}{n_{\text{val}}} \sum_{i=1}^{n_{\text{val}}} \left( y_i - {y}_{\mathcal{GP},i} \right)^2
			\end{equation}
			where \( y_i = f(\mathbf x_i) \) are high-fidelity outputs and \( {y}_{\mathcal{GP},i} = f_{\mathcal{GP}}(\mathbf{x}_i) \) are the surrogate model predictions.
			
			The use of the MSE as a convergence criterion is justified in controlled benchmark scenarios, where access to the true high-fidelity model enables quantitative validation of surrogate model accuracy on an independent validation set of samples, as in the case of the present study. 
			However, this approach is generally impractical in real-world applications, where high-fidelity model evaluations are computationally expensive or experimentally unfeasible. In such cases, alternative convergence criteria are required to ensure sample efficiency without relying on external validation samples.
			Alternative strategies include monitoring the stabilization of the acquisition function~\cite{frazier2018tutorial2}, the decay of the GP posterior predictive variance~\cite{shahriari2016multi}, or the marginal improvement in surrogate model predictions across successive iterations~\cite{forrester2008engineering}.
			A systematic comparison of these strategies and their impact on surrogate model accuracy and downstream inference is beyond the scope of the present study and is deferred to future work.
			
			\subsection{Parameter Inference via Bayesian Inversion}
			\label{subsec:bobi}
			\noindent
			Once the surrogate model reaches the desired accuracy, parameter inference is performed through BI, as illustrated in Figure~\ref{fig:graphical}. 
			Given an observed quantity of interest \( \bar{y} \in \mathbb{R}^m \), the objective is to infer the input parameters \( \mathbf{x} \in \mathbb{R}^d \) that most plausibly reproduce the observed quantity of interest, while quantifying the associated epistemic uncertainty.
			
			Following Bayes’ theorem (\Cref{eq:posterior_bayes2}), the posterior distribution over the input parameters combines the likelihood and the prior distribution. 
			The MAP estimate is obtained by maximizing the posterior distribution or, equivalently, by minimizing its negative logarithm as in \Cref{eq:map_nll10}.

			Assuming a uniform prior distribution — reflecting lack of informative prior knowledge — and Gaussian observational noise (\Cref{eq:synthetic_data}), the posterior distribution simplifies to a least-squares minimization:
			\begin{equation}
				\label{eq:map_ls}
				\mathbf{x}_{\text{MAP}} = \arg\min_{\mathbf{x}} \left\| \bar{y} - f_{\mathcal{GP}}(\mathbf{x}) \right\|^2 = \arg\min_{\mathbf{x}} LS(\mathbf{x})
			\end{equation}
			where \( LS(\mathbf{x}) \) denotes the least-squares functional, i.e., the squared misfit between surrogate model prediction and observed quantity of interest.
			
			Beyond MAP estimation, uncertainty quantification is achieved by analyzing the full posterior distribution. 
			Under the assumptions of additive Gaussian noise and a uniform prior distribution, the posterior distribution takes the form of an unnormalized exponential of the least-squares functional~\cite{miller2006method}:
			\begin{equation}
				\label{eq:nls}
				p(\mathbf{x} \mid \bar{y}) \propto \exp\left( -\frac{1}{2\sigma^2} LS(\mathbf{x}) \right) =: NLS(\mathbf{x})
			\end{equation}
			where \( NLS(\mathbf{x}) \) represents the so-called Negative Least Squares functional, used here to approximate the unnormalized posterior distribution.
			
			This formulation highlights that the posterior distribution not only identifies the most plausible parameters via the MAP estimate, but also encodes the associated epistemic uncertainty in a probabilistic manner. 
			When the \( NLS(\mathbf{x}) \) is approximately unimodal and locally Gaussian, a Laplace approximation centered at the MAP point can be employed \cite{chiappetta2023sparse}:
			\begin{equation}
				\label{eq:mapsig}
				p(\mathbf{x} \mid \bar{y}) \approx \mathcal{N}(\mathbf{x}_{\text{MAP}}, \boldsymbol{\Sigma}_{\text{MAP}}), \quad \text{with} \quad \boldsymbol{\Sigma}^{-1}_{\text{MAP}} = \nabla^2_{\mathbf{x}} \left[ -\log NLS(\mathbf{x}) \right] \big|_{\mathbf{x} = \mathbf{x}_{\text{MAP}}}
			\end{equation}
			The local Gaussian approximation is computationally efficient and provides valuable insight into parameter uncertainty, enabling the construction of confidence regions and uncertainty-aware predictions.
			
			\subsection{Efficiency of the Proposed Bayesian Framework}
			\label{subsec:efficient}
			\noindent
			Building on Sections~\ref{subsec:bobo} and \ref{subsec:bobi}, the present section qualitatively compares the computational efficiency of the proposed Bayesian framework against the separate use of BO and BI (Table~\ref{tab:bo_bi_comparison}). 
			
			In the first row of \Cref{tab:bo_bi_comparison}, BO guided by the EI acquisition function (BO(EI), \Cref{eq:ei}) concentrates evaluations around incumbent optima. 
			This behaviour favours local exploitation for parameter inference but typically produces a surrogate model that is accurate only in the neighbourhood of the best-so-far point; in addition, BO alone does not return a posterior distribution over the parameters, so parameter uncertainty is not quantified.
			
			In the second row, BO driven by the UCB acquisition function (BO(UCB), \Cref{eq:ucb}) emphasises exploration through uncertainty-aware sampling and generally yields a surrogate model that is accurate across the explored parameter domain with a limited budget of high-fidelity model evaluations.
			\mihi{
				The notion of global surrogate accuracy here refers to the benchmark settings considered in the present study and does not imply general scalability to multi-dimensional parameter spaces.
			}
			Nevertheless, BO in isolation still provides no posterior distribution for parameter uncertainty.
			
			In the third row, BI with a MAP estimator (BI(MAP), see \Cref{eq:map_log_likelihood}) directly addresses the inverse problem and yields a posterior distribution that quantifies epistemic uncertainty. 
			However, it does not construct a surrogate model and therefore does not reduce the cost of subsequent predictions.
			
			In the proposed framework (fourth row of \Cref{tab:bo_bi_comparison}), BO with a large exploration parameter (BO(UCB, $\,\kappa\!\gg\!0$)) rapidly constructs a GP surrogate model over the admissible parameter domain, after which BI is performed on the surrogate model via the least-squares formulation BI(MAP--LS), see \Cref{eq:map_ls}. 
			This coupling delivers an accurate MAP estimate together with a posterior distribution over the parameters at negligible prediction cost.
			\mihi{
				The reported efficiency gains are assessed with respect to the low-dimensional benchmark problems investigated in the present work.
			}
			A uniform prior over the admissible bounds is adopted to avoid biasing the parameter inference towards specific regions of the parameter space and to assign equal plausibility to all admissible values, intentionally reflecting broader epistemic uncertainty than an informative Gaussian prior distribution.
			
			\mihi{
				Overall, the unified BO(UCB, $\,\kappa\!\gg\!0$)+BI(MAP--LS) strategy combines fast surrogate model construction with parameter inference and uncertainty quantification, achieving higher efficiency than BO or BI used separately for the class of inverse problems examined in the present study.
			}
			
			\begin{table}[ht!]
				\centering
				
				\newlength{\HeaderH} \setlength{\HeaderH}{42mm}
				\newlength{\BodyH}   \setlength{\BodyH}{34mm}
				
				\newcommand{\CenterCellH}[1]{%
					\begin{minipage}[c][\HeaderH][c]{\linewidth}\centering #1\end{minipage}}
				\newcommand{\CenterCellB}[1]{%
					\begin{minipage}[c][\BodyH][c]{\linewidth}\centering #1\end{minipage}}
				
				\setlength{\tabcolsep}{4.5pt}
				\renewcommand{\arraystretch}{1.12}
				
				\begin{adjustbox}{max width=\textwidth, max totalheight=\textheight, keepaspectratio}
					\begin{tabular}{|C{4.0cm}|C{4.0cm}|C{4.0cm}|C{3.4cm}|C{3.4cm}|}
						\hline
						\CenterCellH{\includegraphics[width=.8\linewidth]{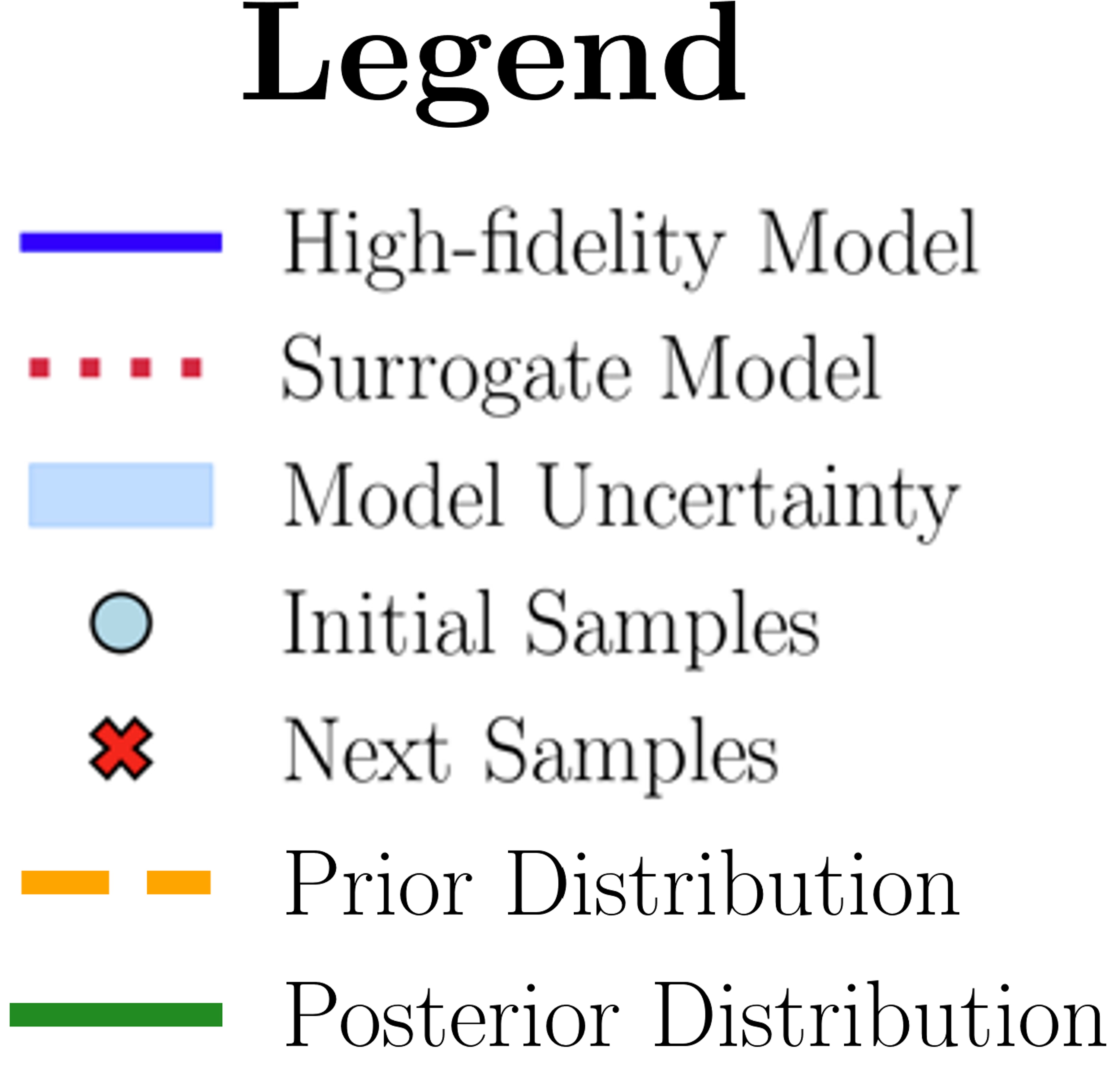}} &
						\CenterCellH{\textbf{Surrogate Model}} &
						\CenterCellH{\textbf{Model Uncertainty}} &
						\CenterCellH{\textbf{Parameter Inference}} &
						\CenterCellH{\textbf{Parameter Uncertainty}} \\
						\hline
						
						\CenterCellB{\textbf{BO(EI)}} &
						\CenterCellB{\includegraphics[width=1.\linewidth]{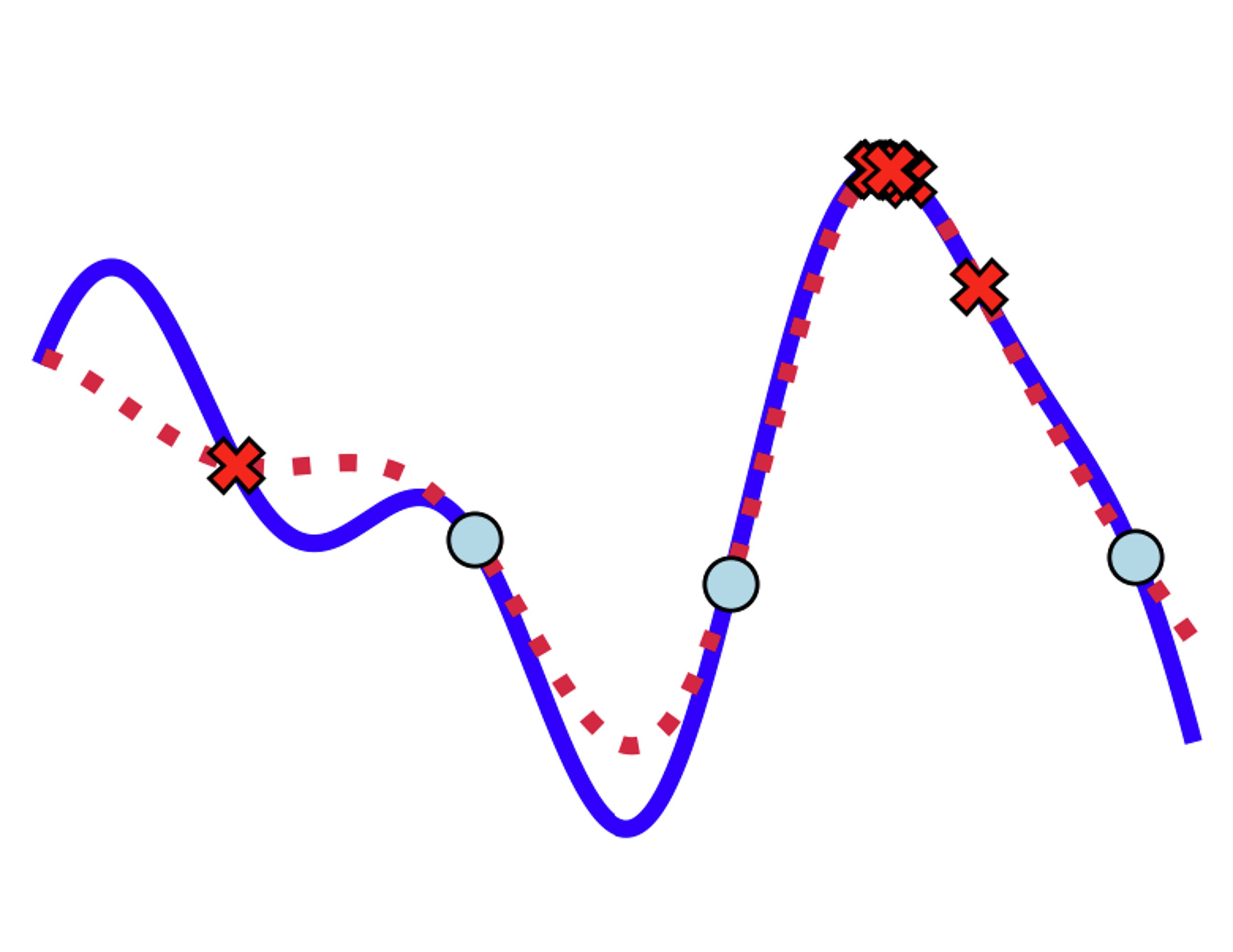}} &
						\CenterCellB{\includegraphics[width=1\linewidth]{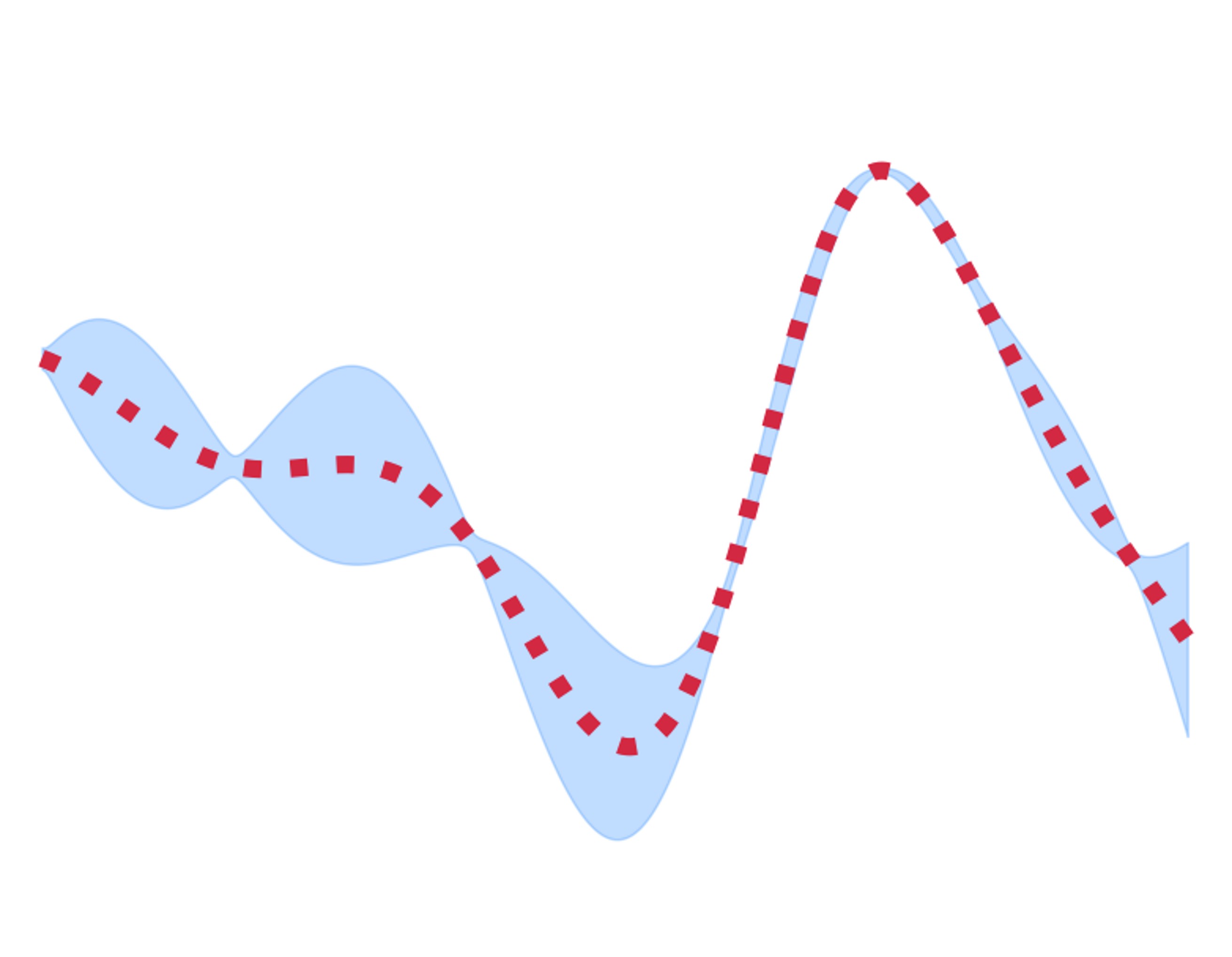}} &
						\CenterCellB{\includegraphics[width=.9\linewidth]{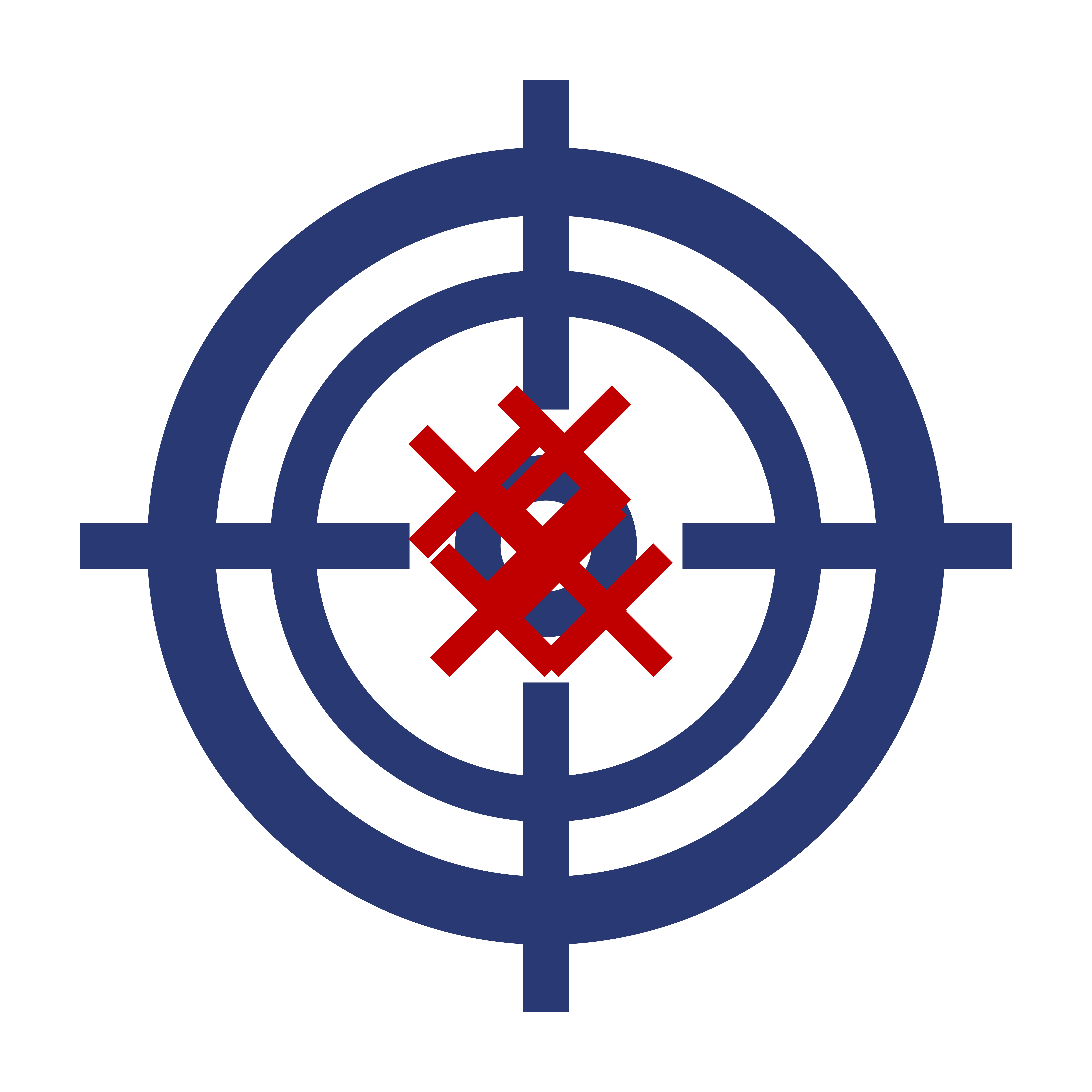}} &
						\CenterCellB{\emph{None}} \\
						\hline
						
						\CenterCellB{\textbf{BO(UCB)}} &
						\CenterCellB{\includegraphics[width=1\linewidth]{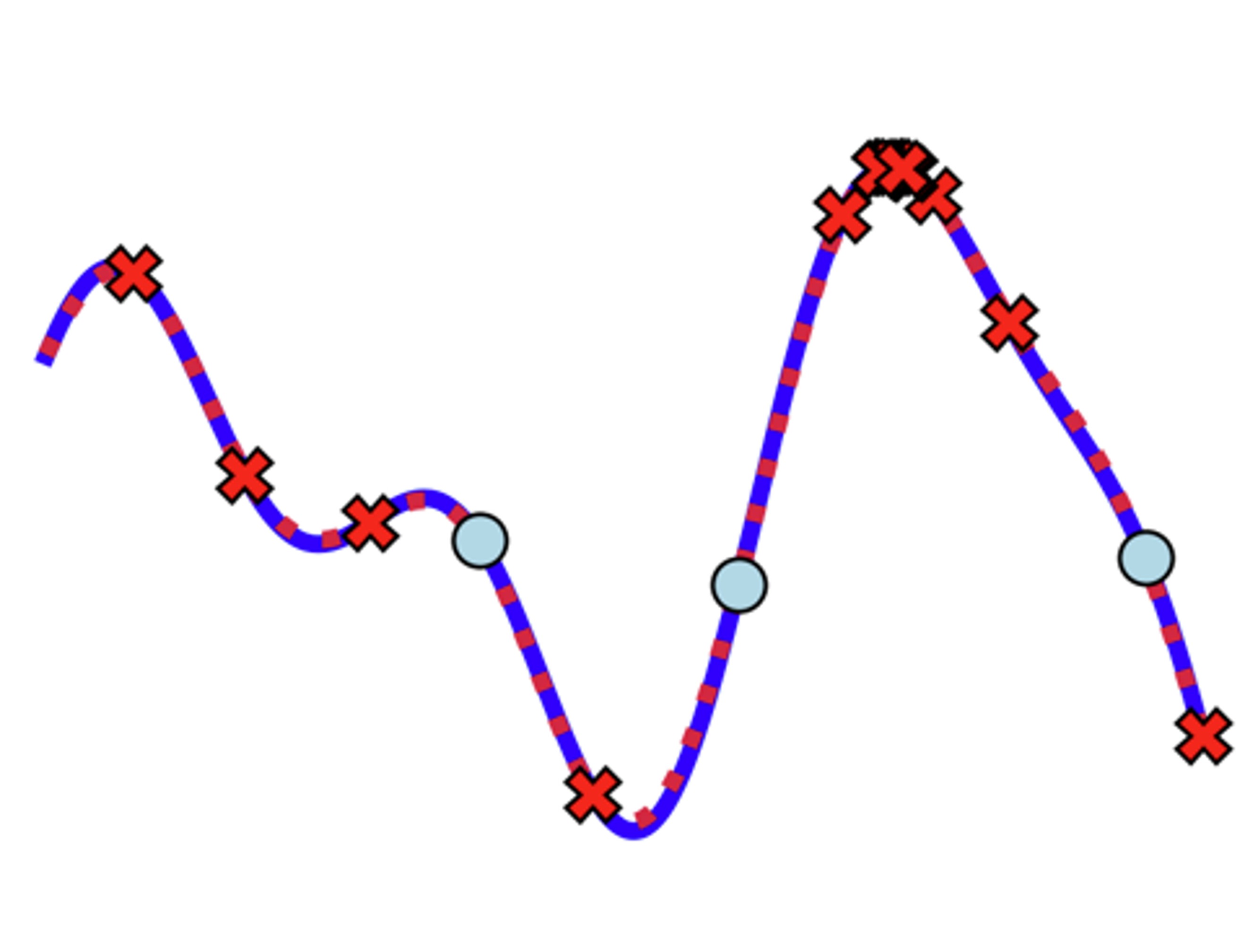}} &
						\CenterCellB{\includegraphics[width=1\linewidth]{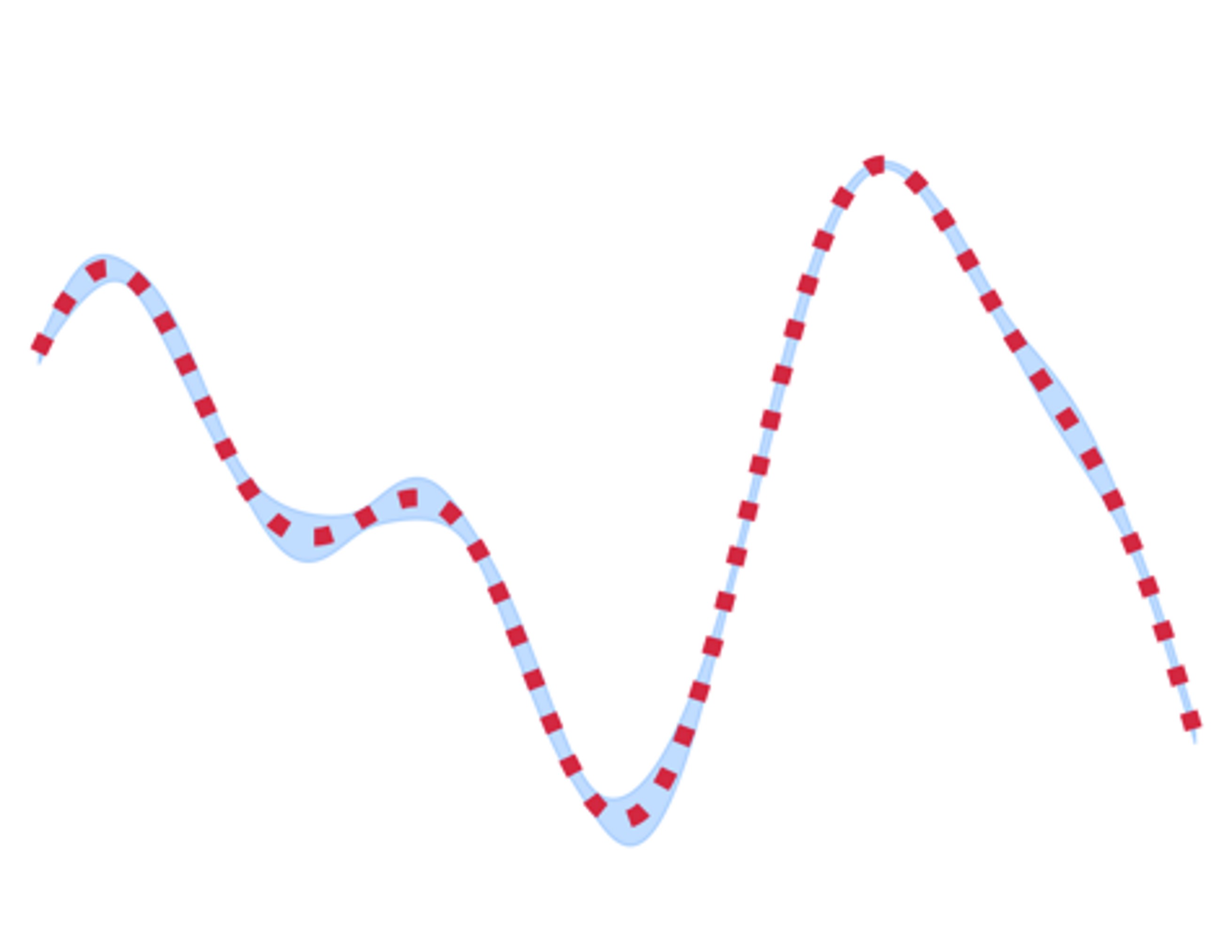}} &
						\CenterCellB{\includegraphics[width=.72\linewidth]{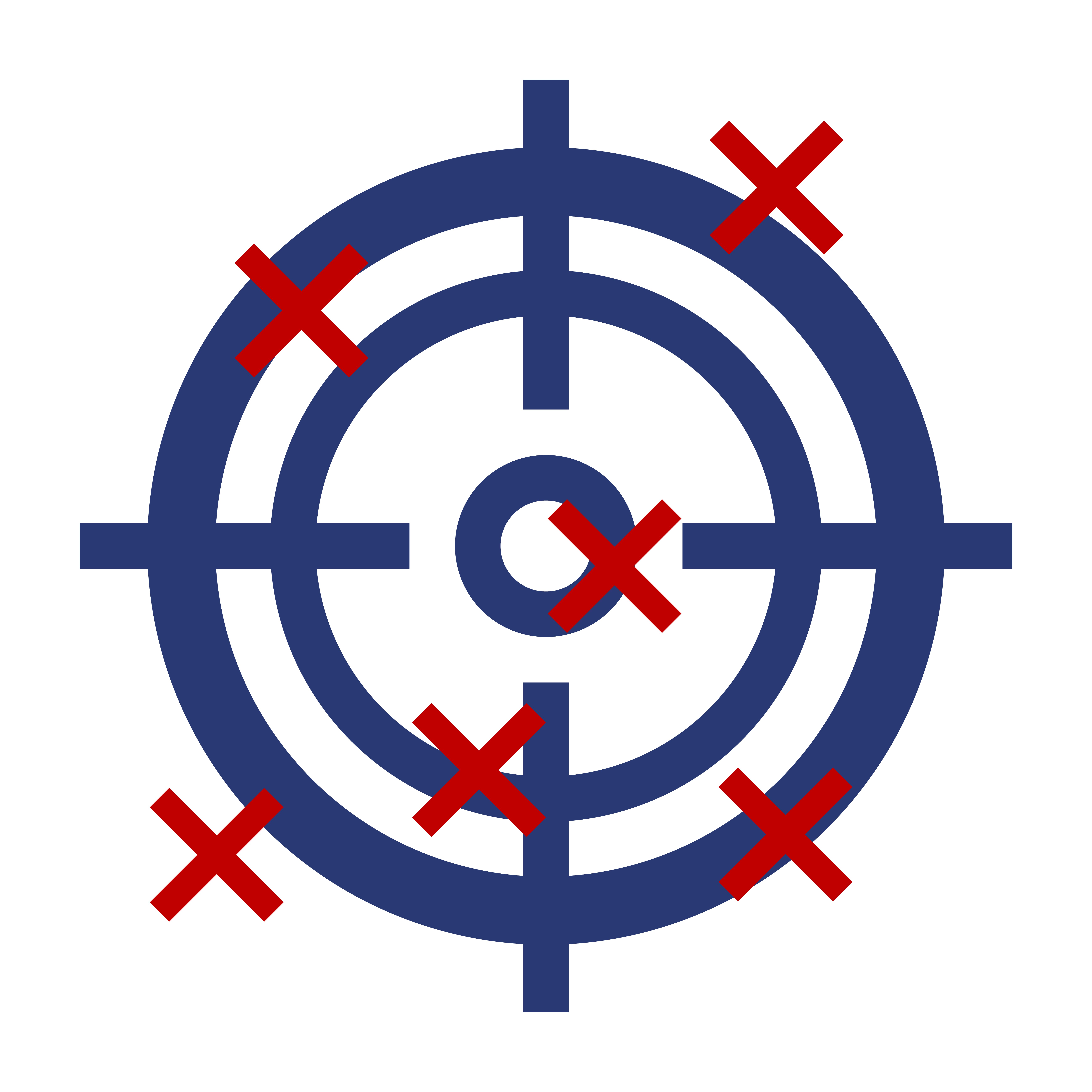}} &
						\CenterCellB{\emph{None}} \\
						\hline
						
						\CenterCellB{\textbf{BI(MAP)}} &
						\CenterCellB{\emph{None}} &
						\CenterCellB{\emph{None}} &
						\CenterCellB{\includegraphics[width=.9\linewidth]{a3}} &
						\CenterCellB{\includegraphics[width=1\linewidth]{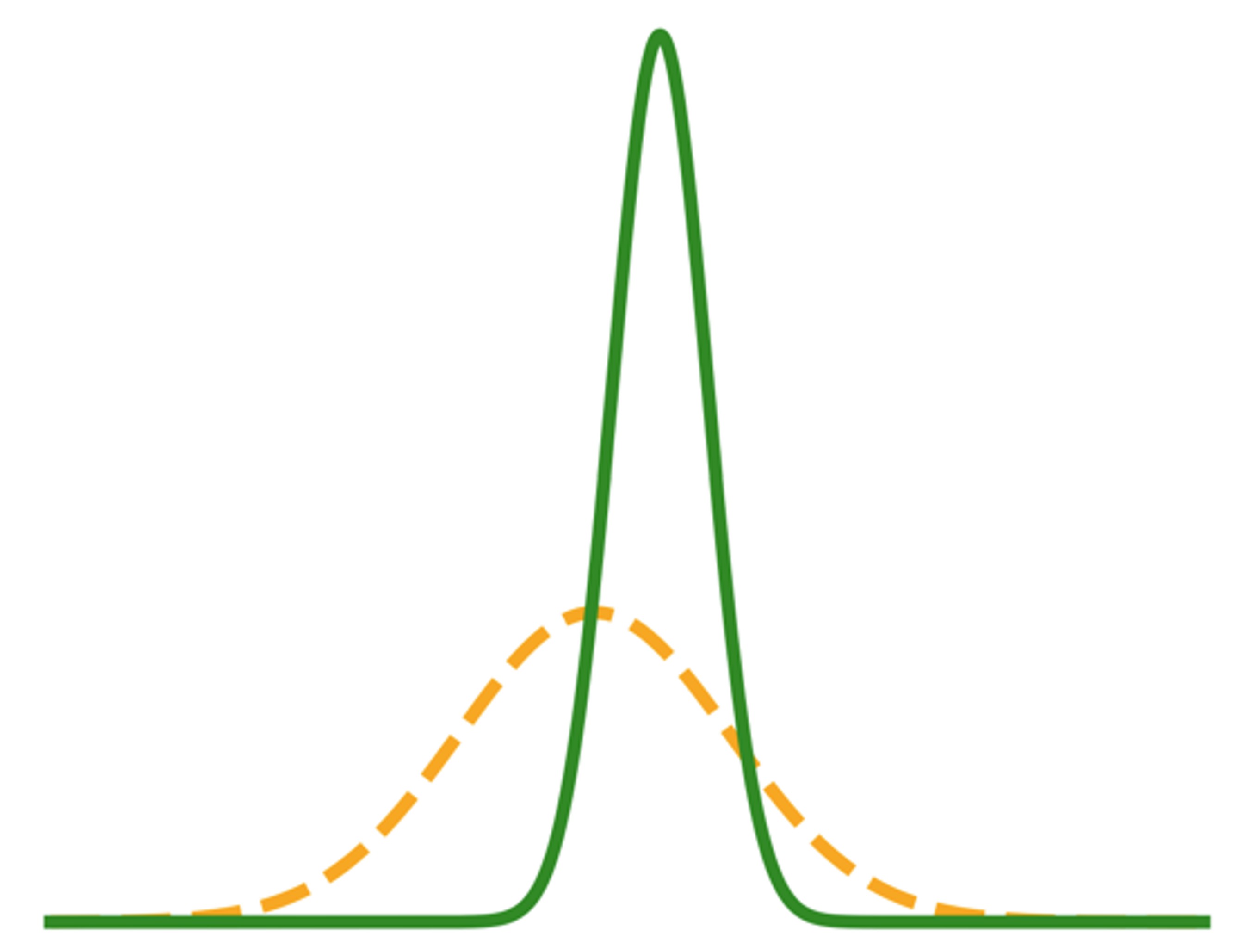}} \\
						\hline
						
						\CenterCellB{\makecell{\textbf{Our Approach}\\[-2pt]\textbf{BO(UCB $\boldsymbol{\kappa\gg 0}$)}\\[-2pt]\textbf{+ BI(MAP--LS)}}} &
						\CenterCellB{\includegraphics[width=1\linewidth]{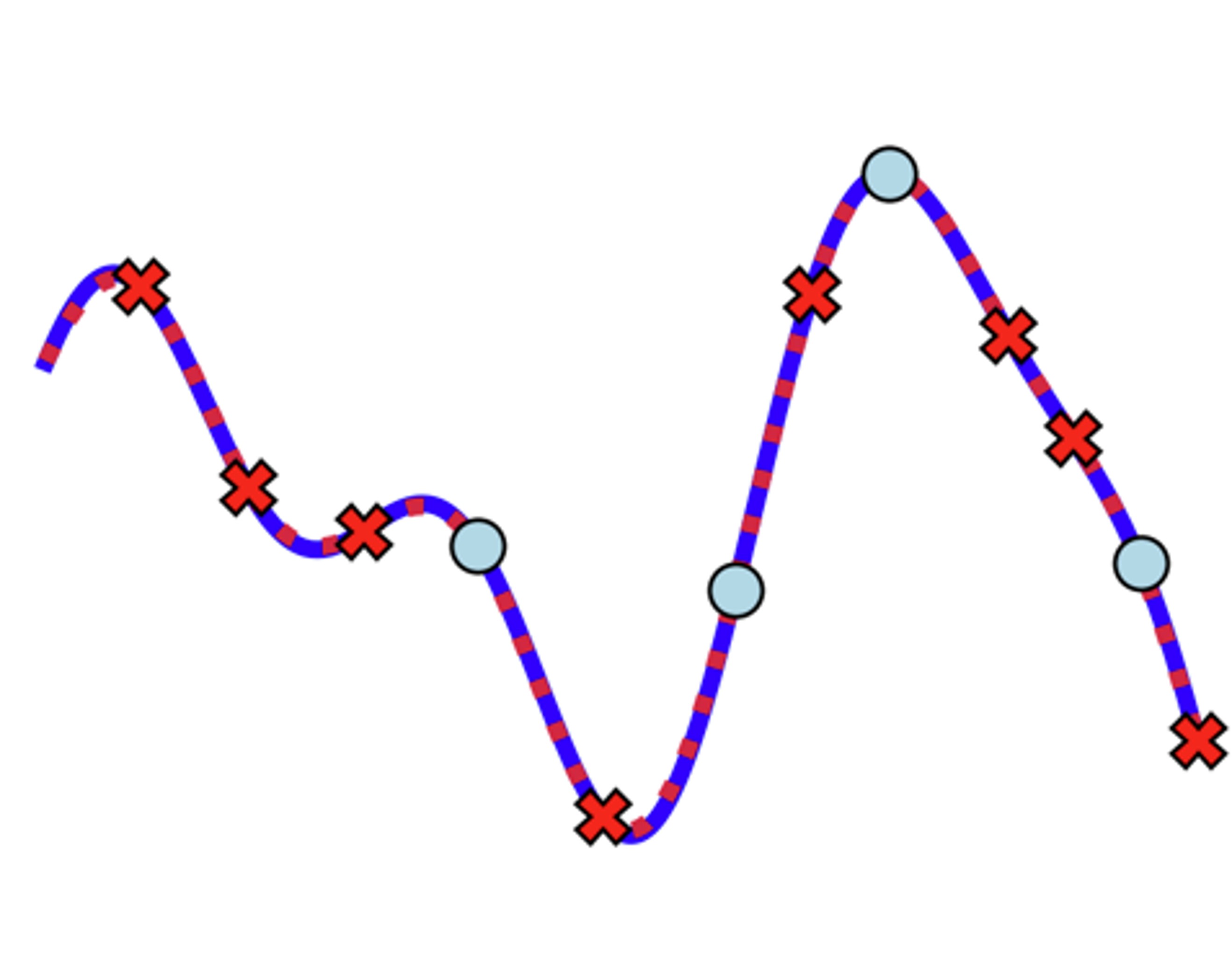}} &
						\CenterCellB{\includegraphics[width=1\linewidth]{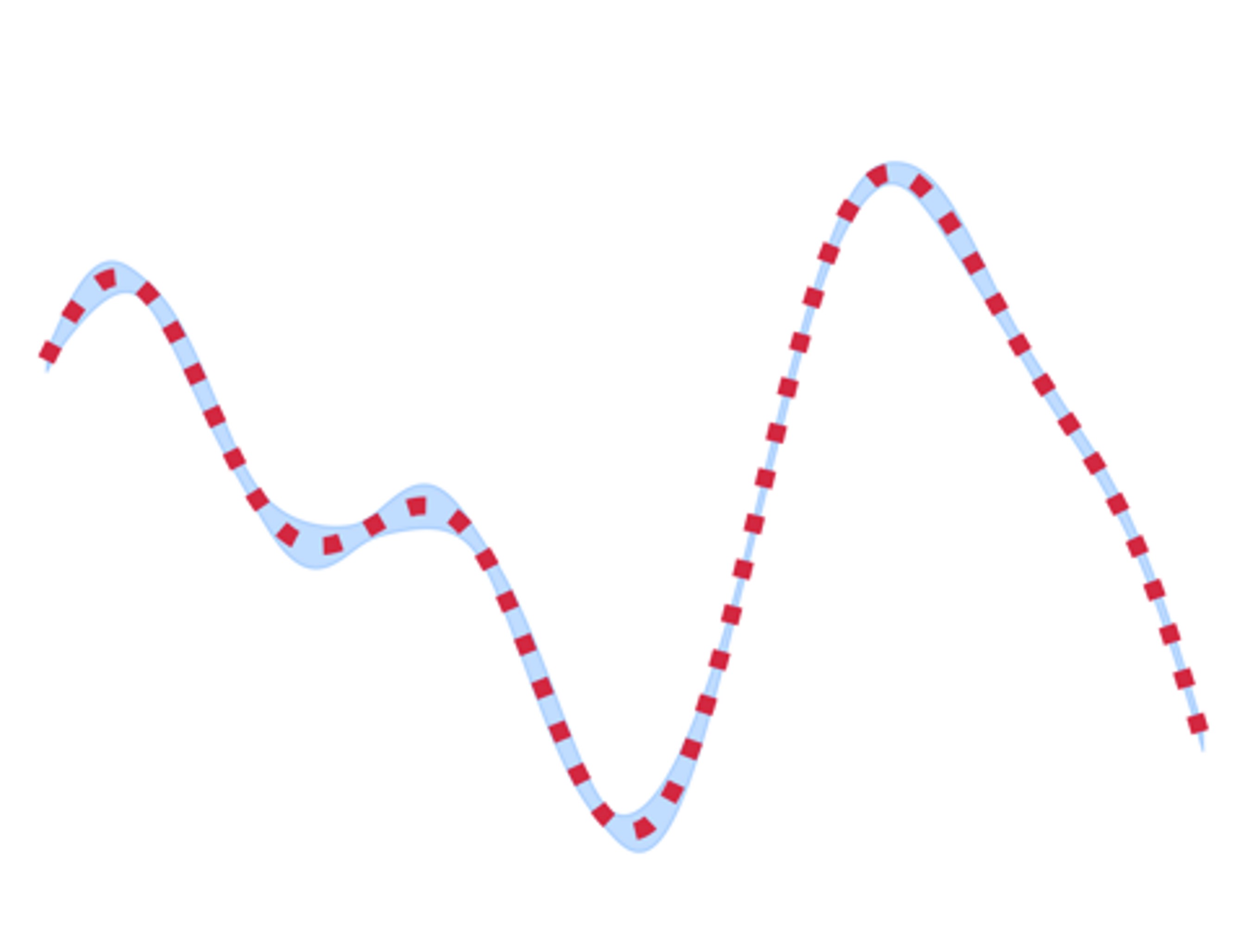}} &
						\CenterCellB{\includegraphics[width=.9\linewidth]{a3}} &
						\CenterCellB{\includegraphics[width=1\linewidth]{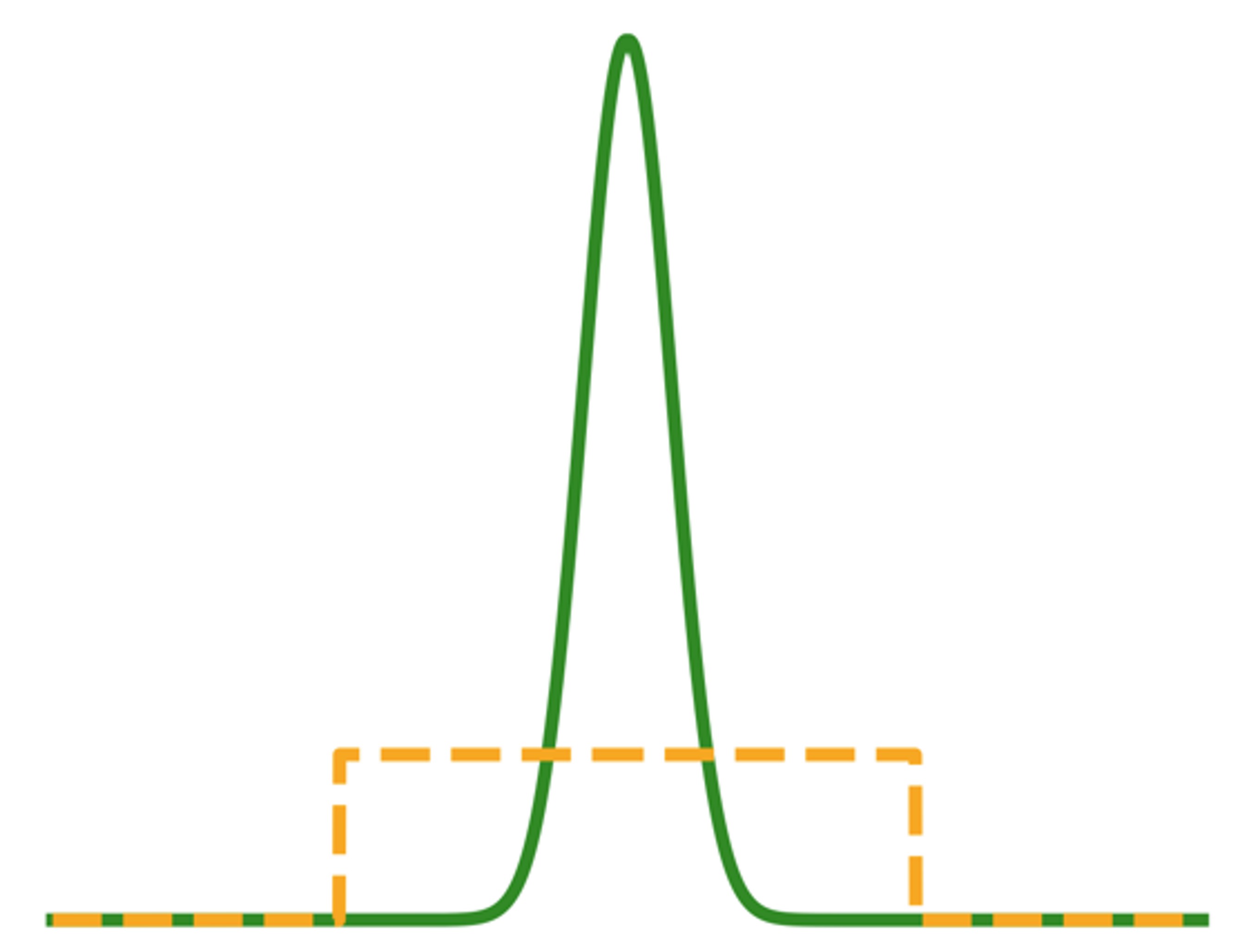}} \\
						\hline
					\end{tabular}
				\end{adjustbox}		
				\caption{Qualitative comparison between using BO and BI separately and the proposed unified framework BO(UCB, $\,\kappa\!\gg\!0$)+BI(MAP–LS). 
				BO(EI) clusters samples around incumbent optima and yields a locally accurate surrogate model with no posterior distribution; 
				BO(UCB) constructs a globally accurate surrogate model yet still provides no posterior distribution; 
				BI(MAP) performs parameter inference and returns a posterior distribution but no surrogate model; 
				the proposed approach (BO(UCB, $\,\kappa\!\gg\!0$)+BI(MAP–LS)) couples rapid surrogate model construction with MAP-LS-based parameter inference and posterior distribution quantification, leading to greater overall efficiency.}
				\label{tab:bo_bi_comparison}
			\end{table}
		
		\section{Numerical Validation on Analytical Benchmarks}
		\label{analytic_benchmark}
		\noindent
		The present section validates the proposed Bayesian framework for inverse problems, introduced in Section~\ref{sec:inverse-design}, through a series of one- and two-dimensional analytical benchmark functions. 
		The analytic benchmarks provide a controlled setting for quantitatively evaluating the framework’s performance in terms of surrogate modeling accuracy, robustness of parameter inference, computational efficiency, and uncertainty quantification.
		
		The section is organized as follows. 
		Section~\ref{subsec:1D} presents the results of the validation for one-dimensional benchmarks, analysing separately the construction of the surrogate model (section~\ref{subsubsec:1Dbo}), parameter inference (section~\ref{subsubsec:1Dbi}) and a concluding discussion of the results (Section~\ref{subsubsec:1Dres}).
		Section~\ref{subsec:2D} extends the validation to two-dimensional benchmarks, following the same structure to ensure consistency between surrogate modeling (Section~\ref{subsubsec:2Dbo}), parameter inference (Section~\ref{subsubsec:2Dbi}) and a concluding discussion of the results (Section~\ref{subsubsec:2Dres}).
		Each case includes an assessment of uncertainty quantification and computational cost, which are critical aspects in real-world inverse problem scenarios.
		
		All numerical experiments are conducted using the open-source library \texttt{UQforPy}, which provides a modular and extensible platform for uncertainty quantification and inverse design. 
		Additional implementation details are provided in~\ref{ref}.
		
			\subsection{One-Dimensional Benchmarks}
			\label{subsec:1D}
\begin{figure}[!ht]
				\centering
				\centering
				\begin{subfigure}{0.24\textwidth}
					\includegraphics[width=\textwidth]{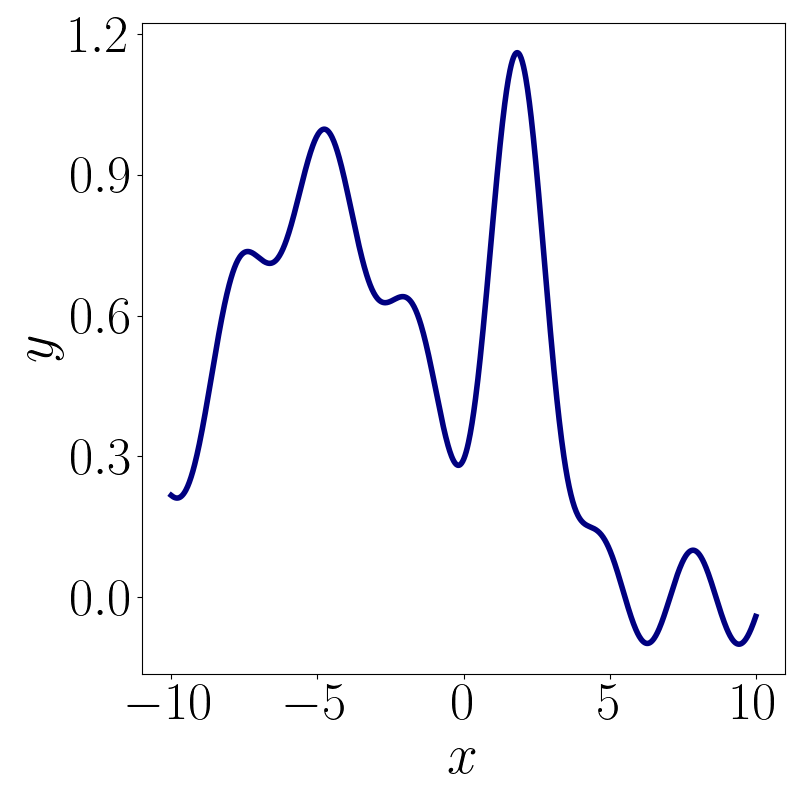}
					\caption{Mixed Gaussian-Periodic.}
					\label{subfig:mixed1d}
				\end{subfigure}
				\begin{subfigure}{0.24\textwidth}
					\includegraphics[width=\textwidth]{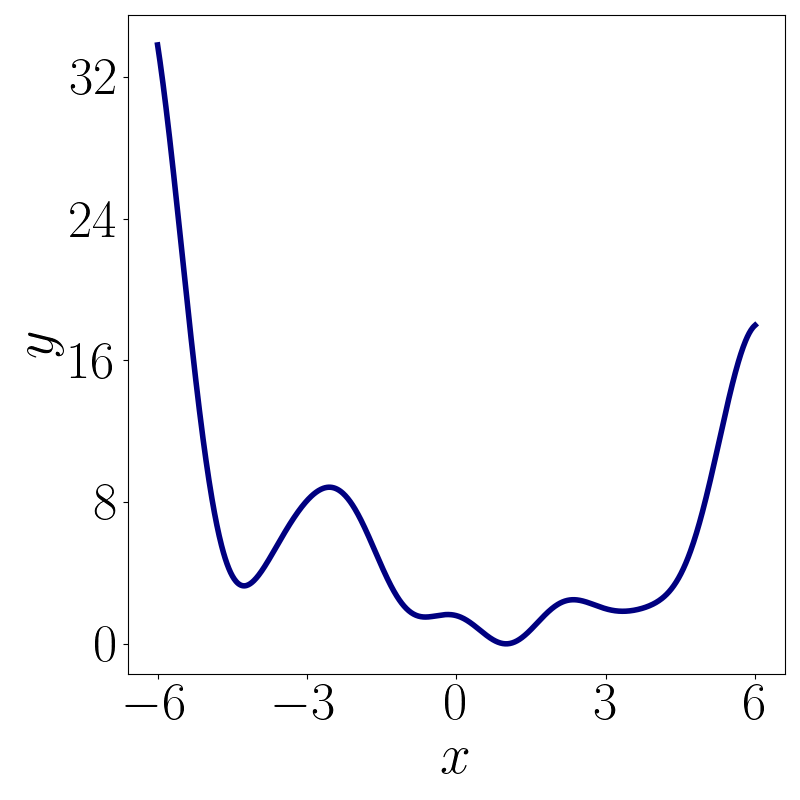}
					\caption{Lévy.}
					\label{subfig:levy1d}
				\end{subfigure}    
				\begin{subfigure}{0.24\textwidth}
					\includegraphics[width=\textwidth]{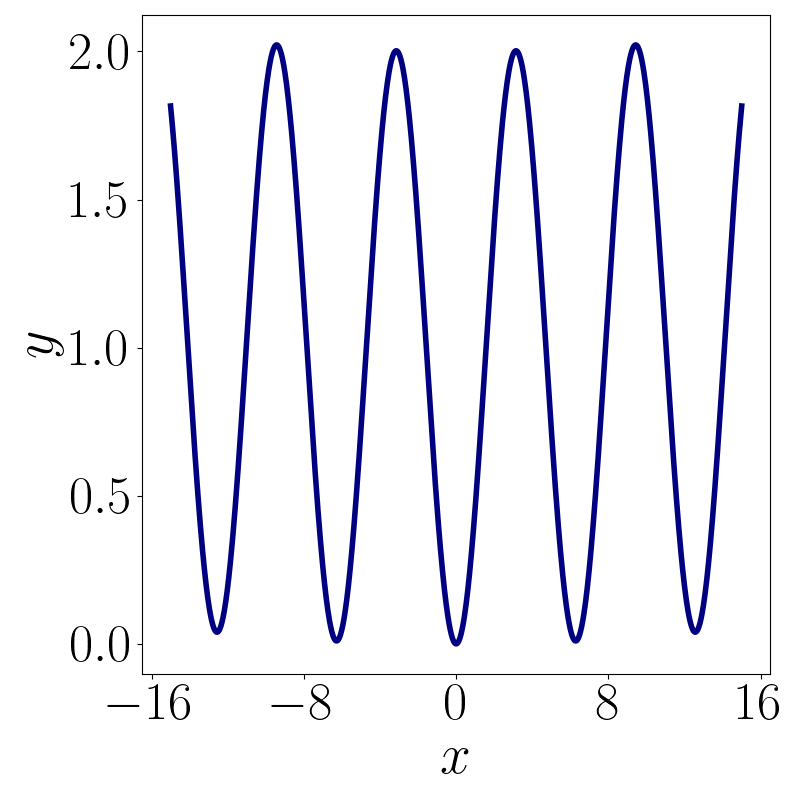}
					\caption{Griewank.}
					\label{subfig:griewank1d}
				\end{subfigure}
				\begin{subfigure}{0.24\textwidth}
					\includegraphics[width=\textwidth]{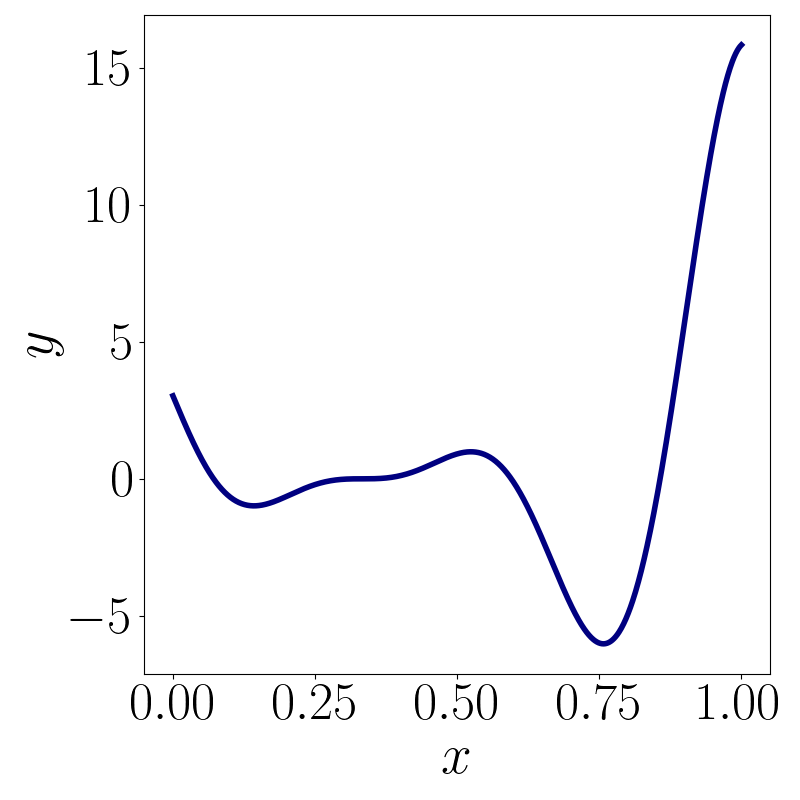}
					\caption{Forrester.}
					\label{subfig:forrester1d}
				\end{subfigure}
				\caption{One-dimensional analytical benchmark functions used to validate the proposed Bayesian framework for inverse problems. The suite includes representative challenges such as smooth nonlinearity, multimodal structure, and fine-scale variability, allowing systematic assessment of surrogate modeling and parameter inference capabilities.}
				\label{fig:Benchmark}
			\end{figure}

			\noindent
			The present section validates the proposed Bayesian framework (\Cref{sec:inverse-design}) using the one-dimensional analytical benchmark functions illustrated in \Cref{fig:Benchmark}. 
			The selected benchmarks capture diverse levels of functional complexity, including strong nonlinearity, multimodality, and high-frequency oscillations, and are widely adopted to assess the performance of surrogate modeling strategies and parameter inference in inverse problems~\cite{stuart2010inverse, idier2013bayesian, dashti2013bayesian, cotter2009bayesian, fitzpatrick1991bayesian}.
			
			The benchmark suite includes four representative functions:
			
			\begin{itemize}
				\item[-] \emph{Mixed Gaussian-Periodic Function}:
				\begin{equation}
					y = \exp\left(-\frac{(x - 2)^2}{2}\right) + 0.9 \exp\left(-\frac{(x + 5)^2}{20}\right) - 0.1 \cos(2x) \quad \text{with} \quad x \in [-10, 10]
				\end{equation}
				combining localized Gaussian peaks with periodic oscillations, resulting in both global smoothness and local high-frequency variations (\Cref{subfig:mixed1d}).
				
				\item[-] \emph{Lévy Function}:
				\begin{equation}
					y = \left( \sin(\pi w) \right)^2 + (w - 1)^2 \left[ 1 + 10 \left( \sin(\pi w + 1) \right)^2 \right] + (w - 1)^4 \left( \sin(2\pi w) \right)^2
				\end{equation}
				where \( w = 1 + \frac{x-1}{4} \) and \( x \in [-6,6] \). 
				The Lévy function presents a rugged, multimodal landscape with many local minima (\Cref{subfig:levy1d}), posing challenges to both global exploration and surrogate model fidelity.
				
				\item[-] \emph{Griewank Function}:
				\begin{equation}
					y = \frac{x^2}{4000} - \cos(x) + 1 \quad \text{with} \quad x \in [-15, 15]
				\end{equation}
				combining a slow quadratic trend with fine-scale periodic modulation (\Cref{subfig:griewank1d}).
				
				\item[-] \emph{Forrester Function}:
				\begin{equation}
					y = (6x - 2)^2 \sin(12x - 4) \quad \text{with} \quad x \in [0, 1]
				\end{equation}
				characterized by smooth global behaviour with sharply localized nonlinearities (\Cref{subfig:forrester1d}).
			\end{itemize}
			
				\subsubsection{Surrogate Modeling via Bayesian Optimization}
				\label{subsubsec:1Dbo}
				\noindent
				The present section assesses the surrogate modeling via BO capabilities of the proposed Bayesian framework (\Cref{sec:inverse-design}) on the one-dimensional analytical benchmark functions introduced in \Cref{fig:Benchmark}, focusing primarily on the Mixed Gaussian–Periodic case (\Cref{fig:Benchmark}a). Since the surrogate modeling procedure is identical for the other benchmark functions (\Cref{fig:Benchmark}b–d), the analysis is limited to the Mixed Gaussian-Periodic benchmark function, while additional results are provided in~\ref{bo1d}. Final surrogate models for all benchmark functions are summarized at the end of the present section.
				
				Surrogate models are constructed via BO using GPs, with uncertainty-aware sampling guided by the UCB acquisition function (\Cref{eq:ucb}). An exploration parameter \( \kappa = 200 \) is used to enforce a strongly exploratory sampling strategy, promoting exploration in regions of high epistemic uncertainty as discussed in \Cref{subsec:bobo,subsec:efficient}. 
				
				The performance of the surrogate models is evaluated in terms of both predictive accuracy and sampling efficiency. The proposed GP-based approach is compared against classical deterministic alternatives: Lagrangian Polynomial (LP) interpolation, Legendre Expansion (LE), and Cubic Spline (CS) interpolation~\cite{werner1984polynomial, alpert1991fast, mckinley1998cubic, knott1999interpolating, najm2009uncertainty}. The two covariance kernels are tested for the GP surrogate models: the RBF kernel (\Cref{eq:rbf}) with length scale \( \ell = 1.0 \) and the Matérn kernel (\Cref{eq:matern}) with smoothness parameter \( \nu = 5/2 \) and the same length scale. All surrogate models are trained on datasets derived from the high-fidelity model, and their accuracy is quantified using the MSE (\Cref{eq:mse}) computed over an independent validation set of \( n_{\text{val}} = 1000 \) samples, which ensures convergence of the reported metrics.
				
\begin{figure}[!ht]
					\centering
					\begin{subfigure}{0.33\textwidth}
						\includegraphics[width=\textwidth]{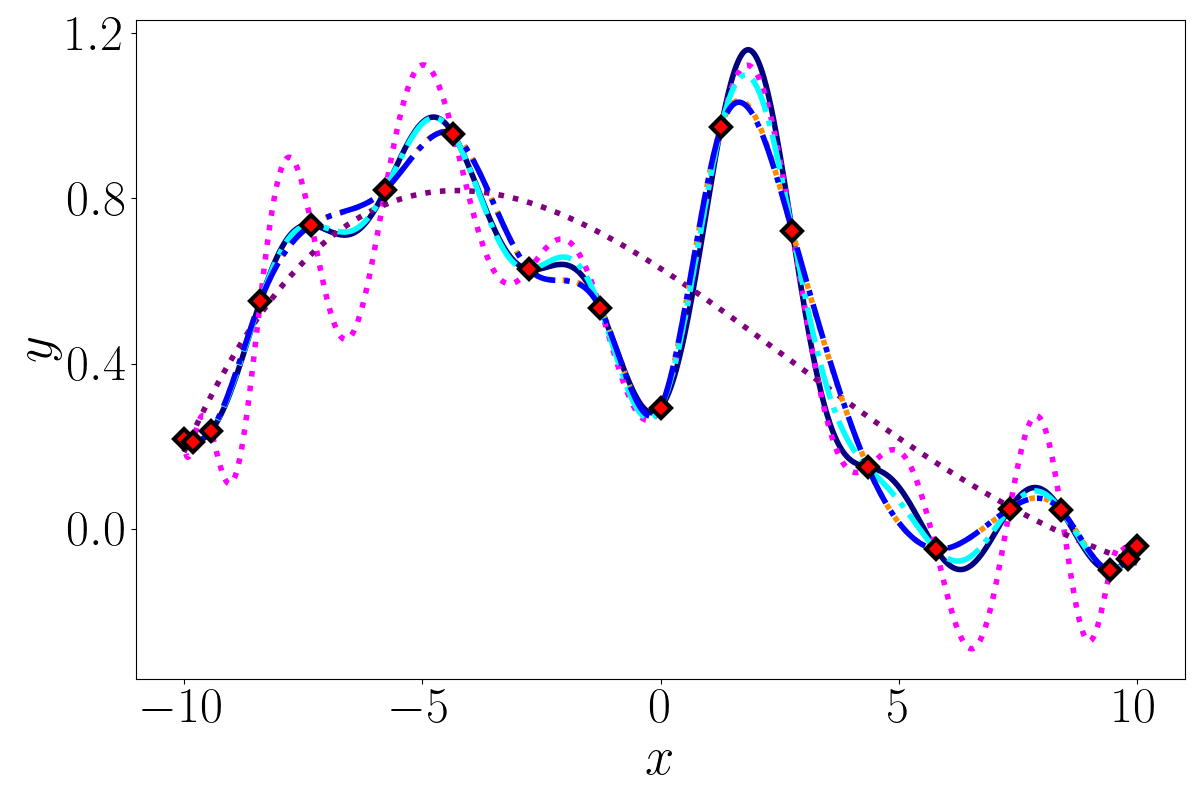}
						\caption{Surrogate model predictions.}
						\label{subfig:mixedCURV}
					\end{subfigure}
					\begin{subfigure}{0.33\textwidth}
						\includegraphics[width=\textwidth]{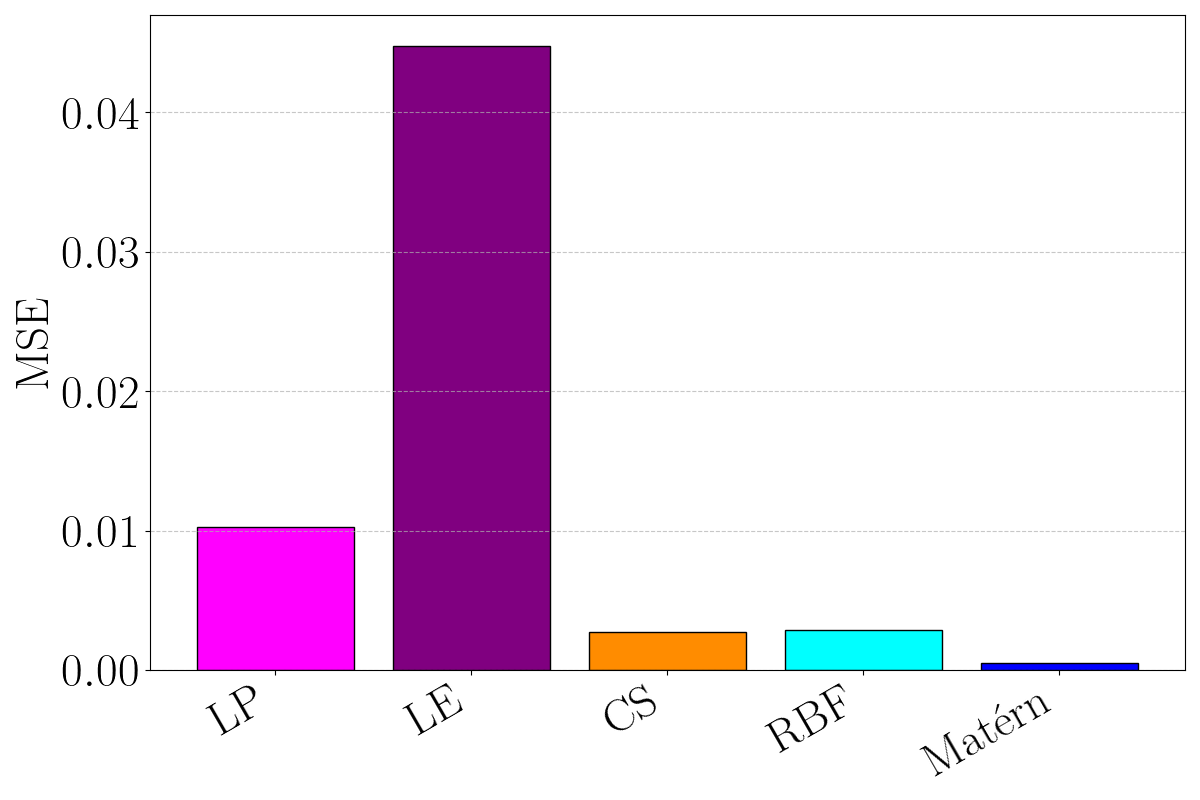}
						\caption{Comparison of MSE.}
						\label{subfig:mixedHIST}
					\end{subfigure}
					\begin{subfigure}{0.31\textwidth}
						\includegraphics[width=\textwidth]{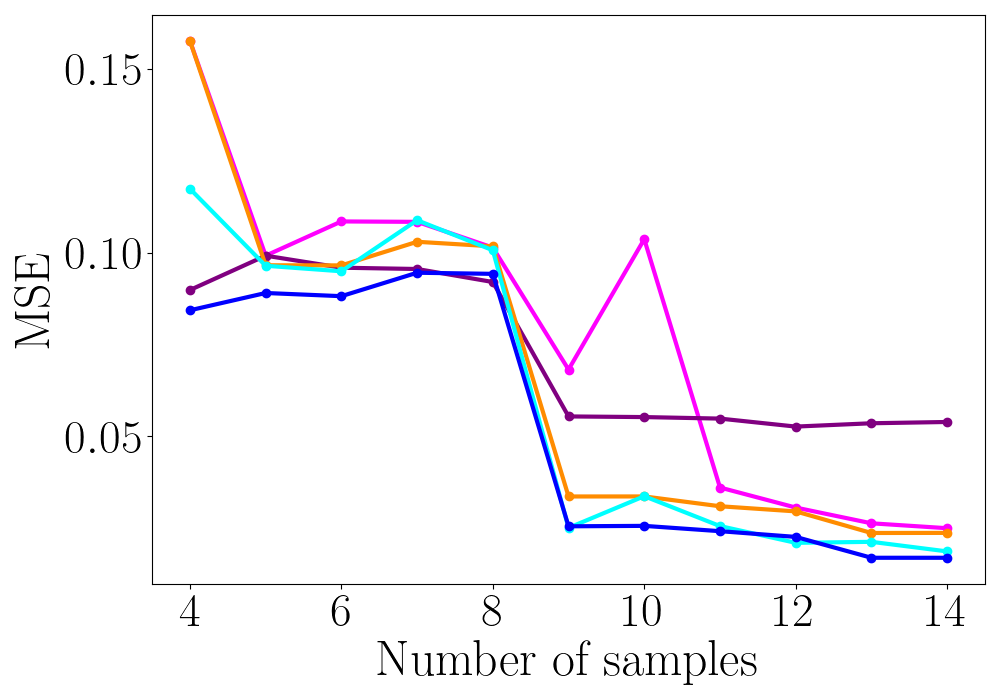}
						\caption{Evolution of MSE.}
						\label{subfig:mixedITER}
					\end{subfigure}
					\caption{Performance comparison of surrogate models for the Mixed Gaussian–Periodic benchmark function.
						(a) Predictive responses obtained with different surrogate models. 
						The solid black line denotes the high-fidelity model; dotted lines correspond to LP (fuchsia), LE (purple), and CS (orange) deterministic alternatives; cyan and blue denote GP-based surrogate models with Matérn and RBF kernels, respectively. Red diamonds indicate number of samples.
						(b) MSE values for 14 number of samples.
						(c) MSE evolution with increasing number of samples.}
					\label{fig:BenchmarkCURV}
				\end{figure}

				\Cref{fig:BenchmarkCURV} illustrates the comparative performance of the surrogate models. As shown in \Cref{fig:BenchmarkCURV}a, GP-based surrogate models successfully capture the multimodal and fine-scale structure of the Mixed Gaussian–Periodic benchmark function, with the Matérn kernel exhibiting enhanced robustness in highly oscillatory regions due to its flexible covariance structure. Among deterministic alternatives, CS interpolation performs reasonably well in smooth regions but lacks the ability to resolve localized nonlinearities. LP interpolation suffers from Runge-type instabilities at the boundaries~\cite{fornberg2007runge}, whereas LE maintains numerical stability but has limited representational capacity. Quantitative comparison via MSE (\Cref{fig:BenchmarkCURV}b) confirms the superior accuracy of GP-based surrogate models, with the Matérn kernel consistently yielding the lowest prediction error. The same conclusions are obtained from the comparison of the convergence behaviour of the surrogate models with increasing number of samples (\Cref{fig:BenchmarkCURV}c).
				
				A key feature of GPs is their ability to provide predictive uncertainty estimates alongside mean predictions. This allows for adaptive sampling strategies focused on areas of high epistemic uncertainty, as formalized by the update rule in \Cref{eq:dataset_update}. 
				
\begin{figure}[!ht]
					\centering
					\begin{subfigure}[b]{0.24\textwidth}
						\includegraphics[width=\textwidth]{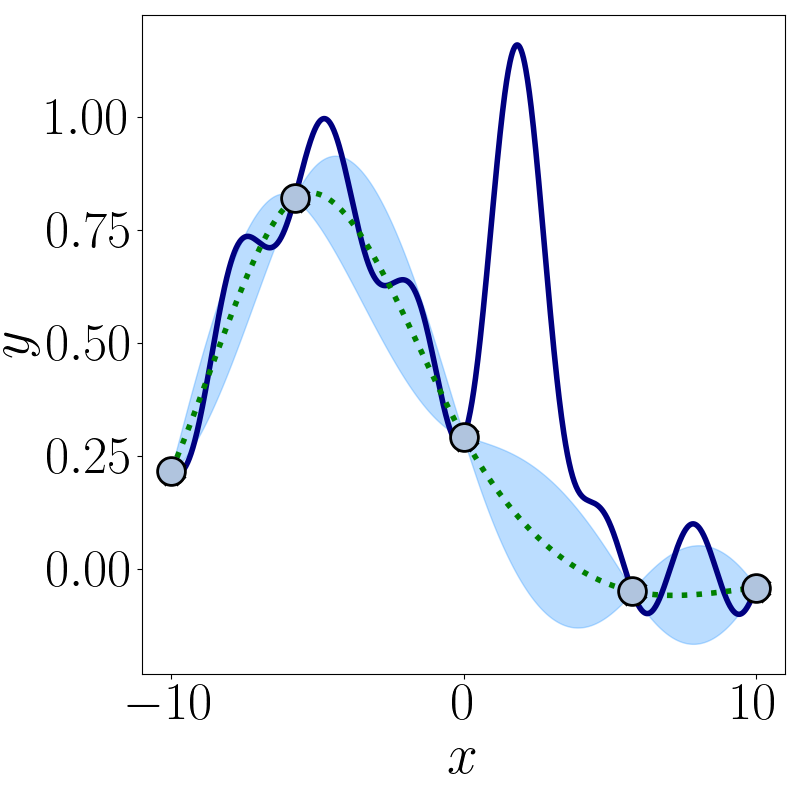}
						\caption{5 number of samples.}
					\end{subfigure}
					\begin{subfigure}[b]{0.24\textwidth}
						\includegraphics[width=\textwidth]{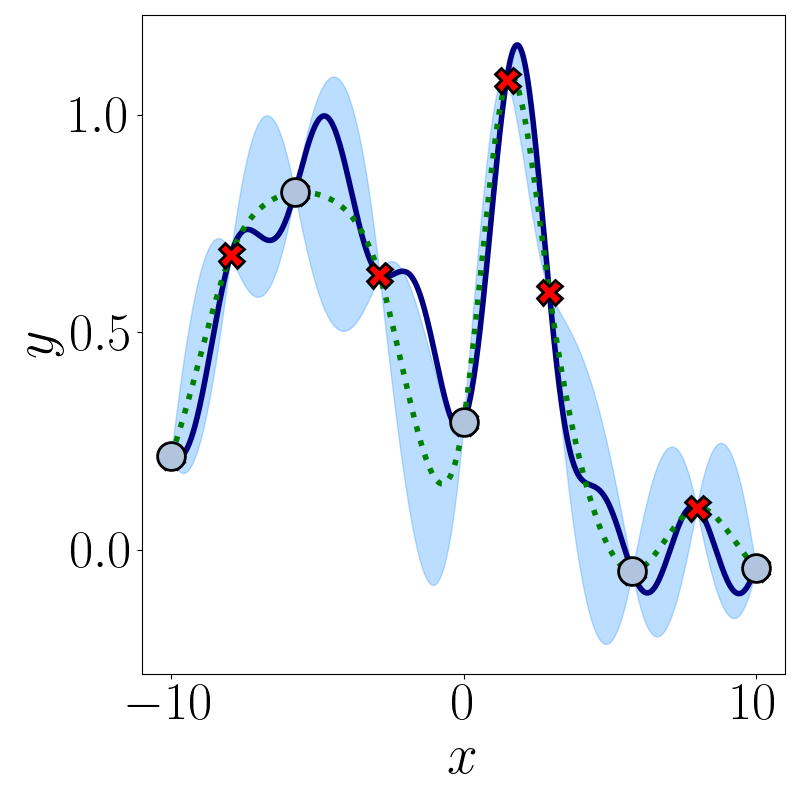}
						\caption{10 number of samples.}
					\end{subfigure}
					\begin{subfigure}[b]{0.24\textwidth}
						\includegraphics[width=\textwidth]{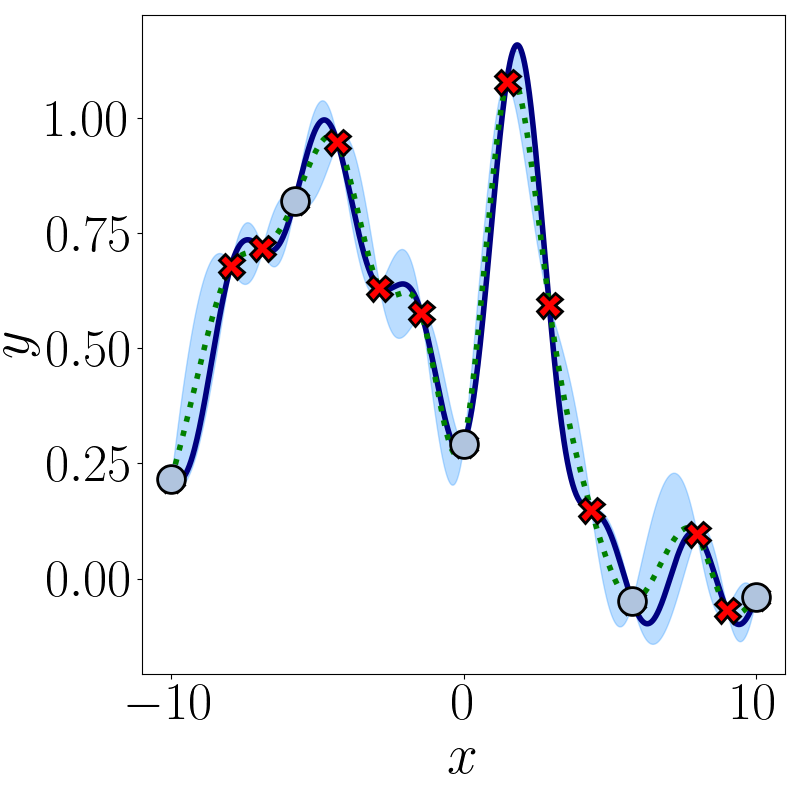}
						\caption{15 number of samples.}
					\end{subfigure}
					\begin{subfigure}[b]{0.24\textwidth}
						\includegraphics[width=\textwidth]{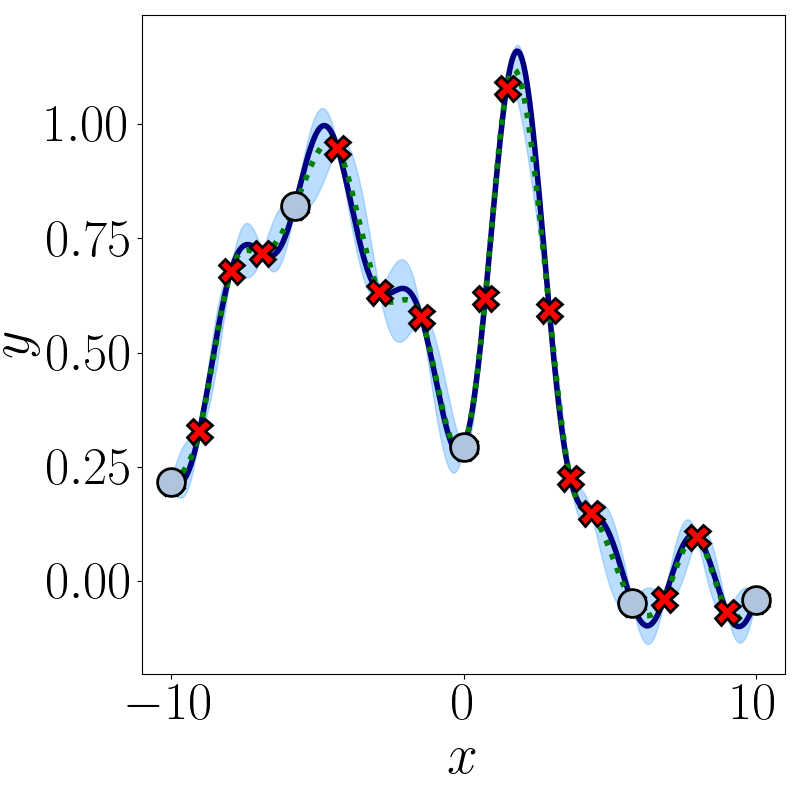}
						\caption{19 number of samples.}
					\end{subfigure}
					\caption{Progressive refinement of the GP surrogate model for the Mixed Gaussian–Periodic benchmark function using the UCB acquisition function.
						Blue solid line: high-fidelity model; green dotted line: GP mean (Matérn); light blue band: 95\% confidence interval. Gray circles: initial number of samples; red crosses: acquired new number of samples.}
					\label{fig:adaptivity}
				\end{figure}

				\Cref{fig:adaptivity} illustrates the progressive refinement of the GP surrogate model for the Mixed Gaussian–Periodic benchmark function. Starting from a sparse initial dataset with 5 number of samples, the surrogate model is iteratively improved by acquiring new number of samples in uncertain regions. This results in rapid uncertainty reduction and improved predictive accuracy, especially in regions with localized variation. The behaviour clearly demonstrates the advantages of uncertainty-aware sampling in surrogate modeling under limited number of samples.
				
\begin{figure}[!ht]
					\centering
					\begin{subfigure}[b]{0.24\textwidth}
						\includegraphics[width=\textwidth]{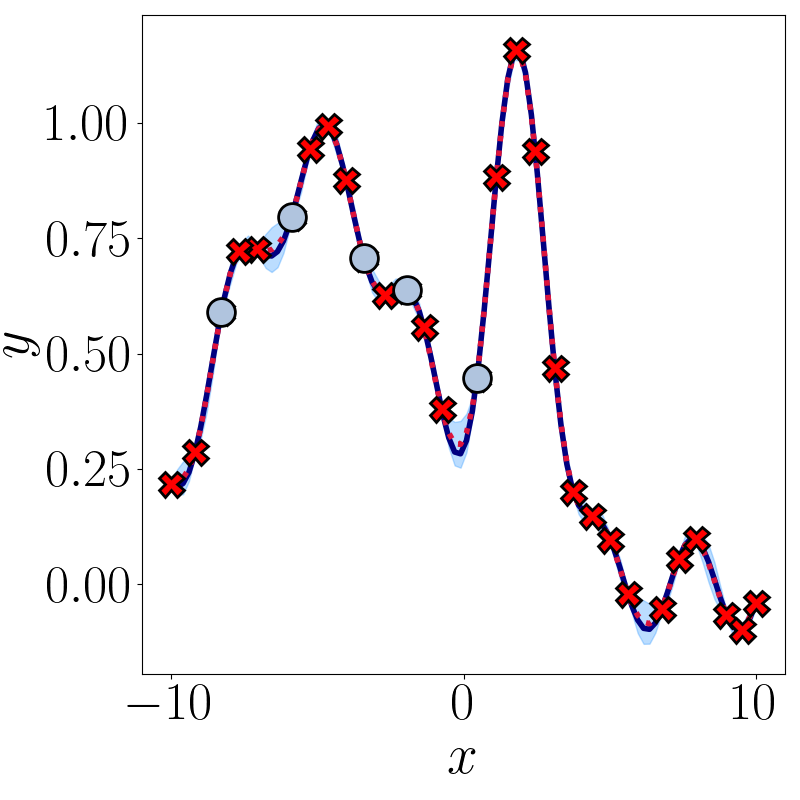}
						\caption{Mixed Gaussian–Periodic.}
					\end{subfigure}
					\begin{subfigure}[b]{0.24\textwidth}
						\includegraphics[width=\textwidth]{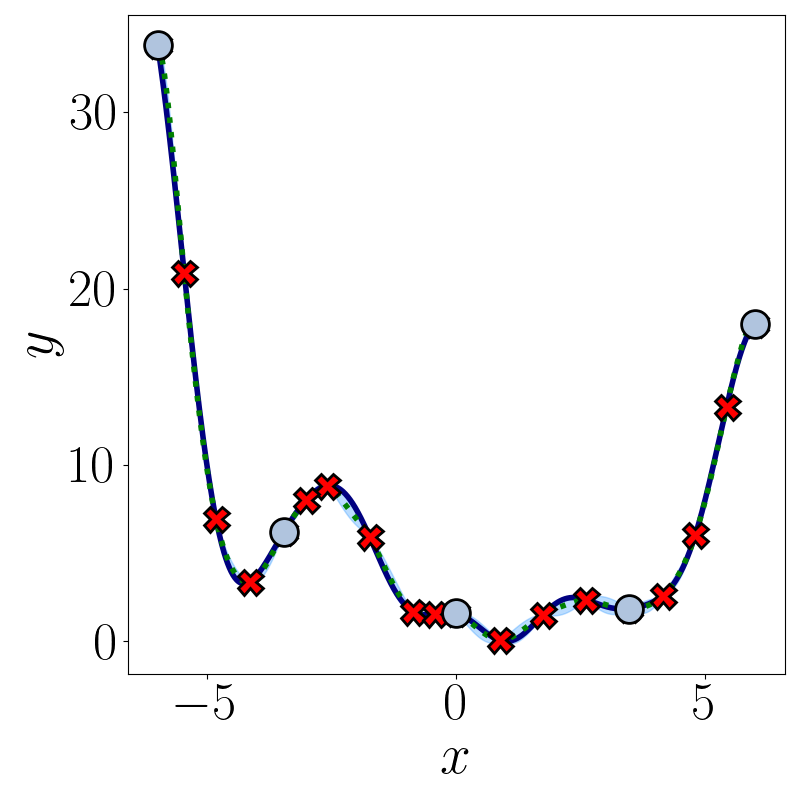}
						\caption{Lévy.}
					\end{subfigure}
					\begin{subfigure}[b]{0.24\textwidth}
						\includegraphics[width=\textwidth]{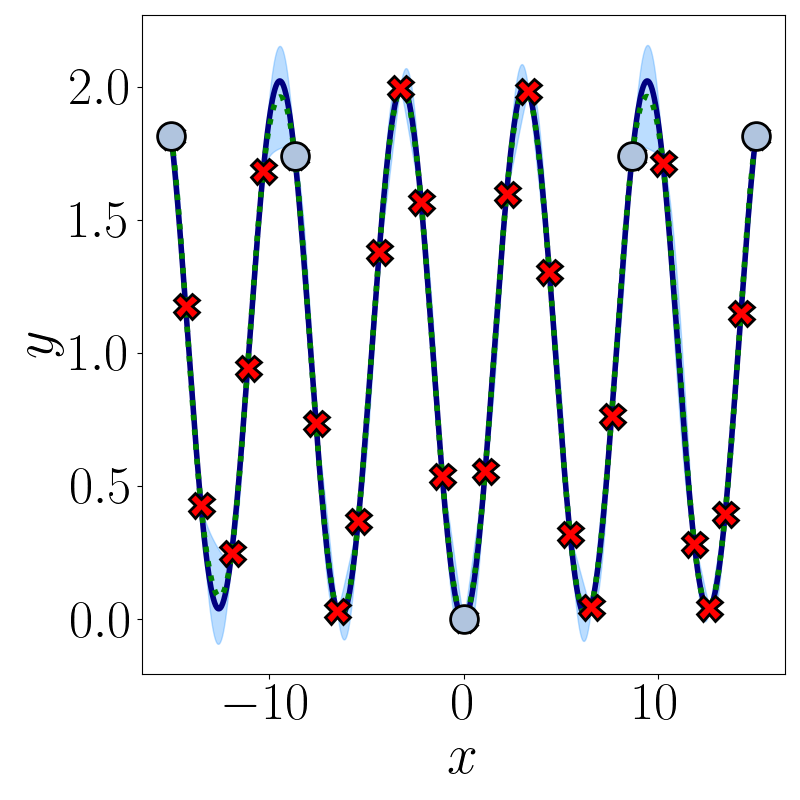}
						\caption{Griewank.}
					\end{subfigure}
					\begin{subfigure}[b]{0.24\textwidth}
						\includegraphics[width=\textwidth]{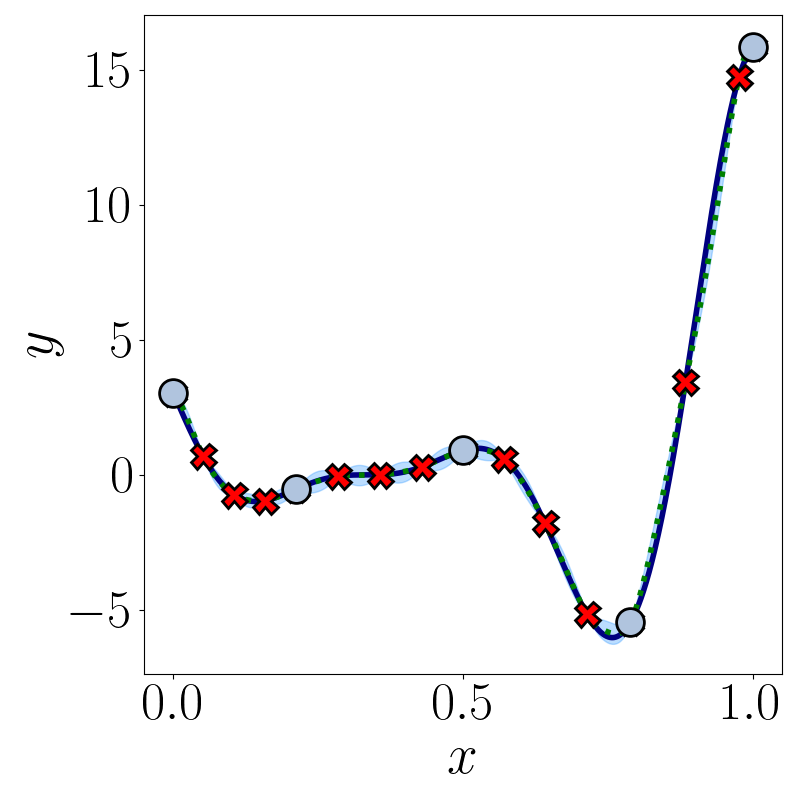}
						\caption{Forrester.}
					\end{subfigure}
					\caption{Final surrogate models for all benchmark functions. Blue solid line: high-fidelity; red dotted line: GP (Matérn); light blue band: predictive uncertainty. Gray circles: initial number of samples; red crosses: acquired new number of samples.}
					\label{fig:bo_conclusion}
				\end{figure}

				Final surrogate models obtained for all benchmark functions are shown in \Cref{fig:bo_conclusion}. The configuration settings used across all benchmark functions are reported in \Cref{tab:surrogate1D}. Despite the diversity in functional complexity — from multimodal behaviour (Mixed Gaussian–Periodic, Lévy), to oscillatory patterns (Griewank), and localized nonlinearities (Forrester) — the proposed framework consistently delivers accurate approximations using a limited number of samples \mihi{within the considered one-dimensional benchmark setting}.
				
				\begin{table}[!ht]
					\centering
					\caption{Configuration settings for surrogate modeling across benchmark functions.}
					\label{tab:surrogate1D}
					\begin{tabular}{ll}
						\hline
						\textbf{Parameter} & \textbf{Value} \\
						\hline
						Initial samples ($n_{\text{init}}$) & 5 \\
						Acquired new samples ($n_{\text{acq}}$) & $15\text{--}20$ \\
						Surrogate model & Gaussian Process  (GP) \\
						Covariance kernel & Matérn $\nu = 5/2$ \\
						Mean function & Zero mean prior \\
						Noise model & Gaussian noise ($\sigma^2 = 10^{-6}$) \\
						Hyperparameters optimization & Log-marginal likelihood maximization \\
						Acquisition function & Upper Confidence Bound (UCB) \\
						UCB exploration parameter & $\kappa = 200$ \\
						Stopping criterion & Validation MSE $< 10^{-3}$ \\
						Validation ($n_{val}$) set & 1000 number of samples \\
						\hline
					\end{tabular}
				\end{table}
				
				The results confirm the accuracy, robustness, and sample efficiency of the proposed surrogate modeling strategy, establishing a reliable foundation for the subsequent parameter inference tasks \mihi{within the scope of low-dimensional benchmark problems}. As anticipated, GP-based surrogate models consistently outperform classical deterministic alternatives, particularly in low-data regimes and in the presence of sharp nonlinearities or oscillatory structures. These outcomes align with theoretical expectations and are presented here to justify the methodological choices made in the development of the proposed Bayesian framework.
				
				In conclusion, it is worth noting that the UCB acquisition function employed in the study is primarily designed for global exploration, aiming to reduce epistemic uncertainty uniformly across the entire input parameter space. While this strategy enables the construction of accurate and well-calibrated surrogate models for uncertainty-aware inference, it may allocate samples to regions not directly relevant to inverse problems.
				In contrast, acquisition functions such as EI (\Cref{eq:ei}) are tailored for local exploitation and are often preferred when the quantity of interest is known a priori. However, in the present context — where the primary goal is to construct a globally accurate surrogate model that faithfully reproduces the high-fidelity response across the parameter space — the use of UCB reflects an emphasis on global model fidelity and robustness in uncertainty quantification.
				
				\subsubsection{Parameter Inference via Bayesian Inversion}
				\label{subsubsec:1Dbi}
				\noindent
				The present section evaluates the capability of the proposed Bayesian framework (\Cref{sec:inverse-design}) to perform parameter inference and quantify epistemic uncertainty via BI for the one-dimensional analytical benchmark functions illustrated in \Cref{fig:Benchmark}. 
				As outlined in \Cref{subsec:bobi}, the objective is to identify the input parameters $\mathbf{x} \in \mathbb{R}^d$ that most likely reproduce a prescribed observed quantity of interest $\bar{y} \in \mathbb{R}^m$, with the MAP estimates defined in \Cref{eq:map_estimation1}. 
				Assuming Gaussian noise and a uniform prior distribution, this task reduces to the minimization of the LS functional (\Cref{eq:map_ls}), evaluated through the surrogate model $f_{\mathcal{GP}}(\mathbf{x})$ introduced in \Cref{eq:gp} and constructed in \Cref{fig:bo_conclusion}. 
				The surrogate models serve as computationally efficient proxies for the high-fidelity model, thereby enabling a systematic and reproducible validation of the inference procedure \mihi{within a controlled low-dimensional benchmark setting}.
				
				We begin by presenting detailed results for the Mixed Gaussian–Periodic benchmark function (\Cref{subfig:mixed1d}) and the Forrester benchmark function (\Cref{subfig:forrester1d}), which collectively illustrate the main features of the proposed methodology. 
				Additional results for the remaining benchmark functions are reported in~\ref{bi1d}.
				
\begin{figure}[!ht]
					\centering
					\begin{subfigure}[b]{0.33\textwidth}
						\centering
						\includegraphics[width=\textwidth]{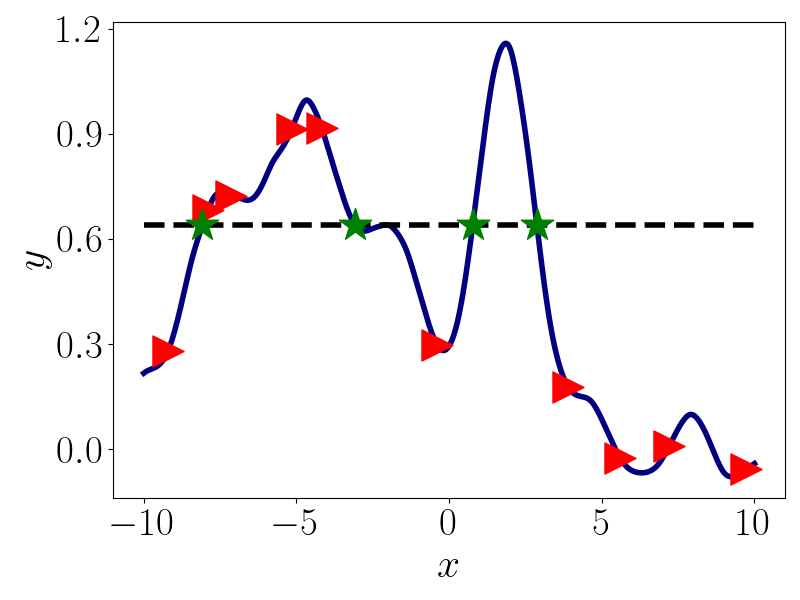}
						\caption{MAP estimates}
						\label{subfig:mapMix}
					\end{subfigure}
					\begin{subfigure}[b]{0.32\textwidth}
						\centering
						\includegraphics[width=\textwidth]{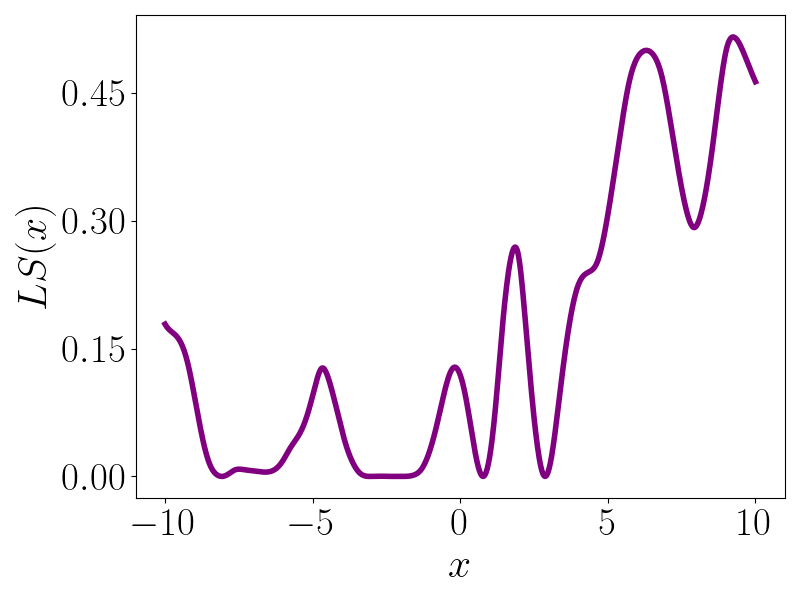}
						\caption{LS profile}
						\label{subfig:LSMix}
					\end{subfigure}
					\begin{subfigure}[b]{0.32\textwidth}
						\centering
						\includegraphics[width=\textwidth]{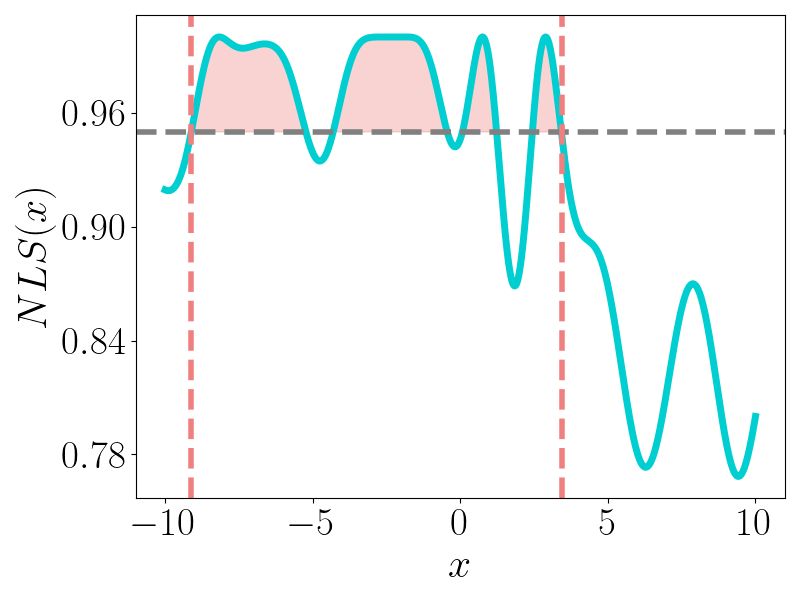}
						\caption{NLS profile}
						\label{subfig:NLSMix}
					\end{subfigure}
					\caption{Parameter inference via BI results for the Mixed Gaussian-Periodic benchmark function.
						(a) MAP estimates obtained from multiple optimization runs. The red triangles denote the multiple initial guess; the observed quantity of interest is indicated with the black dotted horizontal line, while the green stars denote the optimal (MAP estimate) solution. 
						(b) LS profile across the parameter space. 
						(c) NLS profile representing the posterior distribution with the high-probability region marked by the grey dotted horizontal line representing the 0.95-threshold and highlighted by the coral-coloured portion delimited by the dotted vertical lines.
					}
					\label{fig:inverseMixed}
				\end{figure}

				\Cref{fig:inverseMixed} summarizes the parameter inference via BI results for the Mixed Gaussian-Periodic benchmark function (\Cref{fig:Benchmark}a).
				\Cref{subfig:mapMix} illustrates the optimization process, where multiple initial guesses (red triangles) are distributed randomly over the input parameter space to ensure a robust exploration of the parameter landscape.
				The optimization is performed using a limited-memory BFGS algorithm with box constraints (L-BFGS-B), with a maximum of 400 iterations allowed per run. 
				The observed quantity of interest $\bar{y} = 0.63$ is indicated by the black dotted horizontal line, while the green stars denote the optimal solutions $\mathbf{x}_{\text{MAP}}$ identified by minimizing the LS functional.
				
				The results reported in \Cref{fig:inverseMixed}a highlight the presence of four distinct parameter configurations that minimize the LS functional, each corresponding to an independent MAP estimate. 
				The MAP estimates are derived from multiple optimization runs initialized at randomly distributed initial guesses across the parameter space, and collectively indicate the existence of several admissible configurations capable of reproducing the observed quantity of interest. 
				Such multiplicity of MAP estimates reflects the ill-posed nature of the inverse problem and emphasizes the intrinsic multimodality of the underlying LS profile (\Cref{subfig:LSMix}), induced by the highly nonlinear and oscillatory behavior of the Mixed Gaussian-Periodic benchmark function.
				
				The multimodal character of the inverse problem is further elucidated by inspecting the approximation of the posterior distribution, represented by the NLS profile in \Cref{fig:inverseMixed}c. 
				The presence of multiple local maxima confirms that the posterior distribution concentrates around different regions of the parameter space, each associated with a distinct mode of high likelihood. 
				Therefore, the posterior distribution cannot be reliably approximated by a single Gaussian mode --- as typically assumed in Laplace-based approaches (see \Cref{eq:mapsig}) --- due to the strong non-convexity and the presence of multiple solutions~\cite{chiappetta2023sparse}. 
				Any attempt to enforce a unimodal Gaussian approximation would inevitably lead to a severe underestimation of the epistemic uncertainty and a mischaracterization of the parameter landscape.
				
				Despite these challenges, the NLS functional offers a valuable approximation for assessing the structure of the posterior distribution. 
				In particular, the decay behavior of the NLS profile indicates the presence of high-probability regions in the parameter space. 
				As shown in \Cref{fig:inverseMixed}c, introducing a threshold level \( \mathrm{NLS}(x) = 0.95 \), corresponding to the top 5\% probability level set, allows the identification of a reduced subdomain that preserves the most plausible parameter estimates highlighted by the coral region bounded by the two dashed vertical lines. 
				While the original problem is defined on the entire parameter interval \( [-10, 10] \) (\Cref{subfig:mixed1d}), the high-probability region lies within \( [-7.5, 5] \), resulting in a significant reduction of the admissible parameter space \mihi{for this one-dimensional inverse setting}.
				
\begin{figure}[!ht]
					\centering
					\begin{subfigure}[b]{0.3\textwidth}
						\centering
						\includegraphics[width=\textwidth]{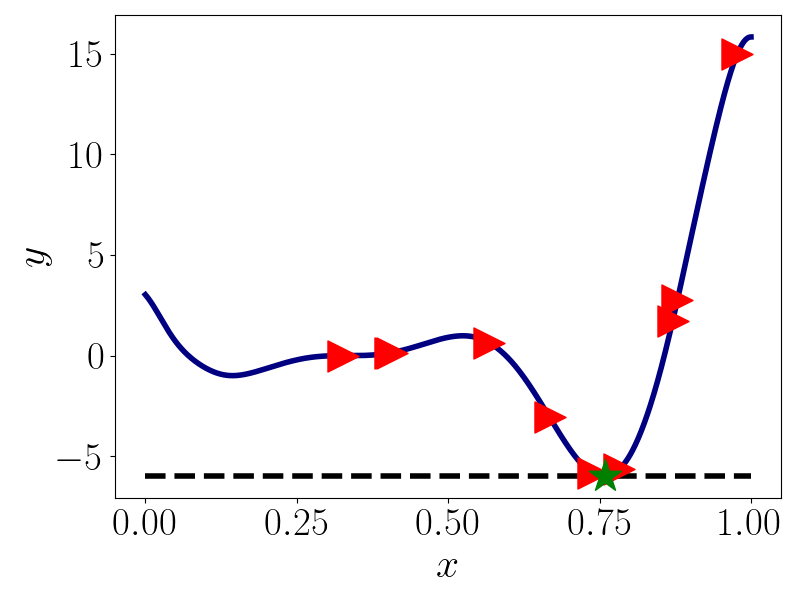}
						\caption{MAP estimates}
						\label{subfig:mapForrester}
					\end{subfigure}
					\begin{subfigure}[b]{0.3\textwidth}
						\centering
						\includegraphics[width=\textwidth]{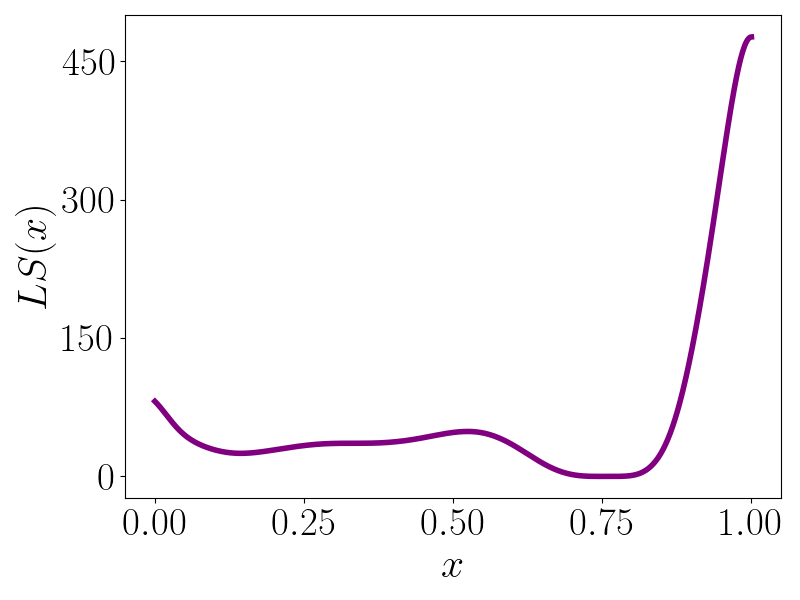}
						\caption{LS profile}
						\label{subfig:LSForrester}
					\end{subfigure}
					\begin{subfigure}[b]{0.3\textwidth}
						\centering
						\includegraphics[width=\textwidth]{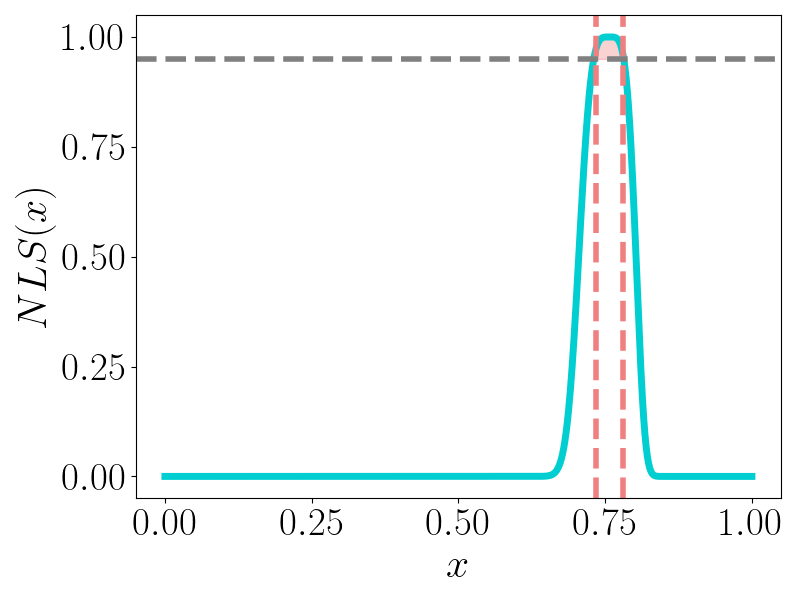}
						\caption{NLS profile}
						\label{subfig:NLSForrester}
					\end{subfigure}
					\caption{Parameter inference via BI results for the Forrester benchmark function.
						(a) MAP estimates obtained from multiple optimization runs. The red triangles denote the multiple initial guess; the observed quantity of interest is indicated with the black dotted horizontal line, while the green stars denote the optimal (MAP estimate) solution. 
						(b) LS profile across the parameter space. 
						(c) NLS profile representing the posterior distribution approximation with the high-probability region marked by the grey dotted horizontal line representing the 0.95-threshold and highlighted by the coral-coloured portion delimited by the dotted vertical lines.}
					\label{fig:inverseForrester}
				\end{figure}

				\Cref{fig:inverseForrester} presents the results of the parameter inference via BI procedure for the Forrester benchmark function, introduced in \Cref{fig:Benchmark}d. 
				\Cref{subfig:mapForrester} shows the MAP estimate obtained from the LS minimization with an observed quantity of interest \(\bar y = -6.02\), while \Cref{subfig:LSForrester} and \Cref{subfig:NLSForrester} show the corresponding LS and NLS profiles across the parameter space.
				
				The LS profile for the Forrester function exhibits a single well-defined global minimum, as clearly visible in \Cref{subfig:LSForrester}. 
				The optimization procedure consistently identifies the same MAP estimate, indicating a well-posed and uniquely identifiable inverse problem. 
				The corresponding NLS profile (\Cref{subfig:NLSForrester}) shows a sharp, unimodal peak centered at \( \approx 0.76 \), with a narrow standard deviation of approximately \( \approx 0.02 \), confirming the strong concentration of the posterior distribution around the inferred solution.
				
				The Gaussian-like shape of the NLS profile representing the posterior distribution approximation and the absence of secondary modes suggest that the epistemic uncertainty associated with the inferred parameter is minimal, and that the NLS provides a robust and confident MAP estimate of the input parameters. 
				Similar to the Mixed Gaussian-Periodic case (\Cref{subfig:NLSMix}), the high-probability region is indicated by the black horizontal dotted line and highlighted by the coral-coloured portion delimited by the dotted vertical lines, as shown in \Cref{subfig:NLSForrester}. 
				The figure illustrates how, for the Forrester function, the admissible parameter space is drastically reduced: from the initial interval \([0,1]\) to the inferred high-probability region \( [0.735, 0.78] \), \mihi{demonstrating the effectiveness of BI-based parameter space reduction in a low-dimensional inverse problem}.
				
				\subsubsection{Discussion}
				\label{subsubsec:1Dres}
				\noindent
				The numerical validation results presented in Sections~\ref{subsubsec:1Dbo} and~\ref{subsubsec:1Dbi} highlight the computational and inferential benefits of the proposed Bayesian framework, which integrates surrogate modeling via BO with parameter inference via BI \mihi{in one-dimensional benchmark settings}. 
				\mihi{A complementary discussion further motivating the adopted parameter inference strategy is provided in~\ref{appendix:mcmc}.}
				
				The surrogate models constructed using uncertainty-aware sampling strategies accurately reproduce the key features of all one-dimensional analytical benchmarks — including multimodality (Mixed Gaussian–Periodic, Lévy), high-frequency oscillations (Griewank), and sharp localized nonlinearities (Forrester) — while requiring only a limited number of high-fidelity model evaluations. 
				For example, in the Mixed Gaussian–Periodic benchmark function, the GP surrogate model based on a Matérn kernel successfully captures the complex high-fidelity model with fewer than 20 number of samples (\Cref{fig:bo_conclusion}a). Similarly, the surrogate model for the Forrester benchmark function achieves high predictive accuracy with only 15 number of samples (\Cref{fig:bo_conclusion}d).
				
				Once trained, the surrogate models enable an efficient {application} of BI. The LS minimization for MAP estimation is performed directly on the GP surrogate model, eliminating the need for repeated high-fidelity model evaluations. 
				In the case of the Mixed Gaussian–Periodic benchmark function (\Cref{fig:inverseMixed}a), the LS minimization recovers multiple MAP estimates from independent optimization runs, revealing the multimodal structure of the inverse problem. 
				Although the multiplicity of local minima prevents the identification of a single optimal parameter configuration, the NLS profile (\Cref{fig:inverseMixed}c) still allows for an informative posterior distribution analysis by isolating the high-probability regions of the parameter space. 
				This effectively reduces epistemic uncertainty by constraining the admissible parameter space to a smaller, well-defined subregion, even in the absence of a unique MAP estimate.
				
				By contrast, the Forrester benchmark function illustrates a well-posed inverse problem with a uniquely identifiable MAP parameter configuration (\Cref{fig:inverseForrester}a). 
				The LS profile presents a single dominant minimum (\Cref{subfig:LSForrester}), and the corresponding NLS profile is sharply unimodal, with high posterior distribution concentration around the MAP estimate (\Cref{subfig:NLSForrester}), confirming the reliability of the parameter inference analysis.
				
				In both cases, the decoupling between surrogate model construction via BO and the parameter inference via BI yields a substantial computational advantage over traditional inversion strategies as anticipated in \Cref{tab:bo_bi_comparison}. 
				Directly applying BI on the high-fidelity model would require hundreds of high-fidelity evaluations per optimization run, especially in multimodal settings. 
				In contrast, operating on the surrogate model reduces the cost of each LS minimization to negligible levels, enabling rapid exploration of the parameter space and efficient uncertainty quantification \mihi{for the low-dimensional benchmarks considered in the present section}.

			\subsection{Two-Dimensional Benchmarks}
			\label{subsec:2D}
\begin{figure}[!ht]
				\centering
				\begin{subfigure}[b]{0.24\textwidth}
					\centering
					\includegraphics[width=\textwidth]{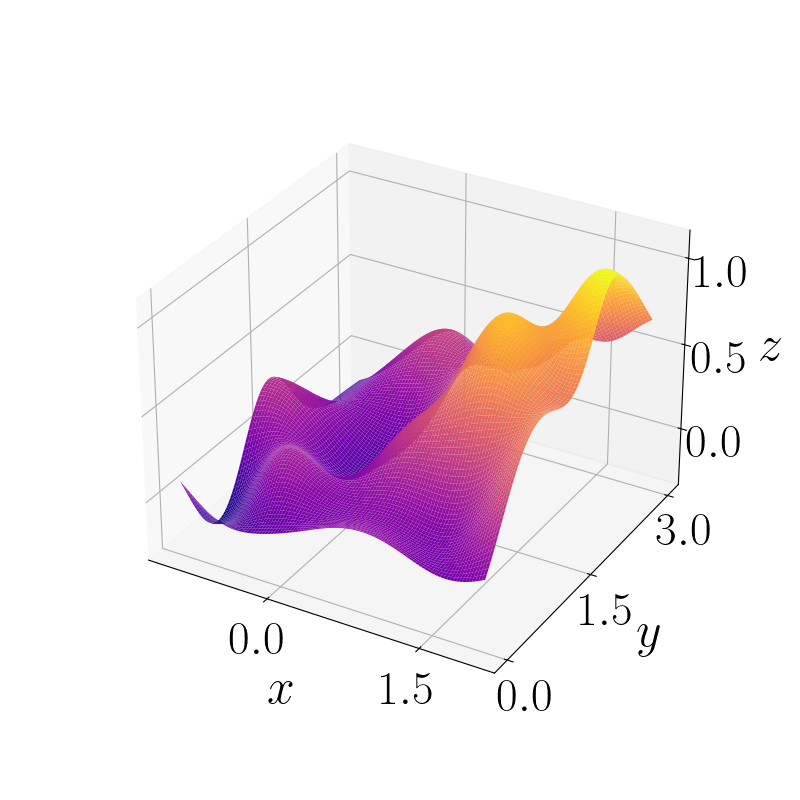}
					\caption{}
					\label{subfig:hfm_mix_surface}
				\end{subfigure}
				\begin{subfigure}[b]{0.24\textwidth}
					\centering
					\includegraphics[width=\textwidth]{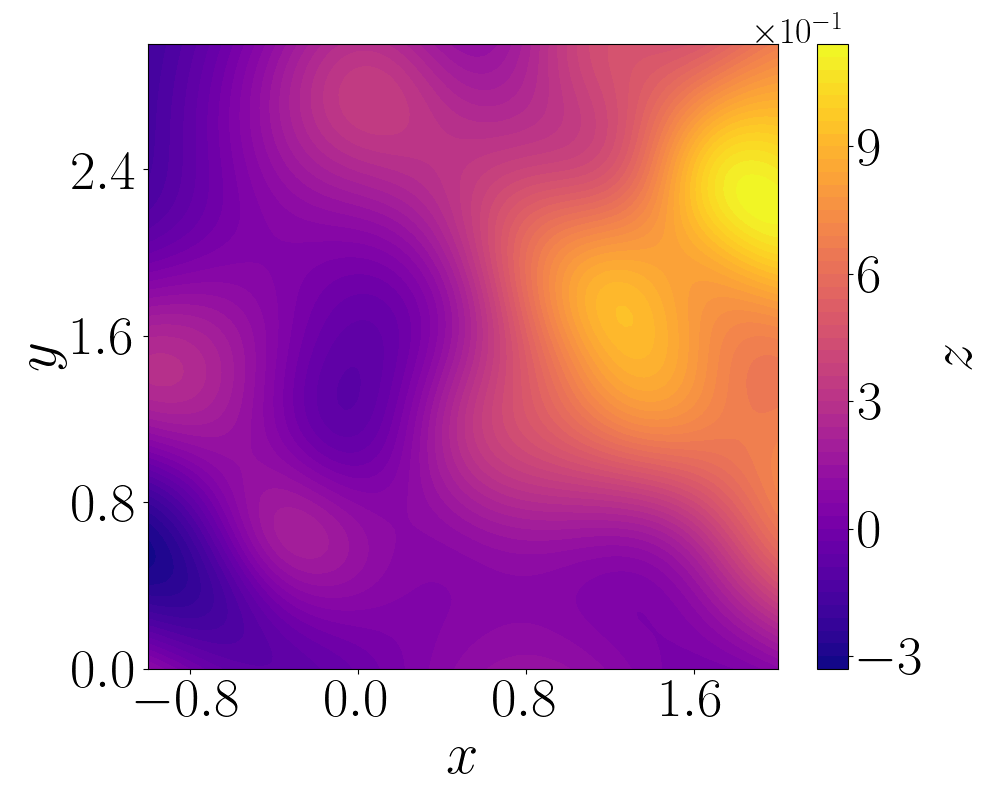}
					\caption{}
					\label{subfig:hfm_mix_contour}
				\end{subfigure}
				\begin{subfigure}[b]{0.24\textwidth}
					\centering
					\includegraphics[width=\textwidth]{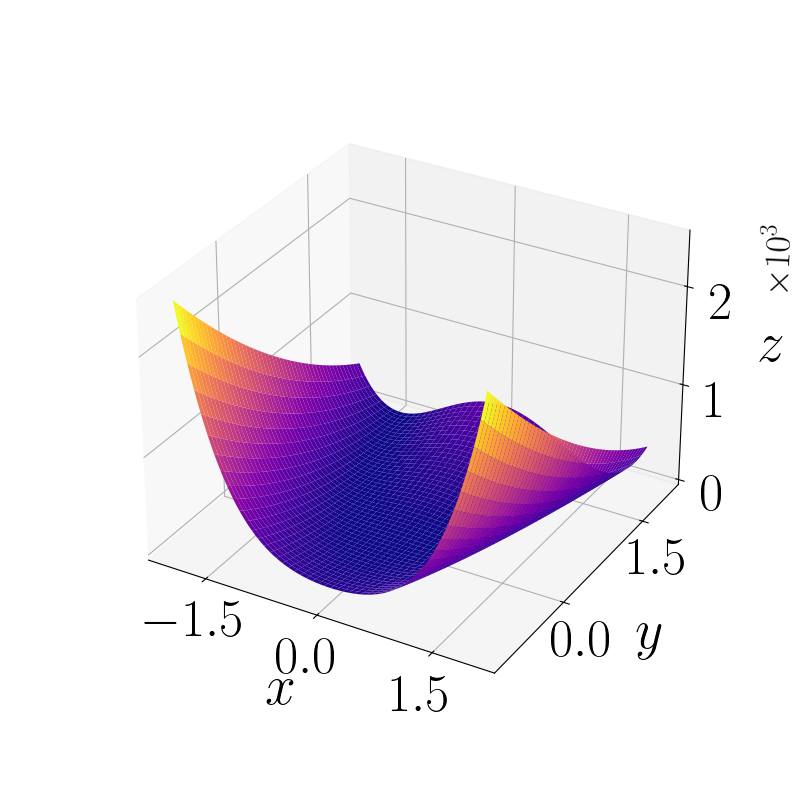}
					\caption{}
					\label{subfig:hfm_rosen_surface}
				\end{subfigure}
				\begin{subfigure}[b]{0.24\textwidth}
					\centering
					\includegraphics[width=\textwidth]{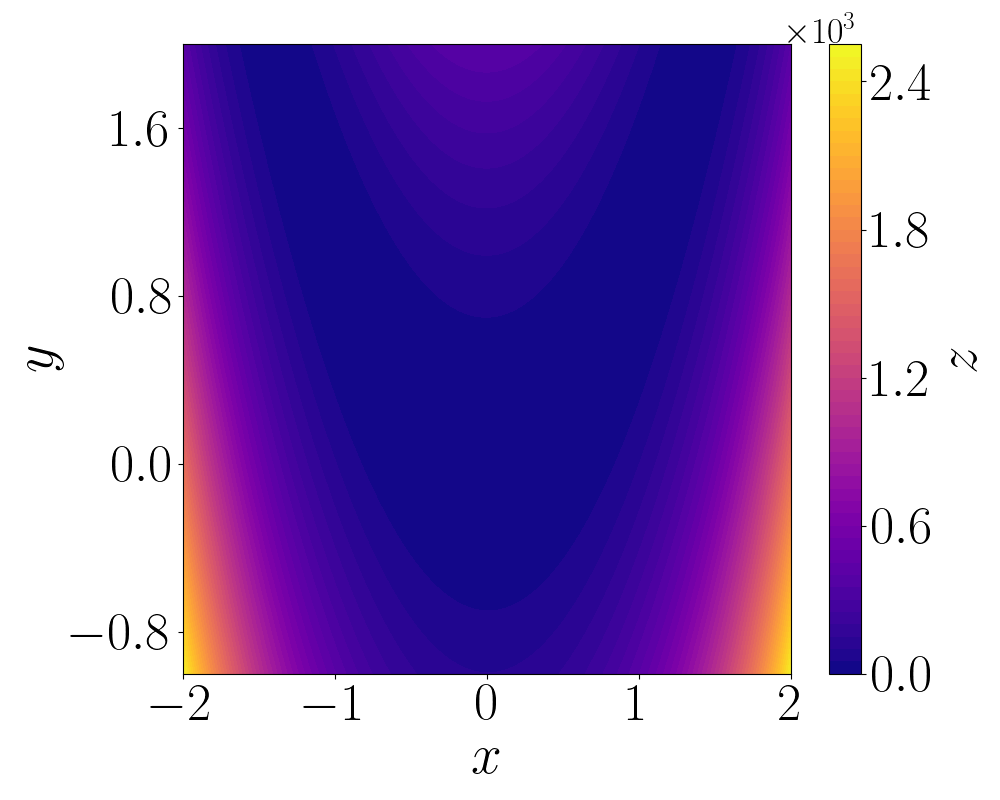}
					\caption{}
					\label{subfig:hfm_rosen_contour}
				\end{subfigure}
				\caption{
					Two-dimensional analytical benchmark functions used to validate the proposed Bayesian framework for inverse problems. 
					(a)–(b): Mixed Gaussian-Periodic function — high-fidelity model surface and contour plots. 
					(c)–(d): Rosenbrock function — high-fidelity model surface and contour plots.}
				\label{fig:hfm_2d}
			\end{figure}

			\noindent
			The proposed Bayesian framework for inverse problems, introduced in~\Cref{sec:inverse-design}, is further validated through two-dimensional analytical benchmark functions, shown in~\Cref{fig:hfm_2d}. 
			\mihi{These benchmarks are employed to investigate the framework’s behaviour in the presence of coupled parameters, increased problem complexity with respect to the two-dimensional setting, and non-trivial functional landscapes, while enabling a controlled and interpretable assessment of surrogate modeling and parameter inference under uncertainty.}
			
			The selected benchmark functions are:
			
			\begin{itemize}
				\item[-] \textit{Mixed Gaussian–Periodic function}, defined over the domain \( x \in [-1, 2] \), \( y \in [0, 3] \), as
				\begin{equation}
					\label{eq:2dmixed}
					f(x, y) = \exp\left( -\frac{(x - 2)^2 + (y - 2)^2}{2} \right) + 0.2 \cos(3x) \sin(3y) + 0.1 \sin(5x + 5y)
				\end{equation}
				combining a smooth Gaussian peak with periodic perturbations. This configuration introduces both multimodal behavior and localized nonlinearities, posing challenges to surrogate model construction and parameter inference accuracy. The high-fidelity model is illustrated in~\Cref{subfig:hfm_mix_surface} and~\Cref{subfig:hfm_mix_contour}.
				
				\item[-] \textit{Rosenbrock function}, defined over the domain \( x \in [-2, 2] \), \( y \in [-1, 2] \), as
				\begin{equation}
					\label{eq:rosenbrock}
					f(x, y) = (1 - x)^2 + 100(y - x^2)^2
				\end{equation}
				characterized by a narrow, curved valley and steep gradients. This benchmark is representative of ill-conditioned inverse problems with strong anisotropies and nonlinearity, challenging both the surrogate modeling phase and the parameter inference analysis. The associated high-fidelity response is shown in~\Cref{subfig:hfm_rosen_surface} and~\Cref{subfig:hfm_rosen_contour}.
			\end{itemize}			
			
				\subsubsection{Surrogate Modeling via Bayesian Optimization}
				\label{subsubsec:2Dbo}
				\noindent
				The present section reports the surrogate modeling results obtained via BO for the two-dimensional benchmark functions introduced in \Cref{fig:hfm_2d}.
				\mihi{The two-dimensional setting is considered here to analyse the effect of coupled parameters and increased functional complexity with respect to the one-dimensional case, rather than to assess scalability to multi-dimensional parameter spaces.}
				
				As discussed in detail in the one-dimensional setting (\Cref{subsubsec:1Dbo}), the surrogate modeling strategy adopted here is based on GP with a Matérn \(5/2\) covariance kernel (\Cref{eq:matern}) and an exploration parameter \(\kappa = 200\). The observation noise is assumed Gaussian with variance \(\sigma^2 = 10^{-6}\). Surrogate model construction is performed via uncertainty-aware adaptive sampling, where the training dataset is iteratively updated according to the acquisition function defined by the UCB (\Cref{eq:ucb}) and the update rule in \Cref{eq:dataset_update}.
				
				We first analyze the results obtained for the Mixed Gaussian–Periodic benchmark function (\Cref{fig:hfm_2d}a–b), followed by those for the Rosenbrock benchmark function (\Cref{fig:hfm_2d}c–d).
				
\begin{figure}[!ht]
					\centering
					\begin{subfigure}[b]{0.24\textwidth}
						\includegraphics[width=\textwidth]{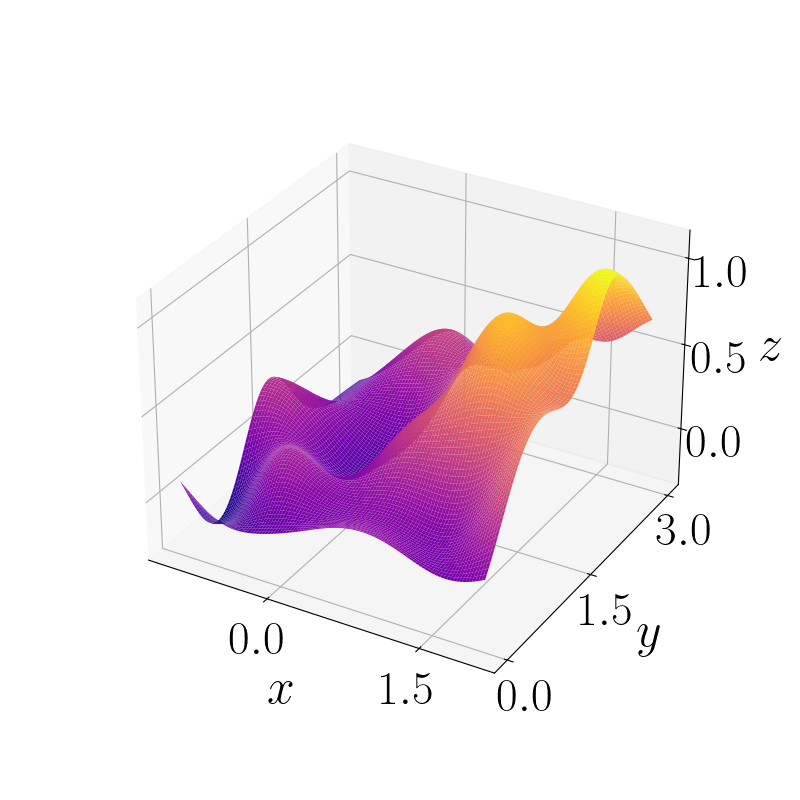}
						\caption{}
						\label{subfig:surface_mix2D}
					\end{subfigure}
					\begin{subfigure}[b]{0.24\textwidth}
						\includegraphics[width=\textwidth]{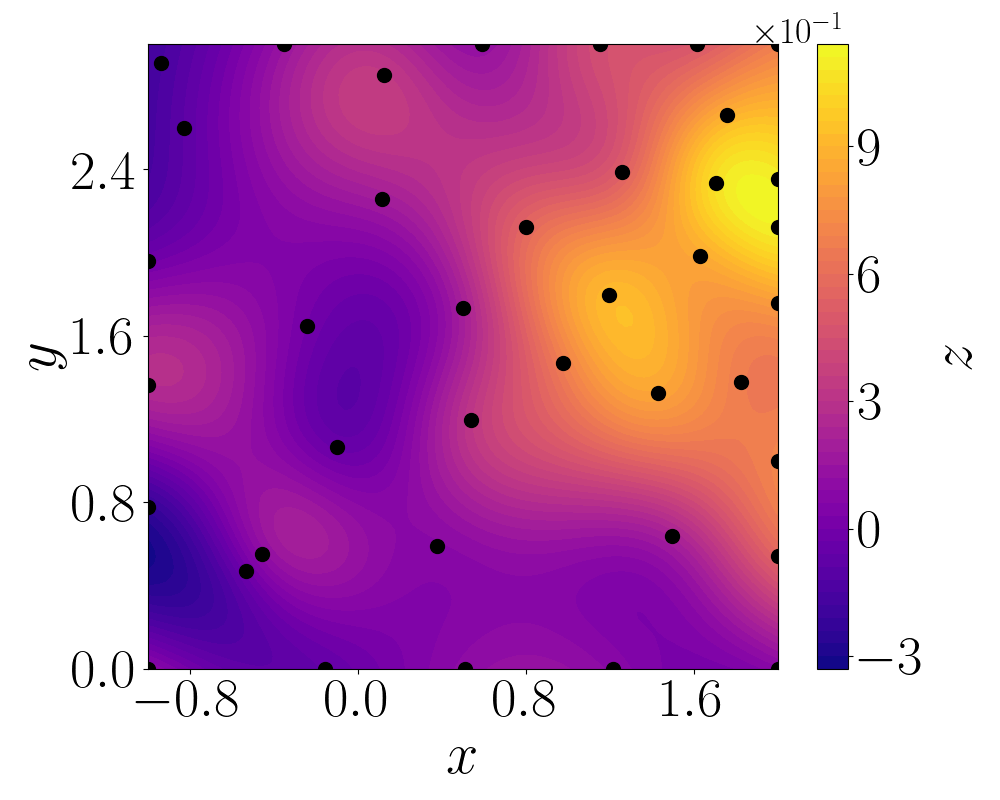}
						\caption{}
						\label{subfig:contour_mix2D}
					\end{subfigure}
					\begin{subfigure}[b]{0.24\textwidth}
						\includegraphics[width=\textwidth]{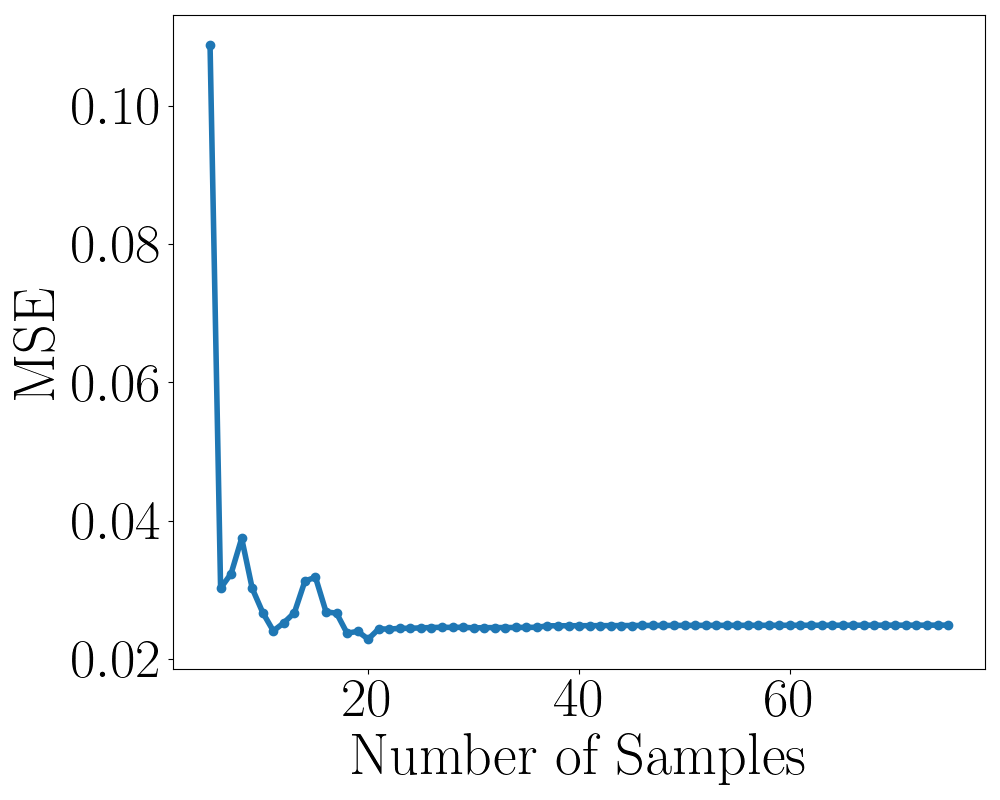}
						\caption{}
						\label{subfig:mse_mix2D}
					\end{subfigure}
					\caption{Surrogate modeling results for the Mixed Gaussian–Periodic benchmark function.
						(a) Final surrogate model surface after 21 high-fidelity evaluations (i.e., number of samples).
						(b) Contour plot with black markers indicating the location of the number of samples.
						(c) Mean squared error (MSE) evaluated on an independent validation set as a function of the number of samples.}
					\label{fig:2D_mix_results}
				\end{figure}

				\Cref{fig:2D_mix_results} reports the results of the surrogate modeling process for the Mixed Gaussian–Periodic benchmark function. The GP surrogate model in \Cref{fig:2D_mix_results}a accurately reconstructs the response surface of the high-fidelity model (\Cref{subfig:hfm_mix_surface} and~\Cref{subfig:hfm_mix_contour}), capturing both the smooth global peak and the superimposed periodic features. The corresponding contour plot in \Cref{fig:2D_mix_results}b shows the distribution of training samples (black markers) acquired during the BO iterations. Starting from an initial design of 5 samples, additional samples are selected in regions of high epistemic uncertainty. The adaptive strategy progressively concentrates samples around the dominant nonlinear features, particularly near the central Gaussian peak, enabling the model to resolve both global and local variations effectively.
				
				The convergence behavior is quantified in \Cref{fig:2D_mix_results}c, where the MSE (\Cref{eq:mse}) evaluated on an independent validation set of \(n_{\text{val}}=1000\) is plotted against the number of samples. A rapid reduction in error is observed, with convergence effectively attained after approximately 21 high-fidelity evaluations, i.e., number of samples. This demonstrates the sample-efficiency of the proposed approach.
				
\begin{figure}[!ht]
					\centering
					\begin{subfigure}[b]{0.24\textwidth}
						\includegraphics[width=\textwidth]{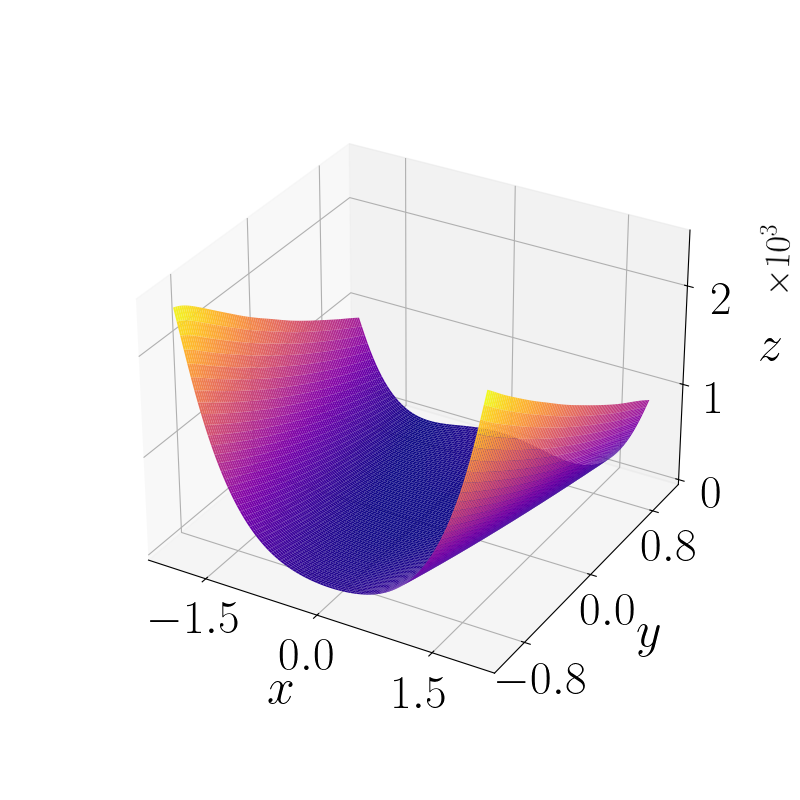}
						\caption{}
						\label{subfig:surface_rosen2D}
					\end{subfigure}
					\begin{subfigure}[b]{0.24\textwidth}
						\includegraphics[width=\textwidth]{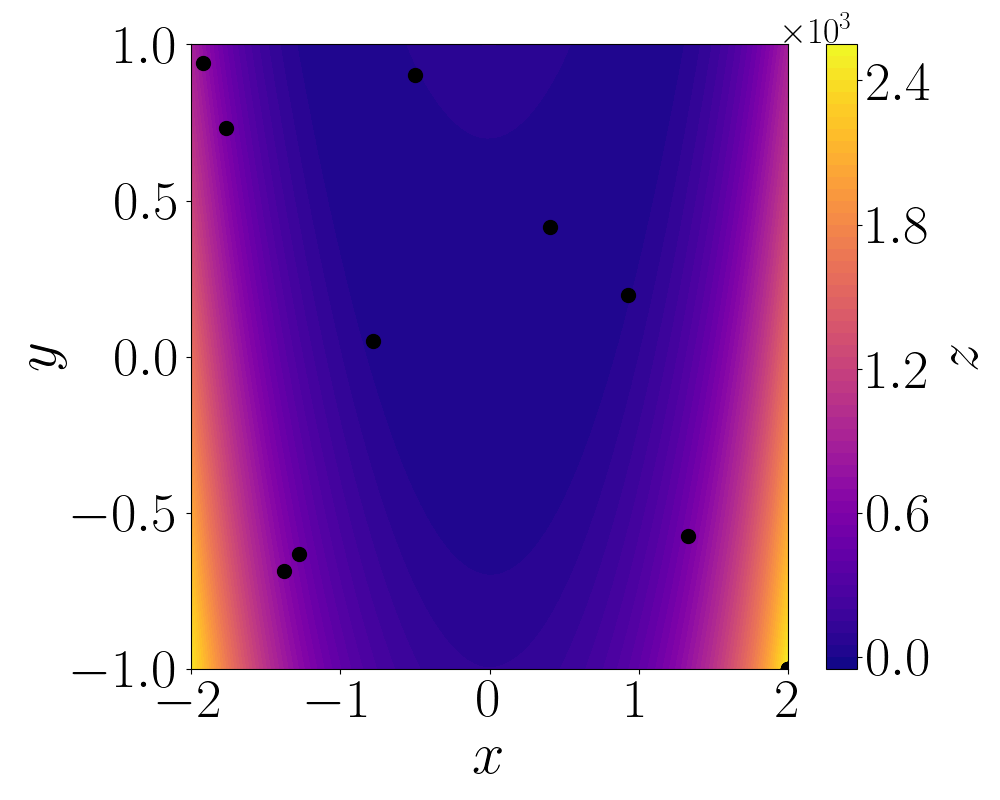}
						\caption{}
						\label{subfig:contour_rosen2D}
					\end{subfigure}
					\begin{subfigure}[b]{0.24\textwidth}
						\includegraphics[width=\textwidth]{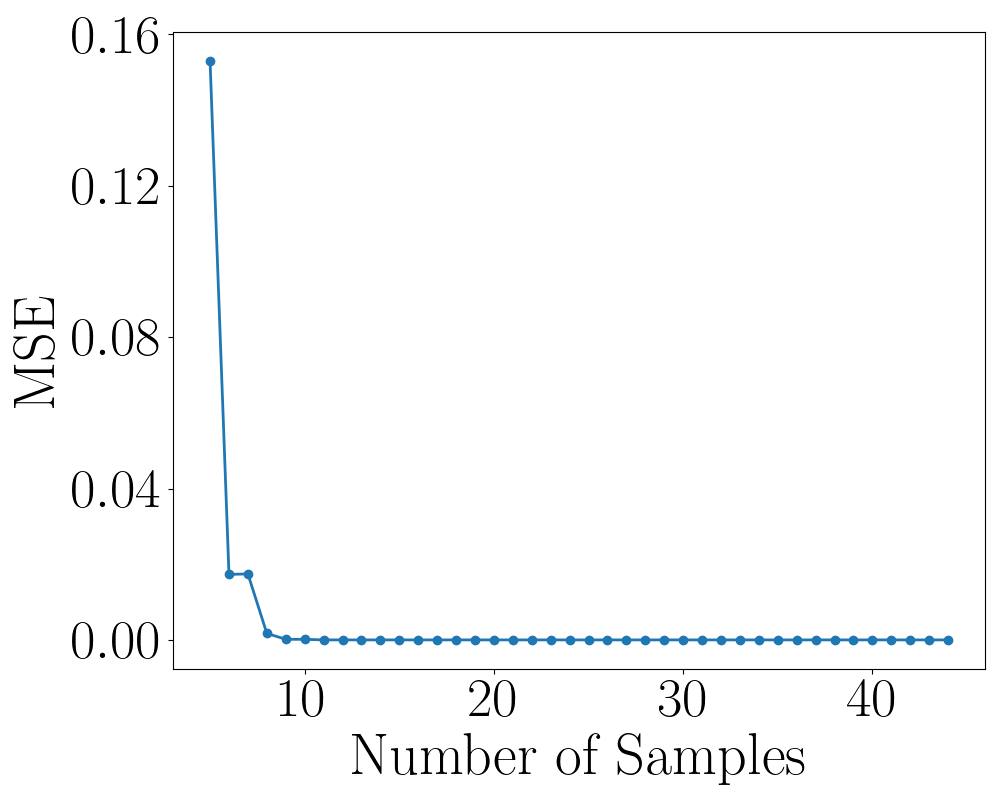}
						\caption{}
						\label{subfig:mse_rosen2D}
					\end{subfigure}
					\caption{Surrogate modeling results for the Rosenbrock benchmark function.
						(a) Final surrogate model surface after 9 high-fidelity evaluations (i.e., number of samples).
						(b) Contour plot with black markers indicating the location of the number of samples.
						(c) Mean squared error (MSE) evaluated on an independent validation set as a function of the number of samples.}
					\label{fig:2D_rosen_results}
				\end{figure}

				\Cref{fig:2D_rosen_results} shows the final results obtained via BO for the Rosenbrock benchmark function. The GP surrogate model in \Cref{fig:2D_rosen_results}a successfully reproduces the key geometric features of the high-fidelity model (\Cref{subfig:hfm_rosen_surface} and~\Cref{subfig:hfm_rosen_contour}), including the valley’s curvature and the steep ascents near the boundary. The spatial distribution of the number of samples in \Cref{fig:2D_rosen_results}b confirms the adaptivity of the sampling procedure: starting from 5 initial number of samples, additional evaluations are directed toward regions of higher epistemic uncertainty, which are predominantly located along the valley path. This leads to a highly efficient resolution of the anisotropic response.
				
				The evolution of the MSE is reported in \Cref{fig:2D_rosen_results}c, showing a sharp decrease followed by early convergence of the surrogate model accuracy. The stopping criterion is met after 9 high-fidelity evaluations, confirming the ability of the proposed framework to construct accurate surrogate models even under strong anisotropy and limited data availability.
				
				\subsubsection{Parameter Inference via Bayesian Inversion} 
				\label{subsubsec:2Dbi}
				\noindent
				The present section reports the results of parameter inference for the two-dimensional analytical benchmarks introduced in \Cref{subsec:2D}, using the BI strategy described in \Cref{subsec:bobi}. 
				\mihi{The analysis is restricted to the two-dimensional setting and is intended to assess the behaviour of the inference procedure under coupled parameters and non-convex response landscapes, rather than to draw conclusions on scalability to multi-dimensional parameter spaces.}
				
				The inverse problems are solved by leveraging the GP surrogate models constructed via BO in \Cref{subsubsec:2Dbo} (see Figures~\ref{fig:2D_mix_results} and~\ref{fig:2D_rosen_results}). For each benchmark function, the analysis includes the computation of the MAP estimate, the evaluation of the LS profile, and the corresponding NLS profile. This allows for a systematic characterization of the admissible parameter space and of the associated epistemic uncertainty. The two cases are discussed separately for the Mixed Gaussian–Periodic function (\Cref{eq:2dmixed}) and the Rosenbrock function (\Cref{eq:rosenbrock}).
				
\begin{figure}[!ht]
					\centering
					\begin{subfigure}[b]{0.24\textwidth}
						\includegraphics[width=\textwidth]{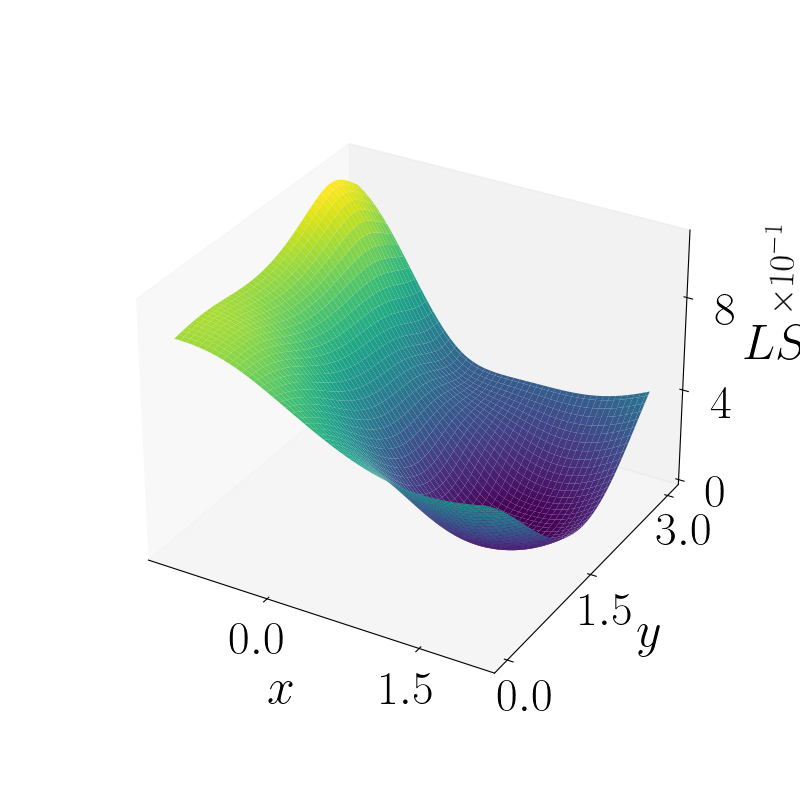}
						\caption{LS surface}
						\label{subfig:ls_surf_mix}
					\end{subfigure}
					\begin{subfigure}[b]{0.24\textwidth}
						\includegraphics[width=\textwidth]{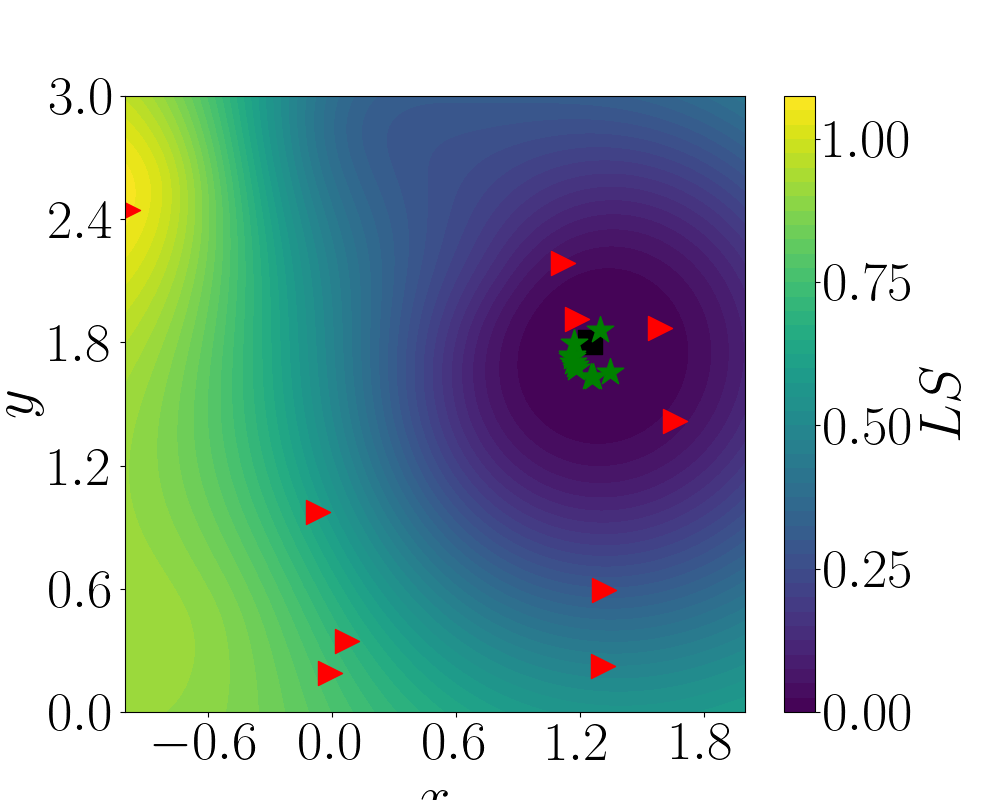}
						\caption{LS contour}
						\label{subfig:ls_contour_mix}
					\end{subfigure}
					\begin{subfigure}[b]{0.24\textwidth}
						\includegraphics[width=\textwidth]{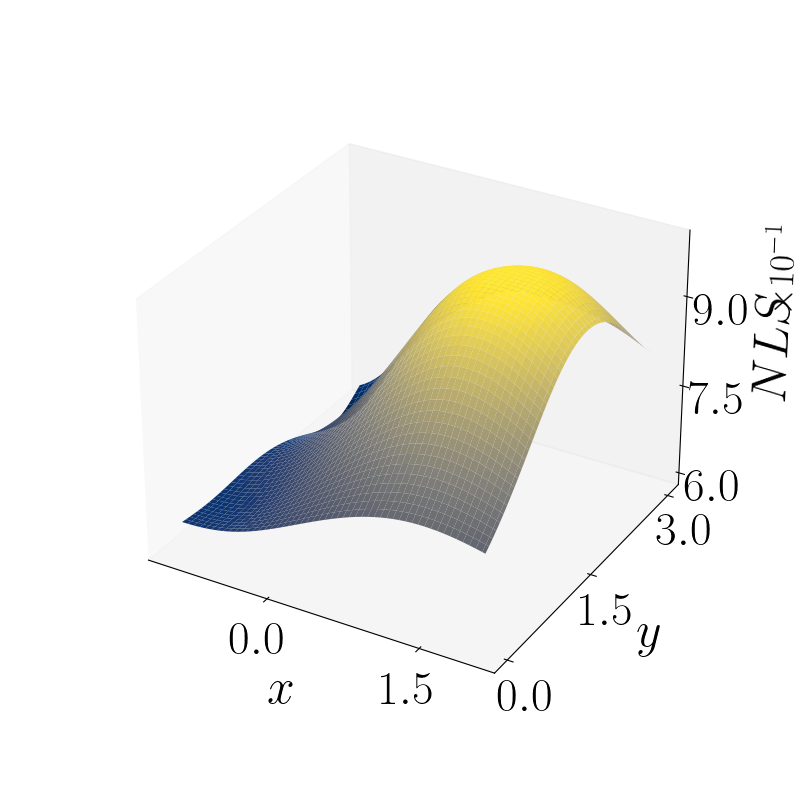}
						\caption{NLS surface}
						\label{subfig:nls_surf_mix}
					\end{subfigure}
					\begin{subfigure}[b]{0.24\textwidth}
						\includegraphics[width=\textwidth]{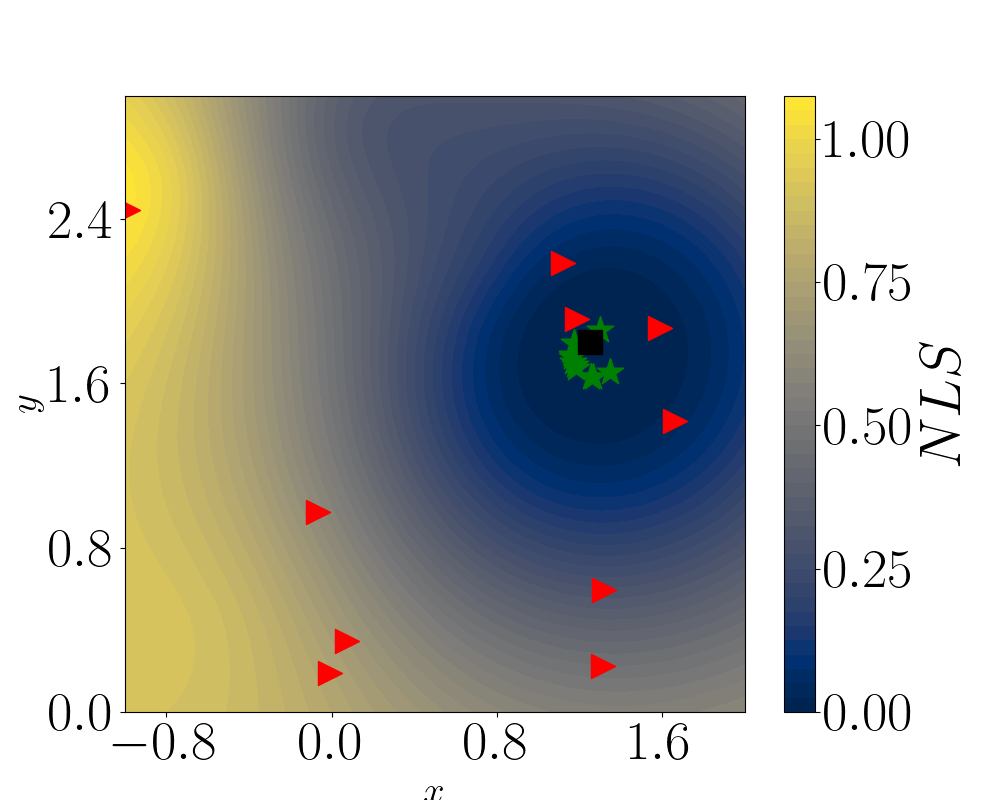}
						\caption{NLS contour}
						\label{subfig:nls_contour_mix}
					\end{subfigure}
					\caption{Parameter inference via BI results for the Mixed Gaussian–Periodic benchmark function. Red triangles: initial guesses; black square: observed quantity of interest; green stars: MAP estimates.}
					\label{fig:bi_mix_results}
				\end{figure}

				\Cref{fig:bi_mix_results} presents the parameter inference via BI results for the Mixed Gaussian–Periodic benchmark function. 
				The LS surface shown in \Cref{subfig:ls_surf_mix} exhibits a non-convex landscape with a dominant basin of attraction, while the corresponding contour plot in \Cref{subfig:ls_contour_mix} highlights a well-defined low-LS region surrounded by nearly concentric level curves. 
				To ensure sufficient exploration of the parameter space, LS minimization is initialized from multiple independent initial guesses (red triangles). 
				Despite the presence of mild irregularities in the LS landscape, most optimization runs converge toward a compact region located near the prescribed observed quantity of interest (black square), 
				\(\bar{y} = 0.63\), yielding several closely clustered MAP estimates (green stars) within the same basin of attraction.
				
				The inferred MAP solutions concentrate around the parameter vector 
				\(\mathbf{x}_{\text{MAP}} = [1.248,\ 1.812]\), 
				minimizing the LS functional with a residual value of approximately \(\mathrm{LS} \approx 0.03\), 
				indicating excellent agreement between the surrogate model prediction and the target observation. 
				The NLS surface and contour plots (\Cref{subfig:nls_surf_mix}, \Cref{subfig:nls_contour_mix}), computed from the LS functional via \Cref{eq:nls}, reveal a sharply peaked posterior distribution centered at \(\mathbf{x}_{\text{MAP}}\). 
				The maximum of the NLS surface reaches approximately \(0.90\), with a rapid decay away from the MAP region. 
				Although the LS landscape exhibits non-convex features, the NLS that approximately represent the posterior distribution remains effectively unimodal and strongly localized, indicating low epistemic uncertainty and strong local parameter identifiability.
				
				Parameter-wise uncertainty is quantified through marginal credible intervals derived from a local Gaussian approximation of the posterior distribution around the MAP estimate. 
				The bounds are computed from the diagonal entries of the covariance matrix \(\boldsymbol{\Sigma}_{\text{MAP}}\), defined as the inverse of the Hessian of the NLS functional evaluated at the MAP point, as in \Cref{eq:mapsig}. 
				This yields an axis-aligned confidence region that approximates the high-probability domain of the posterior and provides interpretable uncertainty bounds for each parameter. 
				The resulting MAP estimate and marginal credible intervals are summarized in \Cref{tab:map_mix}.
				
				\begin{table}[!ht]
					\centering
					\caption{MAP estimate and marginal credible intervals for the Mixed Gaussian–Periodic benchmark function. 
						The target value is \(\bar{y} = 0.63\), the LS residual at the MAP point is approximately \(0.03\), and the posterior peak reaches \(\mathrm{NLS}_{\max} \approx 0.90\).}
					\label{tab:map_mix}
					\begin{tabular}{ccc}
						\hline
						\textbf{Parameter} & \textbf{MAP estimate} & \textbf{Credible interval} \\
						\hline
						\( x \) & 1.248 & [1.158, 1.638] \\
						\( y \) & 1.812 & [1.323, 2.221] \\
						\hline
					\end{tabular}
				\end{table}
				
\begin{figure}[!ht]
					\centering
					\begin{subfigure}[b]{0.24\textwidth}
						\includegraphics[width=\textwidth]{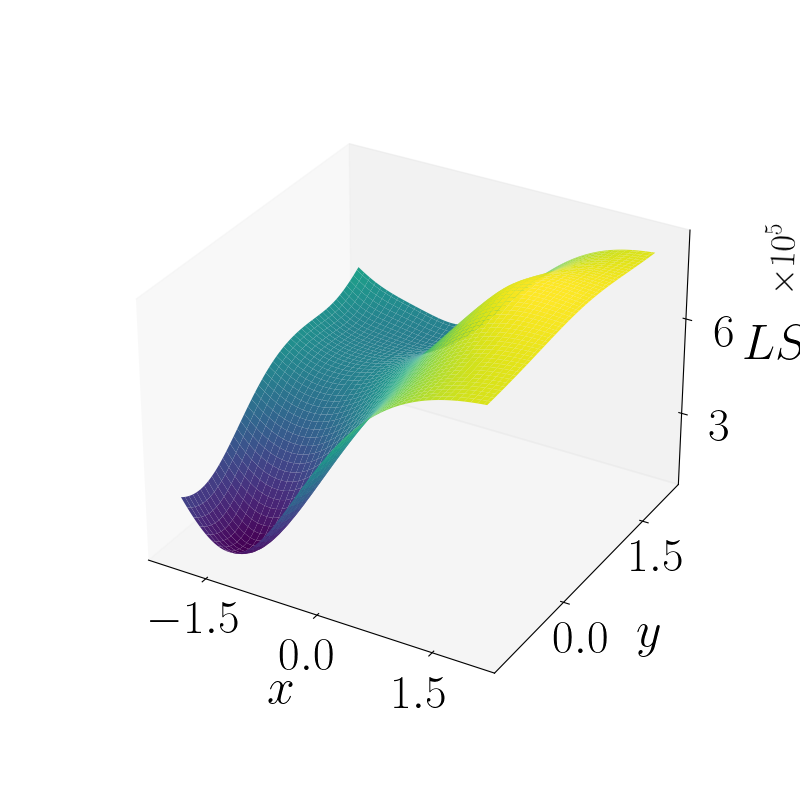}
						\caption{LS surface}
						\label{subfig:rosen_ls_surf}
					\end{subfigure}
					\begin{subfigure}[b]{0.24\textwidth}
						\includegraphics[width=\textwidth]{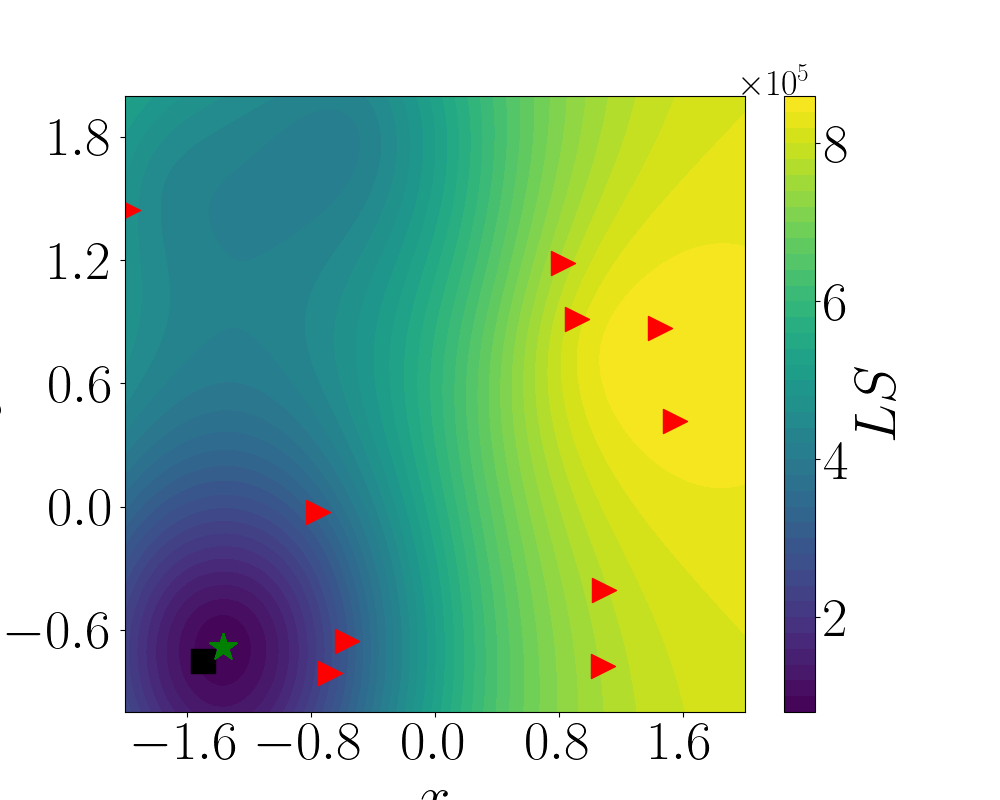}
						\caption{LS contour}
						\label{subfig:rosen_ls_contour}
					\end{subfigure}
					\begin{subfigure}[b]{0.24\textwidth}
						\includegraphics[width=\textwidth]{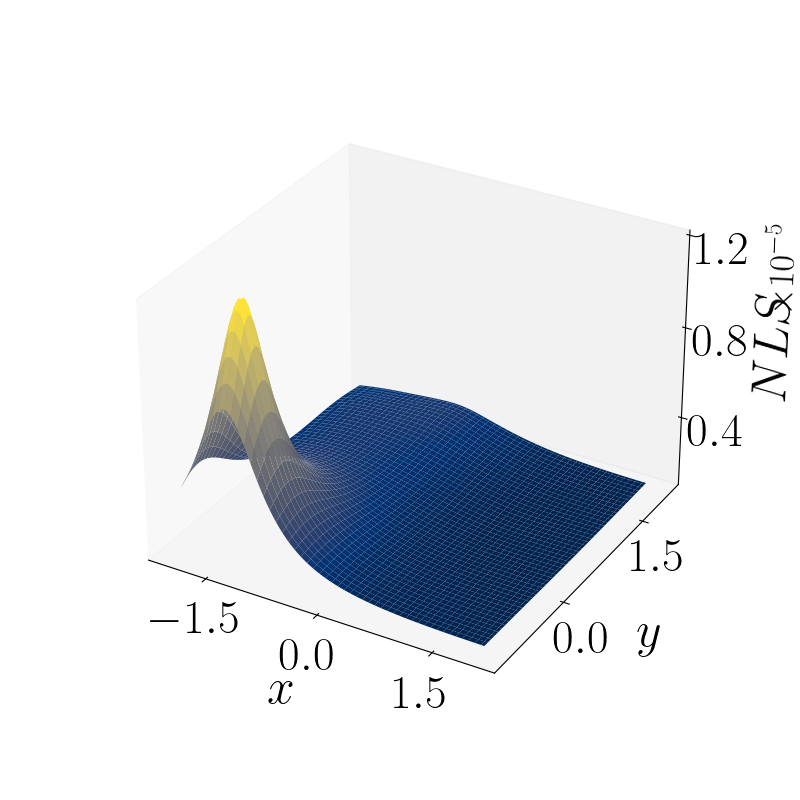}
						\caption{NLS surface}
						\label{subfig:rosen_nls_surf}
					\end{subfigure}
					\begin{subfigure}[b]{0.24\textwidth}
						\includegraphics[width=\textwidth]{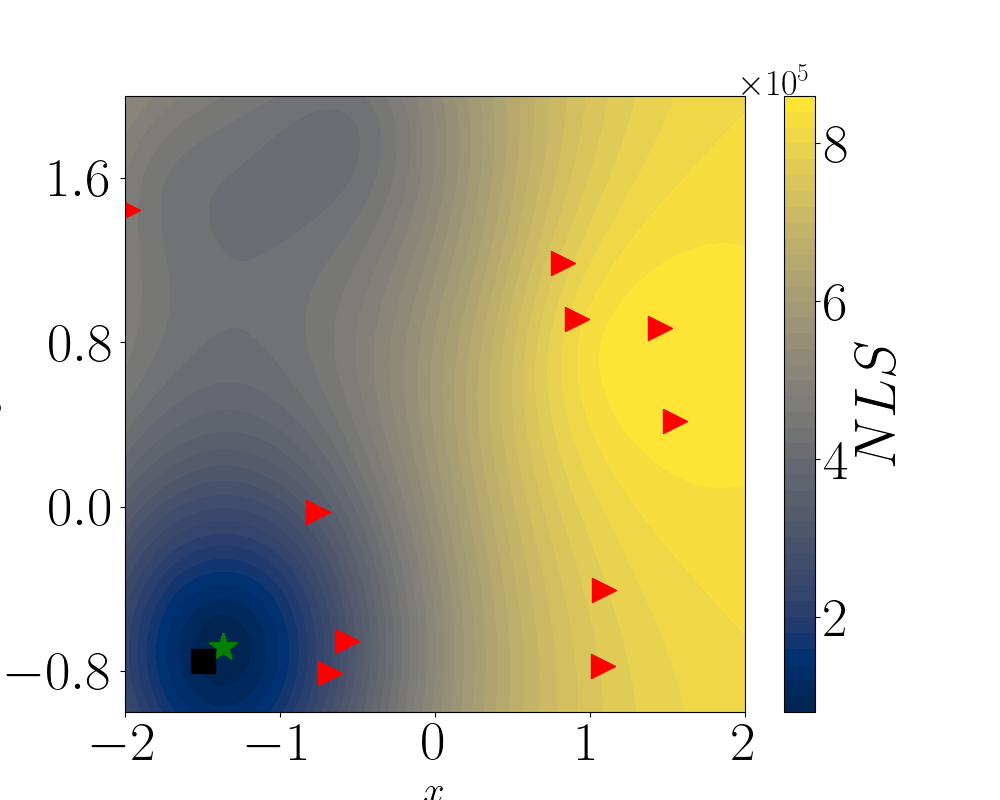}
						\caption{NLS contour}
						\label{subfig:rosen_nls_contour}
					\end{subfigure}
					\caption{Parameter inference via BI results for the Rosenbrock benchmark function. Red triangles: initial guesses; black square: observed quantity of interest; green star: MAP estimate.}
					\label{fig:rosen_bi_results}
				\end{figure}

				\Cref{fig:rosen_bi_results} presents the BI results for the Rosenbrock benchmark function. The LS surface in \Cref{subfig:rosen_ls_surf} captures the characteristic narrow curved valley of the Rosenbrock landscape, with the contour plot in \Cref{subfig:rosen_ls_contour} confirming the presence of a single dominant minimum. Independent optimization runs initialized from multiple starting points (red triangles) consistently converge to this minimum, resulting in a unique MAP estimate (green star) located near \( \mathbf{x}_{\text{MAP}} = [-1.5,\ -0.6] \), in close proximity to the observed quantity of interest (black square).
				
				The corresponding NLS surface (\Cref{subfig:rosen_nls_surf}, \Cref{subfig:rosen_nls_contour}) is sharply unimodal, with the posterior distribution strongly concentrated around the MAP estimate. No secondary modes are detected, indicating excellent parameter identifiability and negligible epistemic uncertainty.
				
				\subsubsection{Discussion}
				\label{subsubsec:2Dres}
				\noindent
				The numerical validation results presented in Sections~\ref{subsubsec:2Dbo} and~\ref{subsubsec:2Dbi} demonstrate the robustness and effectiveness of the proposed Bayesian framework when applied to inverse problems characterized by coupled parameters, non-convex response surfaces, and pronounced anisotropy. 
				
				As in the one-dimensional setting (\Cref{subsec:1D}), the integration of BO and BI proves advantageous in terms of sample efficiency and uncertainty quantification. 
				The surrogate models constructed via GP regression and UCB-based uncertainty-aware sampling accurately approximate the high-fidelity responses using a limited number of evaluations, even in the presence of narrow valleys, steep gradients, and multimodal structures. 
				The availability of a reliable surrogate model, in turn, enables an efficient application of BI, allowing MAP estimation and posterior analysis to be performed at negligible computational cost.
				
				The Mixed Gaussian–Periodic benchmark highlights the framework’s ability to handle non-convex inverse problems with moderately irregular landscapes, where multiple initializations are required to ensure adequate exploration of the parameter space. 
				By contrast, the Rosenbrock benchmark illustrates a well-posed inverse problem with strong anisotropy and a uniquely identifiable solution, for which the posterior distribution is sharply concentrated around the MAP estimate. 
				In both cases, the posterior structure is consistently captured through the surrogate-assisted BI procedure, providing an informative characterization of epistemic uncertainty.
				
				Overall, the two-dimensional results complement the one-dimensional analysis by demonstrating the applicability of the proposed Bayesian framework to inverse problems with increased structural complexity and parameter coupling, while preserving methodological clarity and computational efficiency.

		\section{Conclusions} 
		\label{sec:conclusion} 
		\noindent 
		\mihi{
			The present work introduces a unified Bayesian framework that integrates Gaussian Process (GP) surrogate modeling and Bayesian Inversion (BI) for the efficient solution of inverse problems under uncertainty.
			The framework targets scenarios characterized by computationally expensive high-fidelity models and limited data availability, and provides a consistent probabilistic formulation for robust and interpretable parameter inference.
		}
		
		\mihi{
			During the surrogate modeling phase, GP models are constructed and adaptively refined through Bayesian Optimization (BO), enabling an efficient exploration of the parameter space while accounting for predictive uncertainty.
			This adaptive strategy significantly reduces the number of high-fidelity model evaluations required to achieve accurate surrogate representations, thereby improving computational efficiency without sacrificing reliability.
		}
		
		\mihi{
			Once the surrogate model is established, the inverse problem is addressed within a Bayesian inversion setting, yielding Maximum A Posteriori (MAP) estimates together with posterior distributions of the input parameters.
			This probabilistic characterization enables a rigorous assessment of epistemic uncertainty and provides insight into parameter identifiability and sensitivity within the inverse problem.
		}

		\mihi{
			The proposed framework is validated on a suite of well-established one- and two-dimensional analytical benchmark problems exhibiting nonlinear, multimodal, and non-convex response surfaces.
			These benchmarks provide a controlled and interpretable setting to assess the behaviour of the integrated BO-BI methodology and to isolate its key methodological features.
			The numerical results demonstrate that the unified framework enables efficient surrogate construction, accurate parameter inference, and informative uncertainty quantification within these low-dimensional settings.
		}
		
		\mihi{
			The results highlight the synergistic benefits of combining Bayesian optimization and Bayesian inversion within a coherent framework, allowing uncertainty-aware surrogate modeling and probabilistic inference to be performed in a mutually reinforcing manner.
			This integration supports reliable inverse analysis while maintaining a favourable trade-off between computational efficiency, accuracy, and uncertainty quantification.
		}
		
		\mihi{
			Possible directions for future research include the application of the proposed methodology to inverse problems involving multi-dimensional parameter spaces, the incorporation of multi-fidelity and physics-informed surrogate models, and the development of scalable inference strategies tailored to large-scale computational settings.
			These extensions would further enhance the applicability of the framework to complex engineering problems while preserving its methodological transparency and probabilistic rigor.
		}
		
		\mihi{
			By unifying Bayesian optimization and Bayesian inversion within a single probabilistic workflow, the proposed approach provides a structured and sample-efficient solution strategy for inverse problems in computational mechanics and related fields.
		}

		\section*{Acknowledgments}
		\noindent
		The present work is partially supported by the Italian Ministry of University and Research (MUR) through the DORIAN project (“Sustainable Resilient Safe Environment”) as part of the Ph.D. program at the Department of Civil Engineering and Architecture, University of Pavia. Mihaela Chiappetta, Massimo Carraturo and Ferdinando Auricchio gratefully acknowledge this support.
		
		The present work is also supported by the German Research Foundation (DFG) within the Research Training Group GRK 2868 D³ — project number 493401063. Mihaela Chiappetta, Markus Kästner, and Alexander Raßloff gratefully acknowledge this support.
		
		\section*{Conflict of interest}
		\noindent
		The authors declare that they have no known competing financial interests or personal relationships that could have appeared to influence the work reported in present paper.

		\newpage
		
	\bibliographystyle{unsrtnat}
	\bibliography{reference}

	\newpage
	
	\appendix
	
	\section{Implementation Details: The \texttt{UQforPy} Repository}
	\label{ref}
	\noindent
	All computational components developed in this work are implemented using the open-source \texttt{UQforPy} repository, available at \url{https://bitbucket.org/uqpython/uqforpython/src/main/}. 
	\texttt{UQforPy} provides a modular and extensible platform for uncertainty quantification and inverse problems, supporting a wide range of data-driven modeling tasks in computational science and engineering.
	The codebase is implemented in Python and leverages standard scientific libraries such as NumPy, SciPy, and scikit-learn.
	
	\begin{figure}[!ht]
		\centering
		\includegraphics[width=0.8\textwidth]{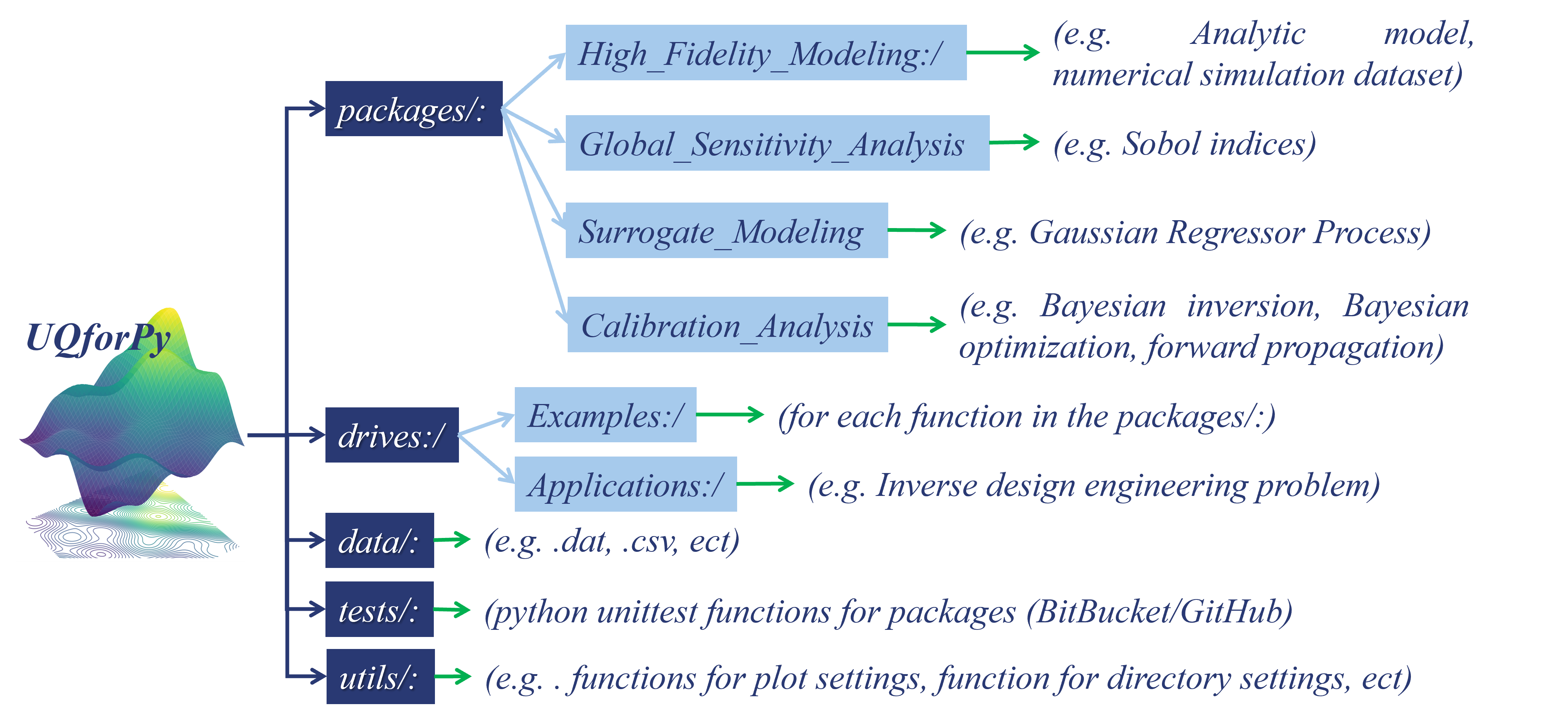}
		\caption{Modular architecture of the \texttt{UQforPy} repository. The framework is organized into self-contained components supporting high-fidelity modeling, surrogate model construction, sensitivity analysis, and parameter inference. Its design enables scalable, reproducible, and customizable workflows for uncertainty-aware inverse problem in engineering applications.}
		\label{fig:gitrepo}
	\end{figure}
	
	As illustrated in Figure~\ref{fig:gitrepo}, the repository is structured into dedicated directories, each encapsulating a specific methodological module:
	
	\begin{itemize}
		\item \texttt{packages/}: Core modules for modeling and inference, comprising:
		\begin{itemize}
			\item \texttt{High\_Fidelity\_Modeling/}: Analytical benchmark functions and simulation routines.
			\item \texttt{Global\_Sensitivity\_Analysis/}: Variance-based sensitivity analysis tools, including Sobol indices~\cite{glen2012estimating}.
			\item \texttt{Surrogate\_Modeling/}: Routines for GP-based surrogate construction and evaluation.
			\item \texttt{Calibration\_Analysis/}: Modules for Bayesian Optimization, Bayesian Inversion, and forward uncertainty propagation.
		\end{itemize}
		
		\item \texttt{drives/}: Driver scripts interfacing with \texttt{packages/}, including:
		\begin{itemize}
			\item \texttt{Examples/}: Standalone demonstrations of individual components.
			\item \texttt{Applications/}: Complete pipelines for inverse problem integrating BO and BI.
		\end{itemize}
		
		\item \texttt{data/}: Standardized datasets (\texttt{.dat}, \texttt{.csv}) for benchmark reproducibility and validation.
		
		\item \texttt{tests/}: Unit testing suite for numerical validation and integration checks (e.g., via Bitbucket Pipelines or GitHub Actions).
		
		\item \texttt{utils/}: Auxiliary utilities for plotting, file I/O, and workflow orchestration.
	\end{itemize}
	
	The \texttt{UQforPy} platform is designed to balance computational efficiency with user flexibility. 
	It supports both exploratory studies and production-grade analyses, making it particularly suited for problems involving expensive high-fidelity evaluations, epistemic uncertainty, and limited data.
	
	While the present work focuses on the integration of BO and BI for inverse problem, the repository is extensible to a broader range of uncertainty quantification techniques, including Monte Carlo simulation, polynomial chaos expansions, and stochastic collocation.
	
	By promoting reproducibility, extensibility, and accessibility, \texttt{UQforPy} aims to support the integration of uncertainty-aware modeling into computational mechanics and engineering workflows. 
	Ongoing developments include advanced Bayesian neural network models, adaptive experimental design strategies, and uncertainty-aware optimization tools.
	These enhancements are expected to further strengthen the capability of \texttt{UQforPy} to tackle complex inverse problems in real-world engineering contexts.
	
	\section{Surrogate Modeling via BO Results for 1D Benchmarks}
	\label{bo1d}
	\noindent
	The present appendix complements the results discussed in Section~\ref{subsubsec:1Dbo} by providing additional validation studies of the surrogate modeling strategy adopted in the present work.
	
	Specifically, systematic tests are presented for the Lévy and Griewank benchmark functions (Figures~\ref{fig:Benchmark}b–d), aimed at evaluating the predictive performance of different surrogate models, acquisition functions, and sampling strategies within the BO framework described in Section~\ref{subsec:bobo}.
	
	These tests reinforce the adoption of Gaussian Process (GP) surrogate models with Matérn covariance kernels, combined with uncertainty-aware sampling via the Upper Confidence Bound (UCB) acquisition function, as the default surrogate modeling configuration within the proposed Bayesian framework.
	
\begin{figure}[!ht]
		\centering
		\begin{subfigure}[b]{0.3\textwidth}
			\includegraphics[width=\textwidth]{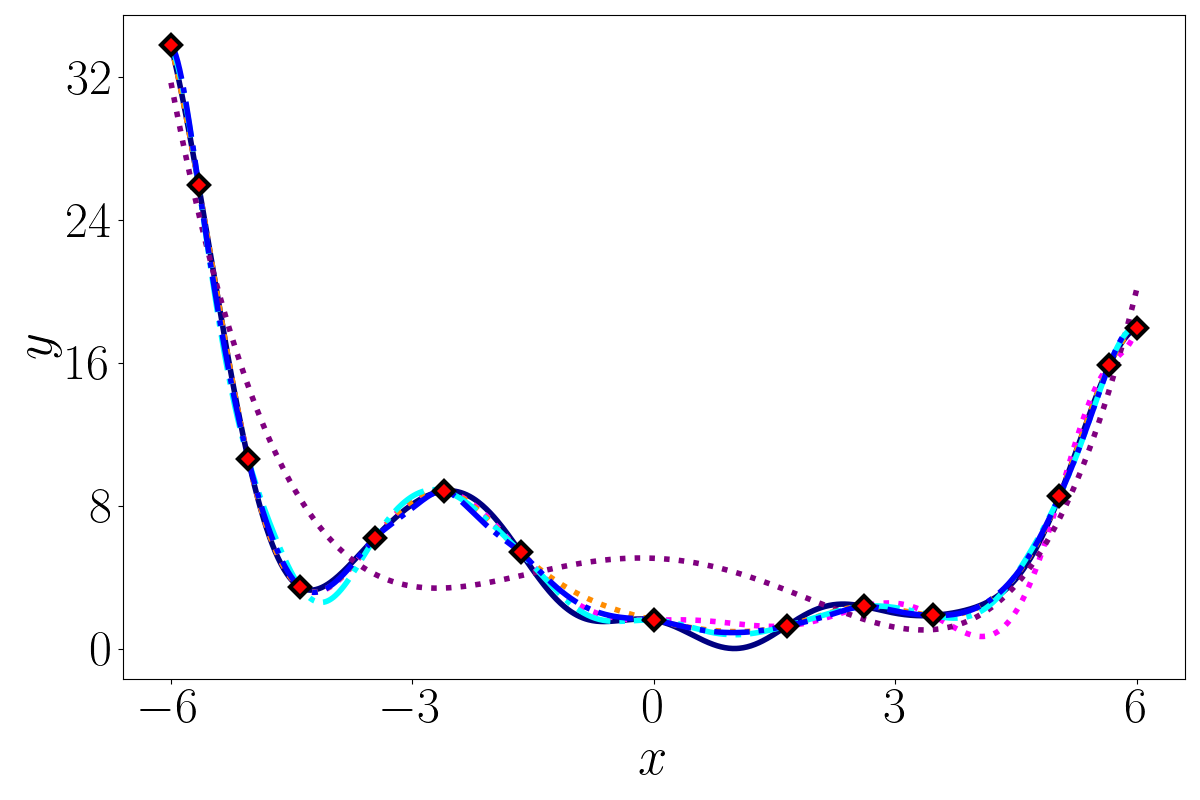}
			\caption{Lévy}
			\label{subfig:levyCURV}
		\end{subfigure}
		\begin{subfigure}[b]{0.3\textwidth}
			\includegraphics[width=\textwidth]{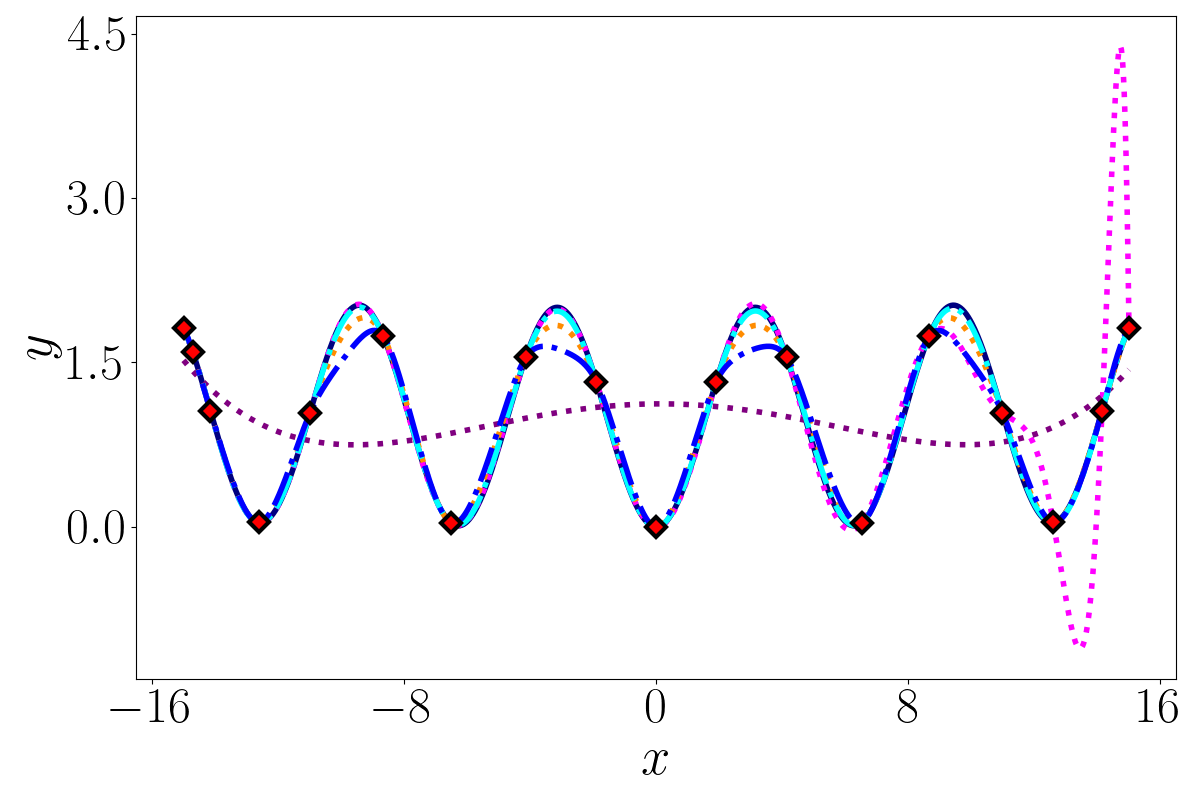}
			\caption{Griewank}
			\label{subfig:griewankCURV}
		\end{subfigure}
		\begin{subfigure}[b]{0.3\textwidth}
			\includegraphics[width=\textwidth]{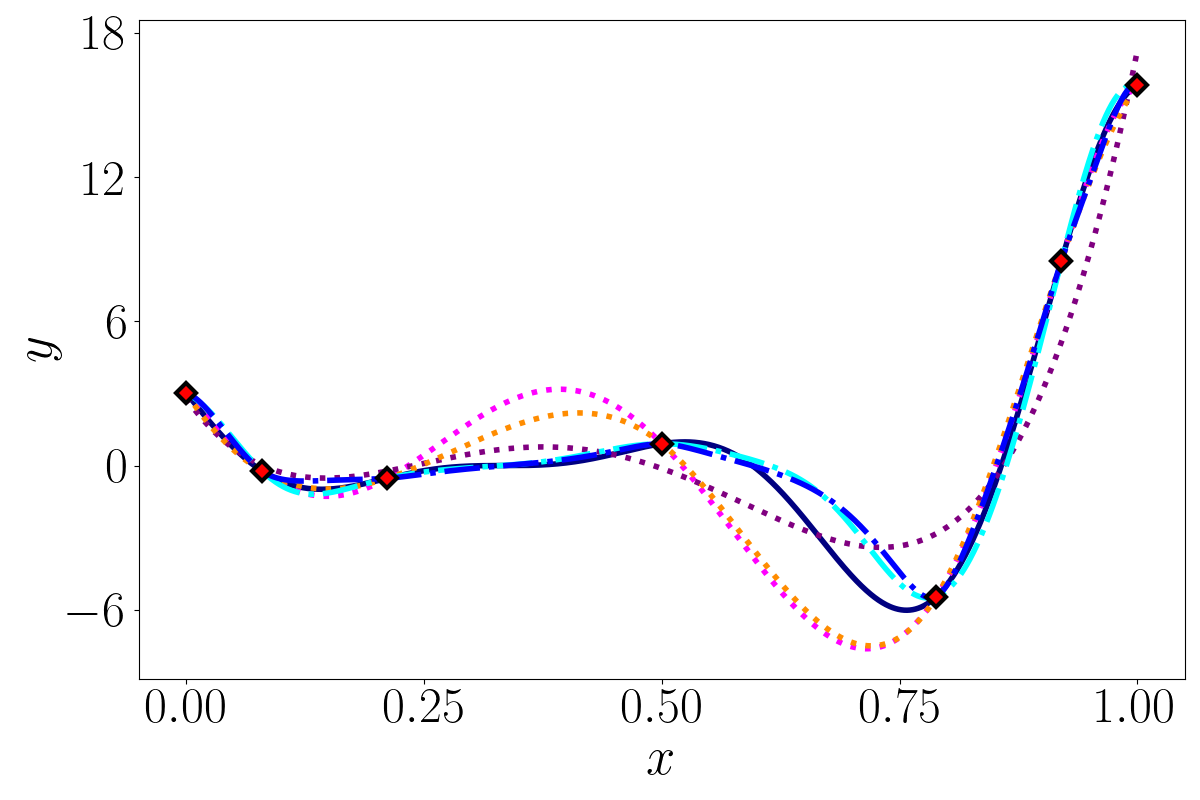}
			\caption{Forrester}
			\label{subfig:forresterCURV}
		\end{subfigure}
		\caption{Predictive responses of different surrogate models for the benchmark functions in \Cref{fig:Benchmark}b–d. 
			The solid black line represents the high-fidelity model; fuchsia, purple, and orange dotted lines correspond to LP, LE, and CS models; cyan and blue lines represent GP surrogate models with Matérn and RBF kernels. Red diamonds denote training samples points.}
		\label{fig:BenchmarkCURVappendix}
	\end{figure}

	Figure~\ref{fig:BenchmarkCURVappendix} reports the predictive responses obtained using different surrogate modeling strategies. The results confirm and extend the conclusions drawn from the Mixed Gaussian-Periodic case (Section~\ref{subsubsec:1Dbo}).
	
	For the Lévy function (Figure~\ref{fig:BenchmarkCURVappendix}a), which exhibits a highly non-convex landscape with multiple local minima, GP models with both Matérn and RBF kernels successfully replicate the complex structural features of the high-fidelity model. The Matérn-based surrogate shows enhanced robustness in resolving localized irregularities, while deterministic models — such as Lagrangian Polynomial (LP), Legendre Expansion (LE), and Cubic Spline (CS) — fail to capture the fine-scale behavior or suffer from poor flexibility in sharp transition zones.
	
	The Griewank function (Figure~\ref{fig:BenchmarkCURVappendix}b), combining a slow global trend with rapid periodic fluctuations, is accurately modeled by GP surrogate models. In contrast, LP models exhibit strong boundary oscillations (Runge's phenomenon~\cite{fornberg2007runge}), while LE models fail to capture the high-frequency content of the benchmark function.
	
	For the Forrester function (Figure~\ref{fig:BenchmarkCURVappendix}c), characterized by smooth global structure and sharp local nonlinearities, the GP-Matérn surrogate model again delivers the most accurate results. Deterministic models tend to over-smooth or misrepresent localized complexity.
	
\begin{figure}[!ht]
		\centering
		\begin{subfigure}[b]{0.3\textwidth}
			\includegraphics[width=\textwidth]{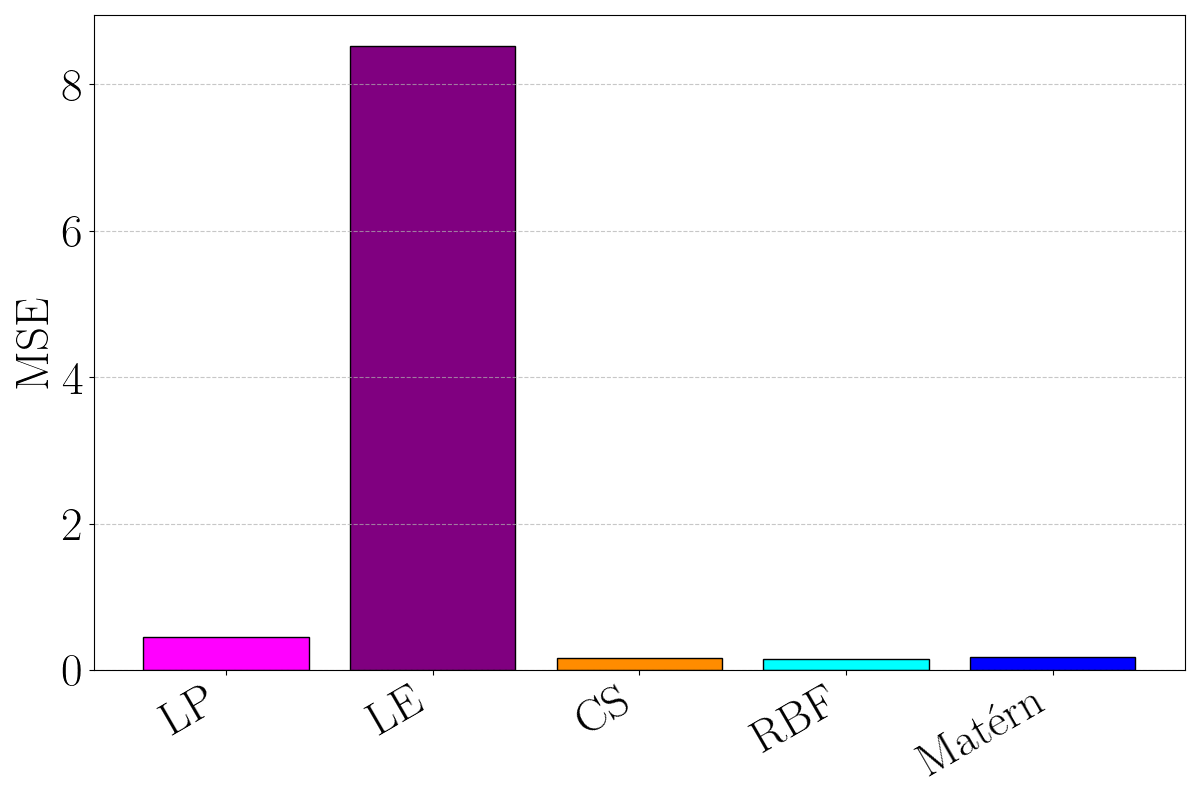}
			\caption{Lévy}
			\label{subfig:levyHIST}
		\end{subfigure}
		\begin{subfigure}[b]{0.3\textwidth}
			\includegraphics[width=\textwidth]{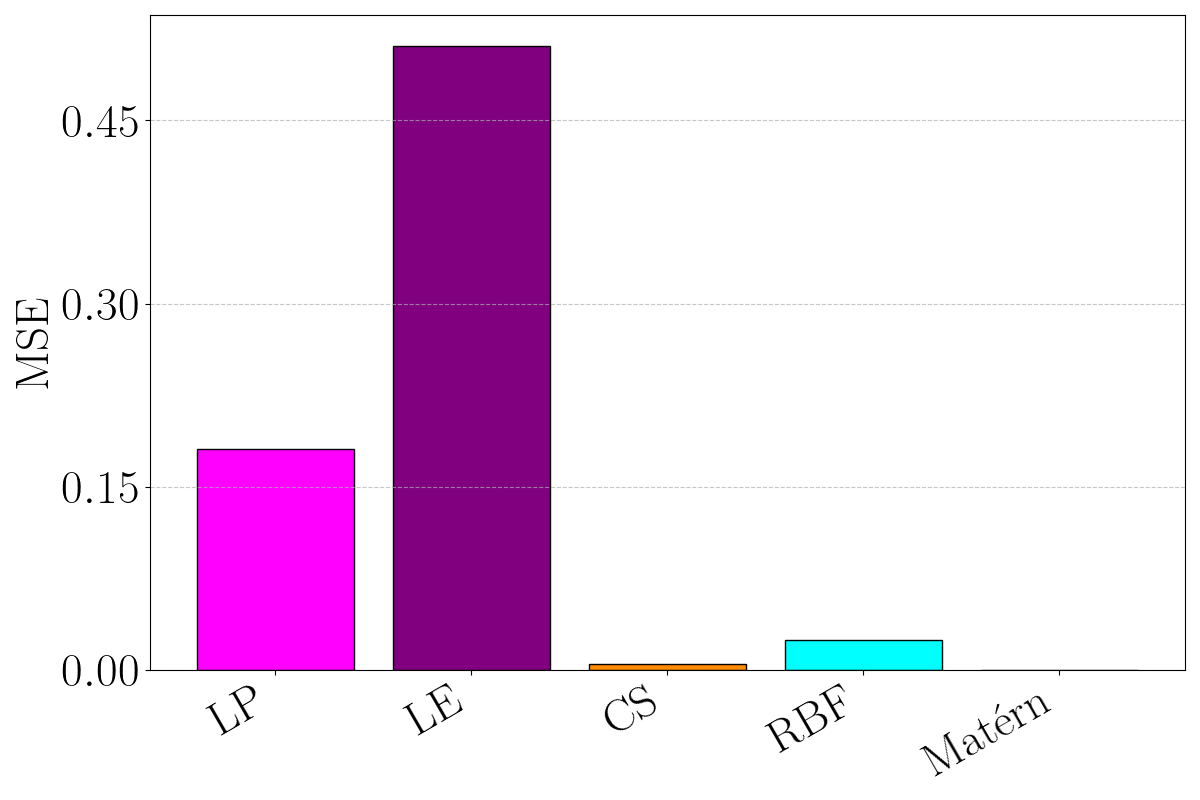}
			\caption{Griewank}
			\label{subfig:griewankHIST}
		\end{subfigure}
		\begin{subfigure}[b]{0.3\textwidth}
			\includegraphics[width=\textwidth]{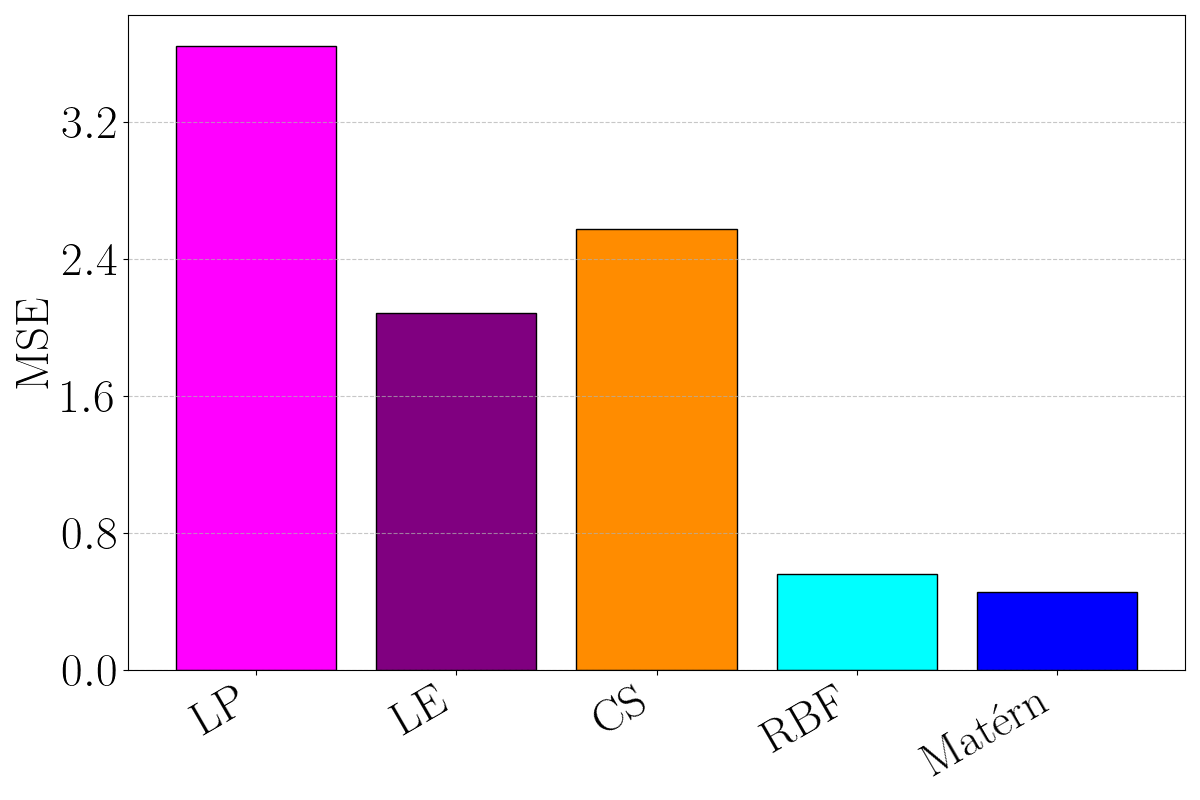}
			\caption{Forrester}
			\label{subfig:forresterHIST}
		\end{subfigure}
		\caption{Mean squared error (MSE) comparison of surrogate models for the benchmark functions in \Cref{fig:Benchmark}b–d.}
		\label{fig:BenchmarkHISTappendix}
	\end{figure}

	These observations are quantitatively confirmed by the MSE values reported in Figure~\ref{fig:BenchmarkHISTappendix}, computed on an independent validation set. GP surrogate models — particularly with the Matérn kernel — consistently outperform deterministic alternatives.
	
	\section{Parameter Inference via BI Results for 1D Benchmarks}
	\label{bi1d}
	\noindent
	The present appendix complements the analysis in Section~\ref{subsubsec:1Dbi} by providing additional validation of the proposed Bayesian framework. Specifically, the parameter inference results obtained via Bayesian Inversion (BI) are reported for the Lévy and Griewank benchmark functions (\Cref{fig:Benchmark}b–c), selected to examine the framework’s capability in the presence of localized nonlinearities and strongly multimodal landscapes.
	
	For each benchmark, the MAP estimate, the Least Squares (LS) functional, and the corresponding Negative Least Squares (NLS) profile are analysed to characterize the structure of the inverse problem and the associated epistemic uncertainty.
	
\begin{figure}[!ht]
		\centering
		\begin{subfigure}[b]{0.3\textwidth}
			\centering
			\includegraphics[width=\textwidth]{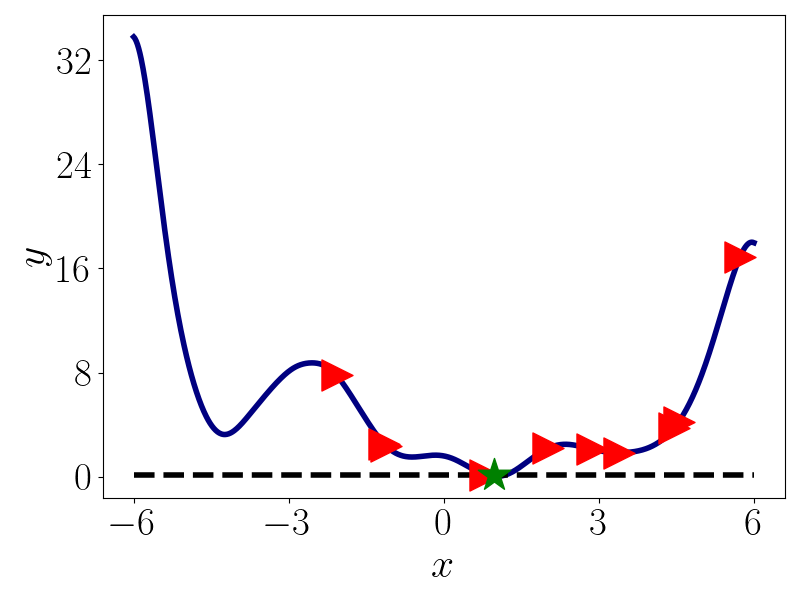}
			\caption{MAP estimates}
			\label{subfig:levy_map}
		\end{subfigure}
		\begin{subfigure}[b]{0.3\textwidth}
			\centering
			\includegraphics[width=\textwidth]{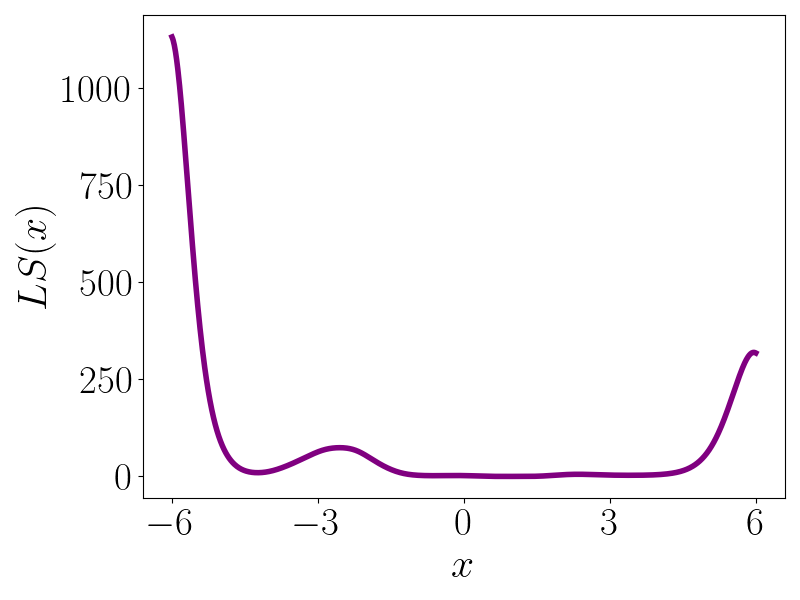}
			\caption{LS profile}
			\label{subfig:levy_ls}
		\end{subfigure}
		\begin{subfigure}[b]{0.3\textwidth}
			\centering
			\includegraphics[width=\textwidth]{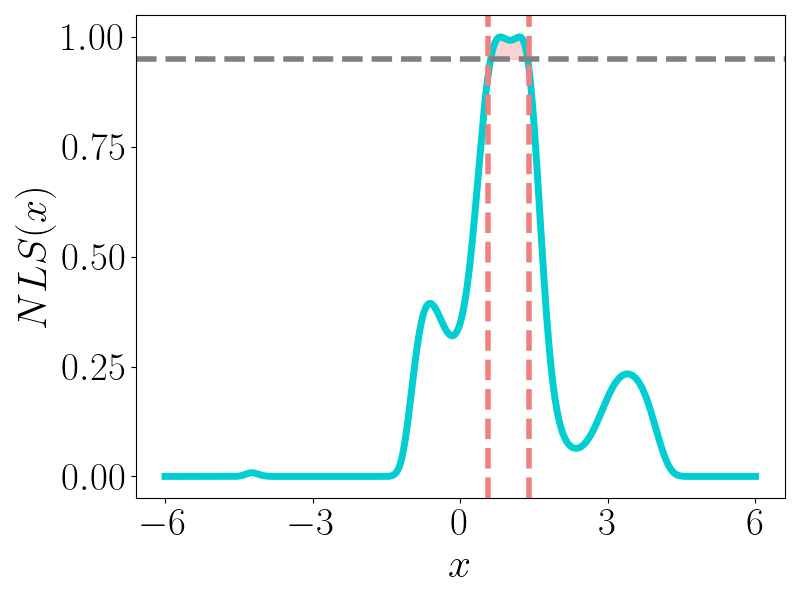}
			\caption{NLS profile}
			\label{subfig:levy_nls}
		\end{subfigure}
		\caption{Parameter inference via BI results for the Lévy benchmark. (a) MAP estimates from multiple optimization runs. (b) LS functional profile. (c) NLS functional profile representing the posterior distribution approximation with high-probability region marked by the grey dotted horizontal line representing the 0.95-threshold and highlighted by the coral-coloured portion delimited by the dotted vertical lines.}
		\label{fig:levy_bi}
	\end{figure}

	\Cref{fig:levy_bi} reports the results for the Lévy benchmark. \Cref{subfig:levy_map} shows the inferred MAP estimates obtained from multiple independent optimization runs, each initialized from a different guesses point in the parameter space. All runs are based on the GP surrogate model constructed via BO (\Cref{fig:bo_conclusion}b), and consistently converge to the same region.
	
	The LS profile (\Cref{subfig:levy_ls}) exhibits a sharp and well-separated minimum, indicating strong local identifiability. The corresponding NLS profile (\Cref{subfig:levy_nls}), computed as in Equation~\eqref{eq:nls}, is sharply unimodal and concentrated around the MAP estimate \( \mu = 0.76 \), with standard deviation \( \sigma \approx 0.02 \). This concentrated posterior distribution reflects the presence of a well-posed inverse problem under the surrogate model.
	
	Minor secondary peaks are visible in the NLS profile, but their amplitudes are negligible relative to the main mode. These features are attributed to residual numerical artifacts or minor secondary minima and do not affect the stability of the MAP solution.
	
	The inferred posterior distribution allows for a significant reduction in the admissible parameter space, effectively isolating a narrow high-probability region. This confirms the robustness of the proposed approach in resolving inverse problems with localized nonlinear responses.
	
\begin{figure}[!ht]
		\centering
		\begin{subfigure}[b]{0.3\textwidth}
			\centering
			\includegraphics[width=\textwidth]{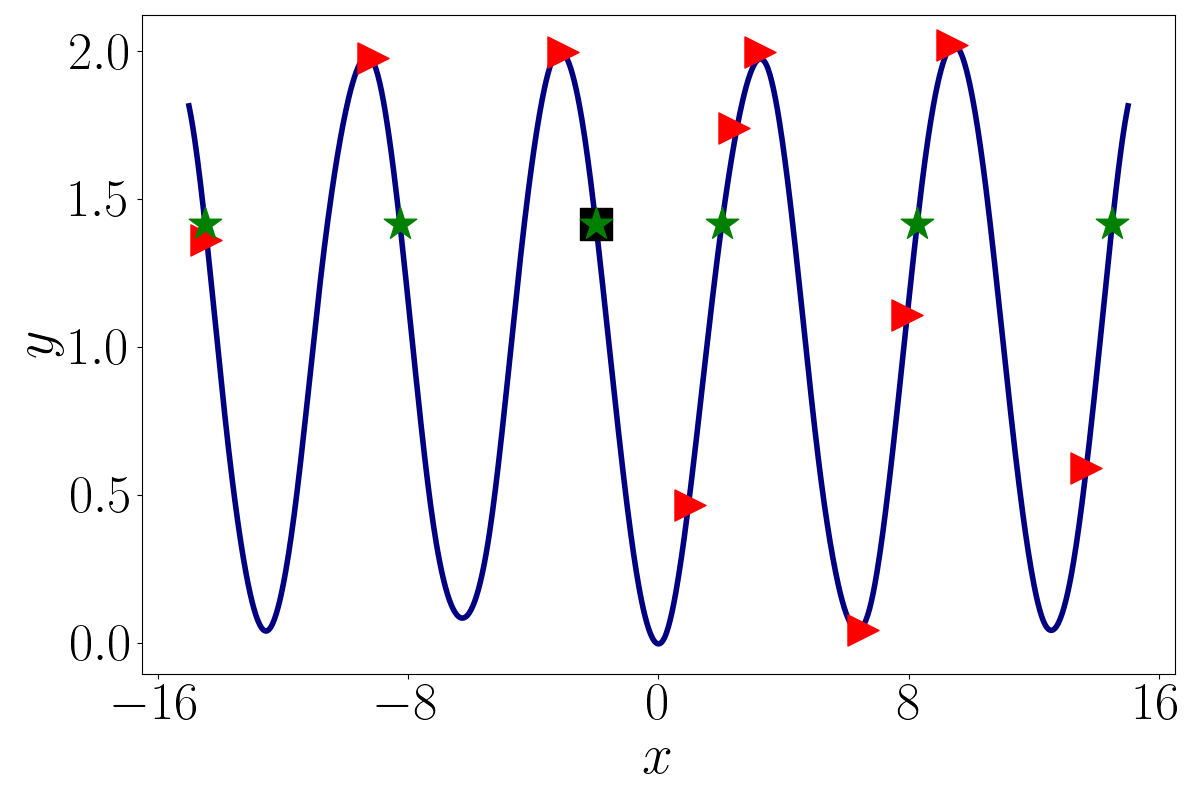}
			\caption{MAP estimates}
			\label{subfig:griewank_map}
		\end{subfigure}
		\begin{subfigure}[b]{0.3\textwidth}
			\centering
			\includegraphics[width=\textwidth]{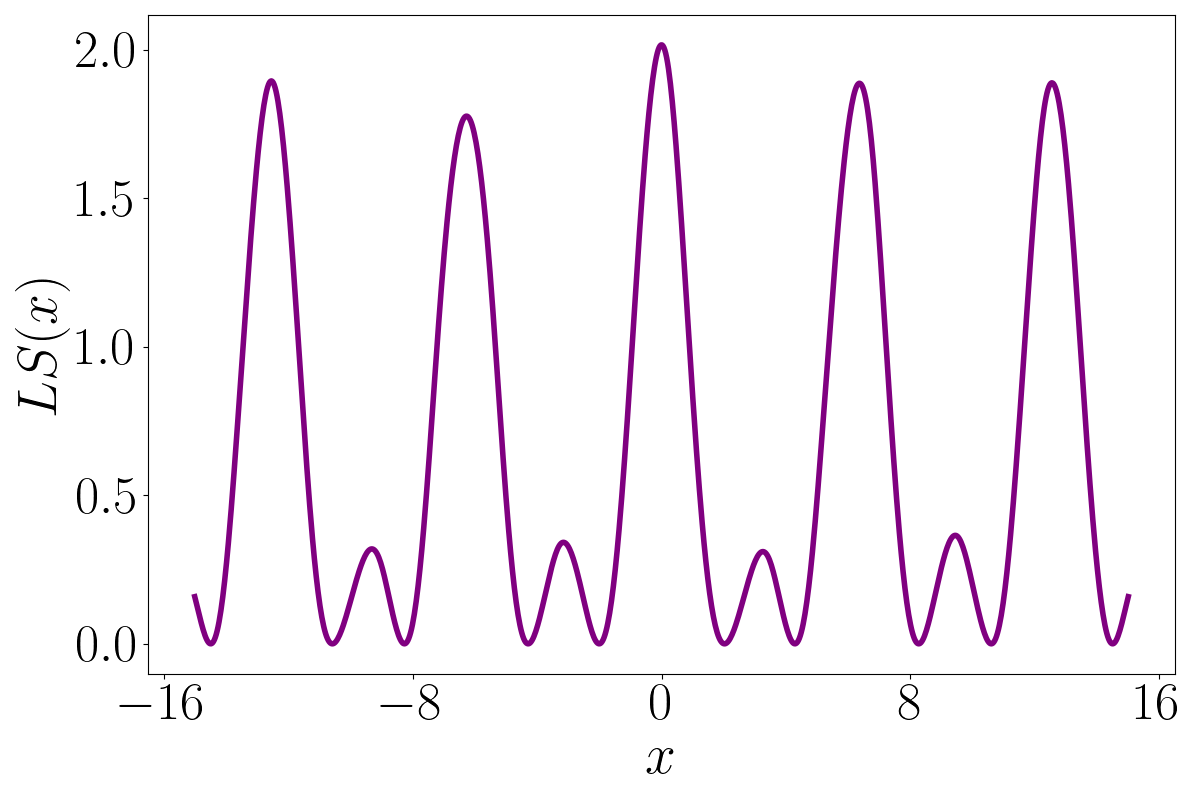}
			\caption{LS profile}
			\label{subfig:griewank_ls}
		\end{subfigure}
		\begin{subfigure}[b]{0.3\textwidth}
			\centering
			\includegraphics[width=\textwidth]{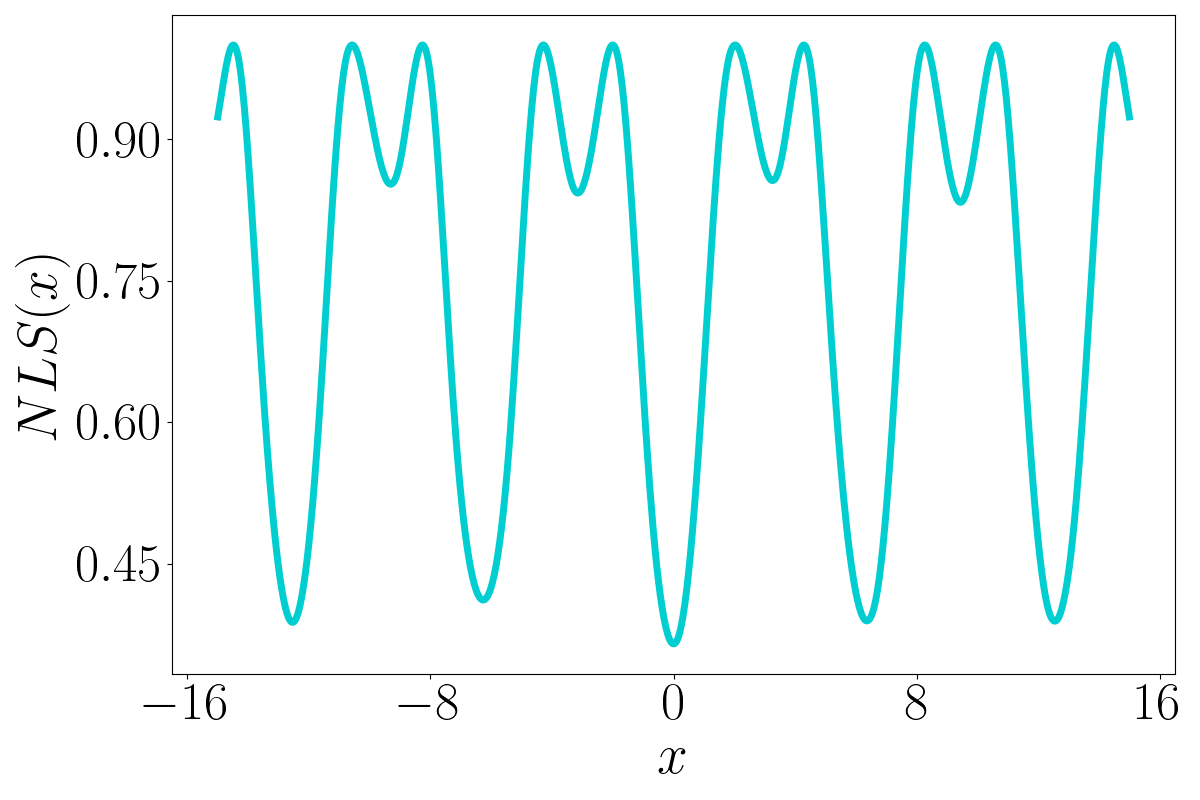}
			\caption{NLS profile}
			\label{subfig:griewank_nls}
		\end{subfigure}
		\caption{Parameter inference via BI results for the Griewank benchmark. (a) MAP estimates from multiple initializations. (b) LS functional profile. (c) NLS functional profile revealing non-identifiability due to periodic structure.}
		\label{fig:griewank_bi}
	\end{figure}

	\Cref{fig:griewank_bi} shows the results for the Griewank benchmark. \Cref{subfig:griewank_map} presents the MAP estimates obtained from optimization runs initialized at different guesses points, using the GP surrogate constructed via BO (\Cref{fig:bo_conclusion}c).
	
	The LS profile (\Cref{subfig:griewank_ls}) reveals a highly multimodal and periodic landscape, with multiple minima of comparable depth. As a result, the NLS profile (\Cref{subfig:griewank_nls}) shows multiple peaks of similar amplitude, corresponding to equally plausible parameter configurations.
	
	This inverse problem is intrinsically non-identifiable in a probabilistic sense. Despite the surrogate model’s accuracy in reconstructing the admissible solution set, the posterior distribution fails to isolate a dominant mode due to the periodic nature of the objective function.
	
	The analyses of the Lévy and Griewank benchmarks highlight the flexibility and diagnostic power of the proposed framework. In the case of the Lévy function, the  posterior distribution represented by the NLS profile is sharply concentrated and uniquely peaked, confirming the well-posedness of the inverse problem. In contrast, the Griewank case demonstrates that the framework can effectively reveal inherent non-identifiability in periodic inverse settings.
	
	In both cases, the proposed Bayesian framework yields an interpretable characterization of the admissible parameter space, supporting robust inference and informed decision-making even in the presence of structural ambiguity.

	\section{\mihi{Supplementary Surrogate-Based Posterior Analysis}}
	\label{appendix:mcmc}
	
	\mihi{The present appendix provides a supplementary posterior analysis based on surrogate-assisted Markov Chain Monte Carlo (MCMC) sampling \cite{brooks2011handbook,andrieu2008tutorial}. 
		It is intended as a clarifying complement rather than a methodological extension of the proposed Bayesian framework (see \Cref{sec:inverse-design,analytic_benchmark}), and illustrates those regimes in which sampling-based inference is required to accurately capture the structure of the posterior distribution. 
		The analysis is conducted using the Gaussian Process (GP) surrogate model constructed via Bayesian Optimization (BO), as discussed in \Cref{analytic_benchmark}, and adopts the one-dimensional Mixed Gaussian-Periodic benchmark function introduced in \Cref{fig:Benchmark}a as a representative reference case for the present analysis.}
	
	\mihi{The objective is to obtain an independent posterior representation in a regime where the inverse problem is strongly non-convex and the resulting posterior distribution is expected to be multimodal. 
		In such settings, local Gaussian summaries around a single maximizer are not appropriate, and sampling-based posterior exploration provides a direct qualitative benchmark for assessing posterior complexity.}
	
\begin{figure}[!ht]
		\centering
		\begin{subfigure}[t]{0.22\textwidth}
			\includegraphics[width=\textwidth]{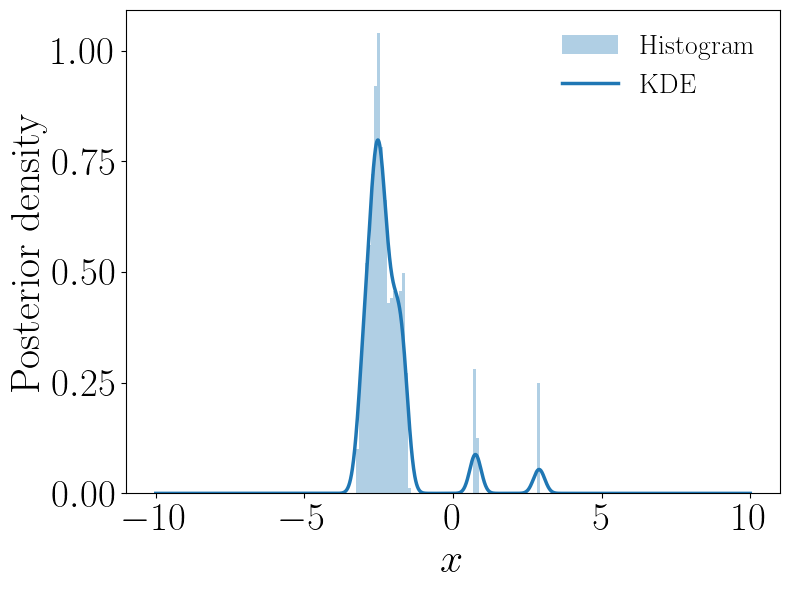}
			\caption{\mihi{Chain 1}}
		\end{subfigure}
		\begin{subfigure}[t]{0.22\textwidth}
			\includegraphics[width=\textwidth]{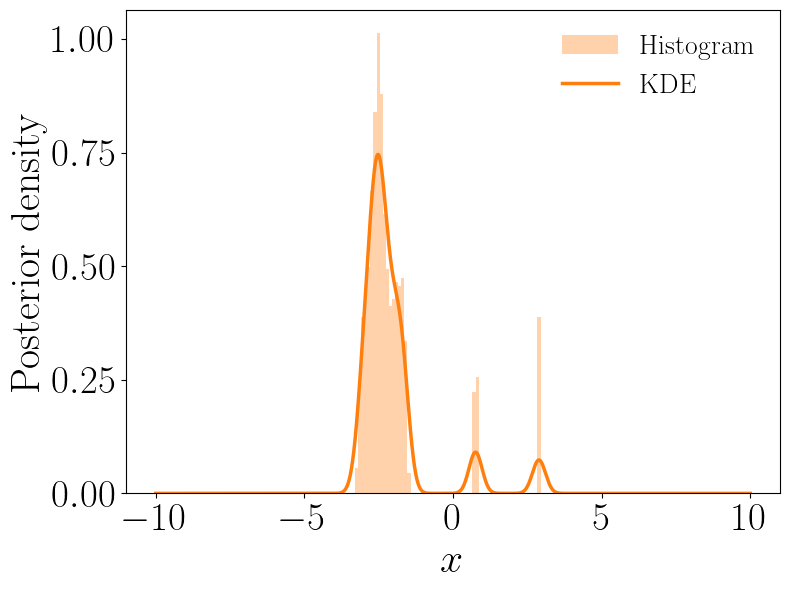}
			\caption{\mihi{Chain 2}}
		\end{subfigure}
		\begin{subfigure}[t]{0.22\textwidth}
			\includegraphics[width=\textwidth]{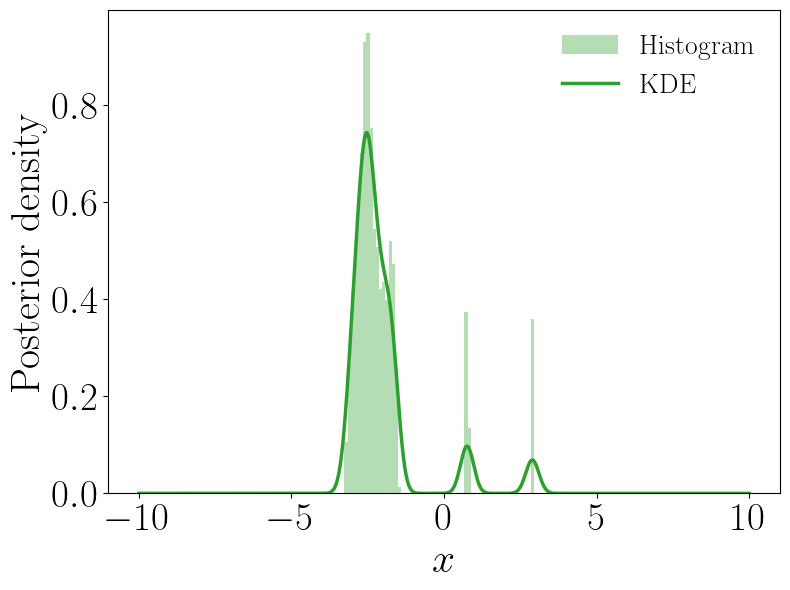}
			\caption{\mihi{Chain 3}}
		\end{subfigure}
		\begin{subfigure}[t]{0.22\textwidth}
			\includegraphics[width=\textwidth]{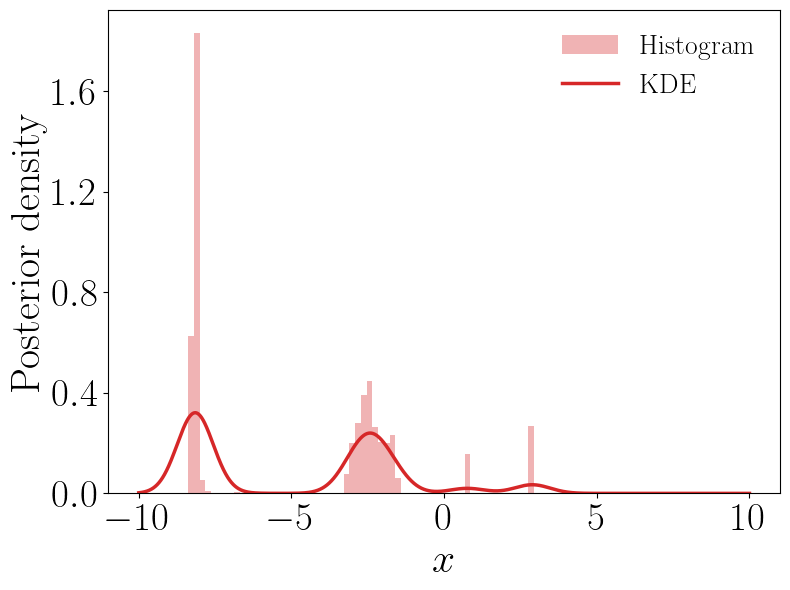}
			\caption{\mihi{Chain 4}}
		\end{subfigure}
		\begin{subfigure}[t]{0.22\textwidth}
			\includegraphics[width=\textwidth]{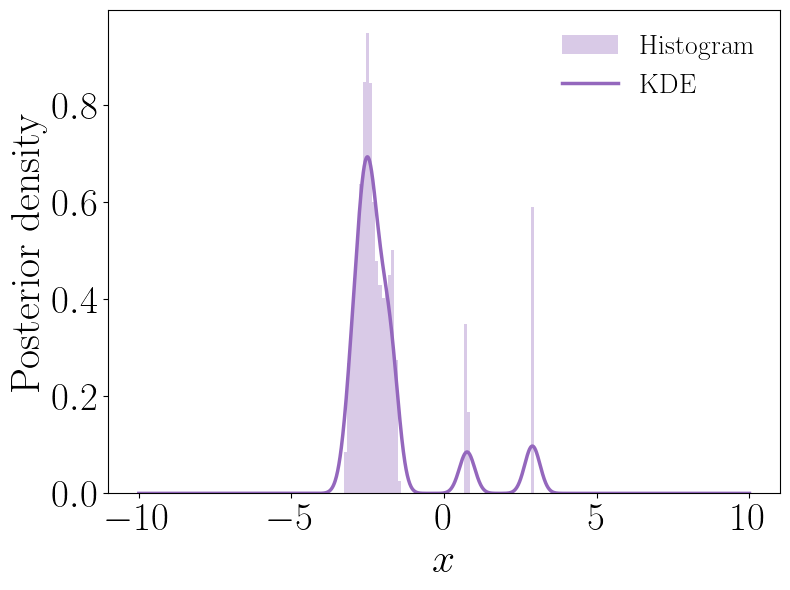}
			\caption{\mihi{Chain 5}}
		\end{subfigure}
		\begin{subfigure}[t]{0.22\textwidth}
			\includegraphics[width=\textwidth]{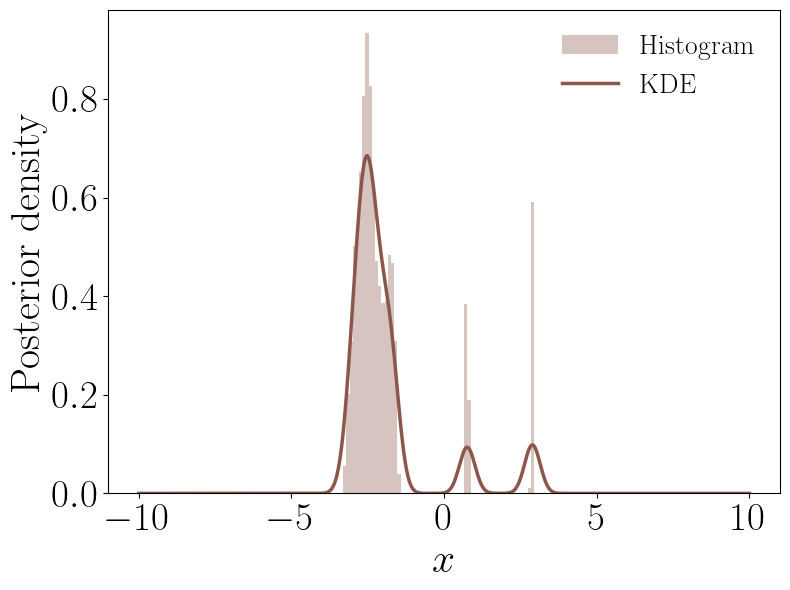}
			\caption{\mihi{Chain 6}}
		\end{subfigure}
		\begin{subfigure}[t]{0.22\textwidth}
			\includegraphics[width=\textwidth]{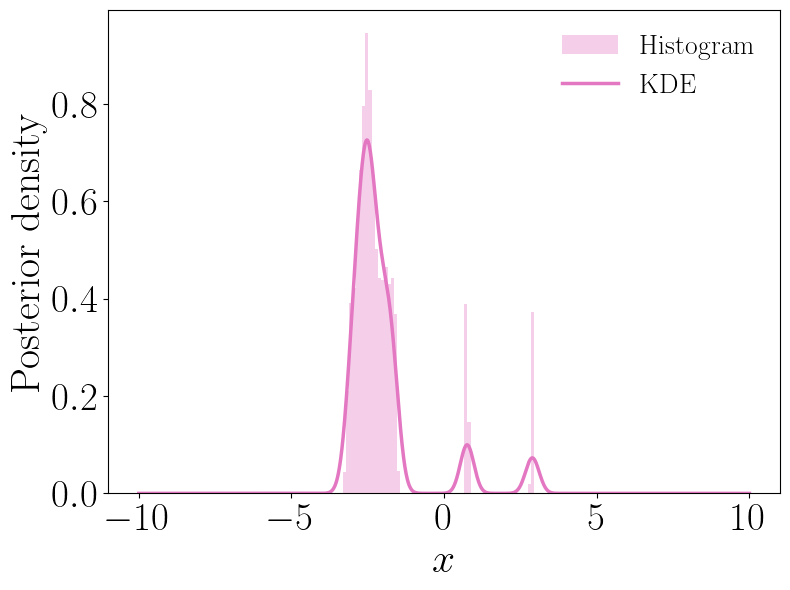}
			\caption{\mihi{Chain 7}}
		\end{subfigure}
		\begin{subfigure}[t]{0.22\textwidth}
			\includegraphics[width=\textwidth]{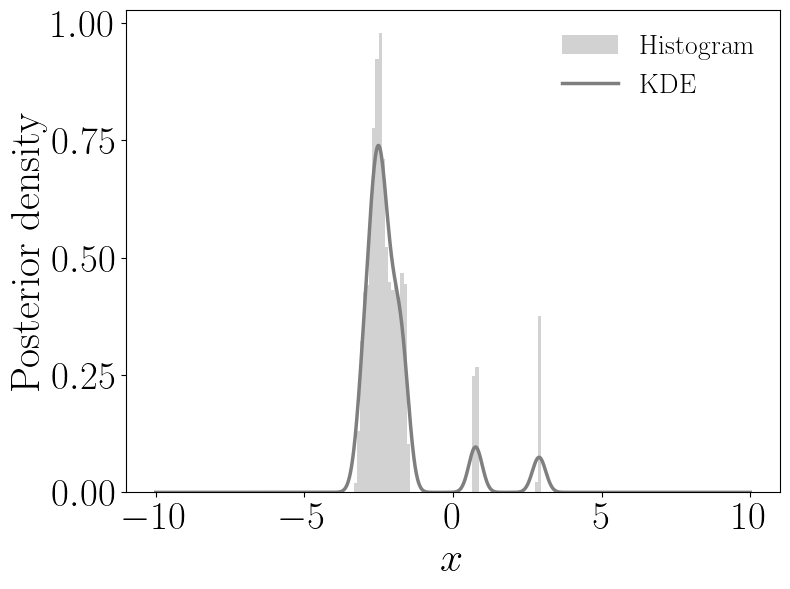}
			\caption{\mihi{Chain 8}}
		\end{subfigure}
		\begin{subfigure}[t]{0.22\textwidth}
			\includegraphics[width=\textwidth]{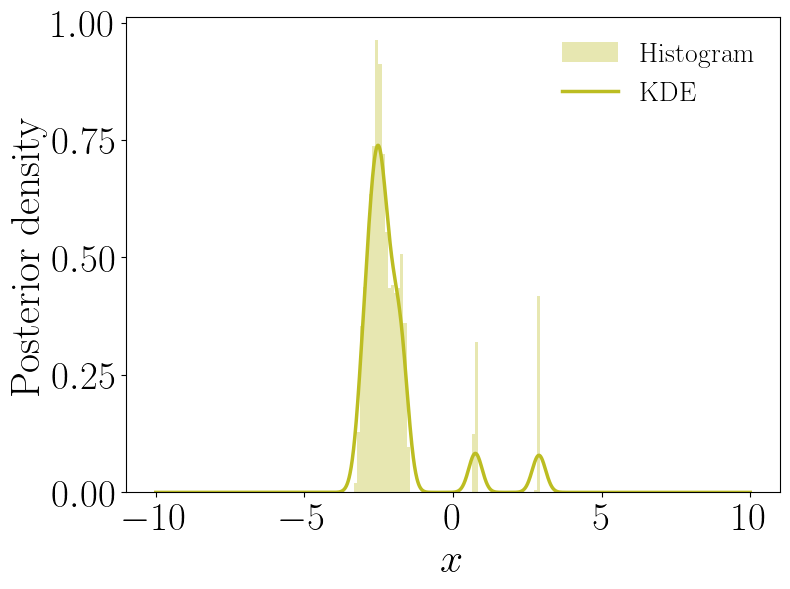}
			\caption{\mihi{Chain 9}}
		\end{subfigure}
		\begin{subfigure}[t]{0.22\textwidth}
			\includegraphics[width=\textwidth]{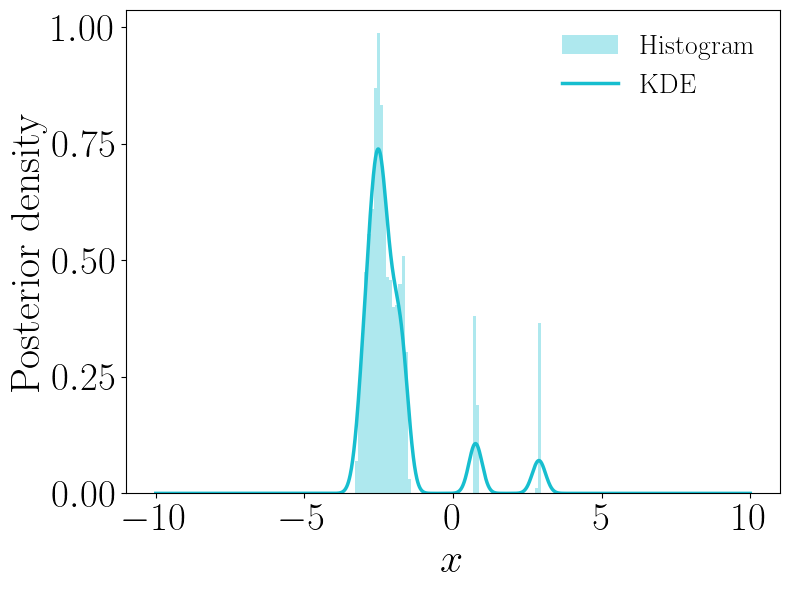}
			\caption{\mihi{Chain 10}}
		\end{subfigure}
		\caption{\mihi{Marginal posterior densities obtained from surrogate-assisted MCMC sampling for the Mixed Gaussian-Periodic benchmark. 
				For each chain, the empirical histogram of the sampled values is shown together with a kernel density estimate. 
				While a dominant posterior mode is consistently identified across chains, secondary modes and low-probability regions are intermittently sampled, leading to noticeable inter-chain variability in the estimated posterior shape.}}
		\label{fig:mcmc_posteriors_chains}
	\end{figure}

	\mihi{The posterior densities estimated from the generated samples are reported in \Cref{fig:mcmc_posteriors_chains}. 
		The results consistently indicate the presence of a dominant posterior region together with additional secondary modes, whose relative prominence varies across runs, reflecting the intrinsically multimodal structure of the underlying inverse problem.}
	
	\mihi{The per-chain posterior estimates reveal a clearly multimodal structure, with a dominant mode robustly captured across most chains and additional secondary modes that are only sporadically explored. 
		As a consequence, for the Mixed Gaussian-Periodic benchmark (\Cref{fig:Benchmark}a), the assumptions underlying local Laplace approximations around a single maximum are not satisfied, as already discussed in Section~\ref{subsubsec:1Dbi}.}
	
\begin{figure}[!ht]
		\centering
		\includegraphics[width=0.85\textwidth]{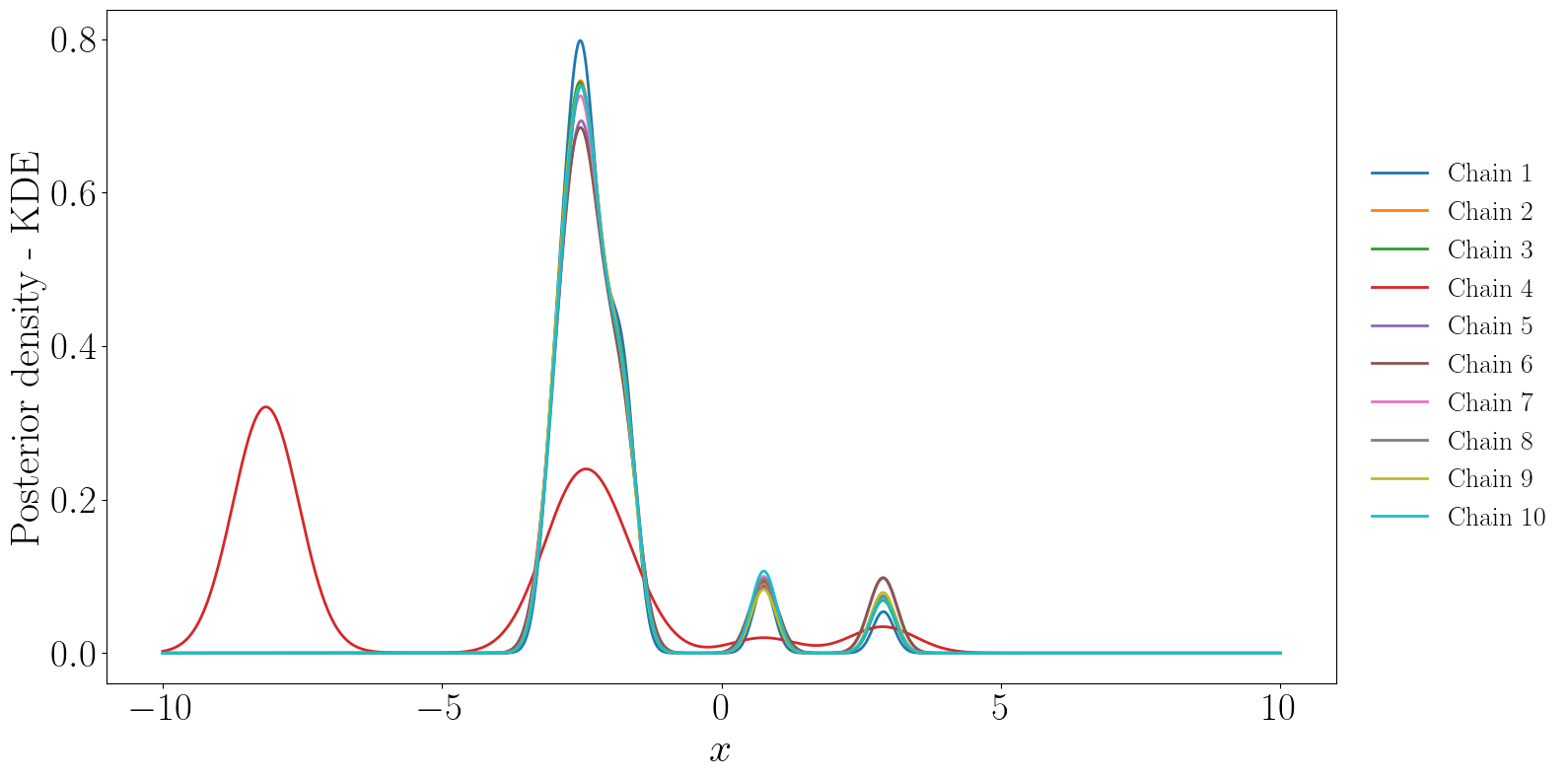}
		\caption{\mihi{Superimposed marginal posterior density estimates obtained from all surrogate-assisted MCMC chains. The overlay highlights the dominant posterior mode shared across chains, together with secondary modes that are sampled with varying frequency, reflecting the multimodal nature of the posterior distribution.}}
		\label{fig:all_pdf}
	\end{figure}

	\mihi{For additional synthesis, the marginal posterior densities of all chains are superimposed in \Cref{fig:all_pdf}, providing a compact visualization of inter-chain variability and mode consistency.}
	
	\mihi{Accordingly, for the Mixed Gaussian-Periodic benchmark, the present manuscript does not enforce a local Laplace approximation around a single MAP estimate. 
		Instead, the negative least-squares (NLS) functional is employed as a computationally inexpensive diagnostic tool to assess posterior structure and to identify regions of high posterior probability, without imposing a unimodal Gaussian approximation that would be theoretically unjustified. 
		This strategy enables a localisation of high-probability intervals while avoiding unnecessary sampling or restrictive assumptions.}
	
	\mihi{While surrogate model evaluations are computationally inexpensive, the effective cost of sampling-based posterior exploration depends on the complexity of the posterior distribution and on the accuracy requirements of the analysis \cite{brooks2011handbook,andrieu2008tutorial}. 
		A detailed investigation of these aspects would require additional modelling choices and would substantially broaden the scope of the present manuscript. 
		Surrogate-assisted MCMC is therefore regarded as a natural extension of the proposed framework, rather than its primary objective.}
	
\begin{figure}[!ht]
		\centering
		\includegraphics[width=0.85\textwidth]{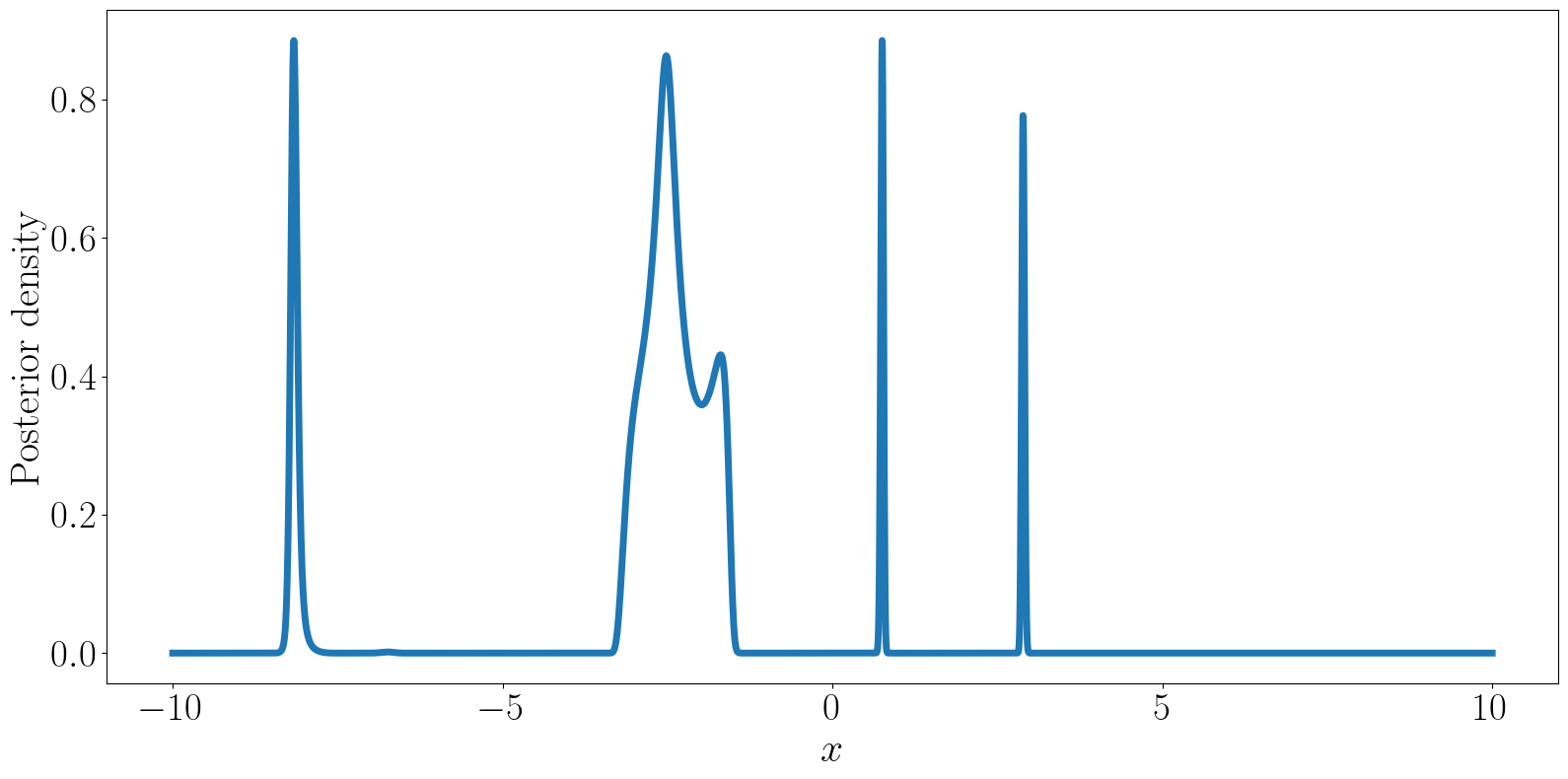}
		\caption{\mihi{Grid-based posterior density for the Mixed Gaussian-Periodic benchmark, obtained by evaluating the surrogate-induced likelihood on a dense uniform discretization of the parameter space and normalizing the resulting values. The plot reveals a clearly multimodal posterior structure with multiple well-separated regions of high probability.}}
		\label{fig:grid_posterior}
	\end{figure}

	\mihi{Finally, for further qualitative comparison, a reference posterior density is constructed on a dense one-dimensional grid covering the admissible parameter domain. 
		Figure~\ref{fig:grid_posterior} reports the resulting grid-based posterior density, obtained by evaluating the surrogate-induced likelihood on a uniform discretization of the parameter space and normalizing the resulting values under a uniform prior. 
		The grid posterior clearly reveals multiple well-separated modes and is fully consistent with the LS and NLS profiles reported in \Cref{fig:inverseMixed} and discussed in Section~\ref{subsubsec:1Dbi}. 
		This deterministic representation provides a complementary global diagnostic of posterior complexity against which the MCMC results can be qualitatively assessed.}
	
	\mihi{Overall, the results presented in this appendix reinforce the central methodological message of the manuscript: local Laplace approximations are employed only when the inverse problem is empirically well-posed and the posterior is effectively unimodal, whereas NLS-based diagnostics are systematically used to assess posterior structure and to identify high-probability regions at negligible additional cost. 
		When these conditions are not met, as in the Mixed Gaussian-Periodic benchmark, sampling-based approaches become necessary. 
		In this sense, the Laplace approximation and MCMC are complementary tools, applicable in different regimes of posterior complexity, rather than interchangeable inference strategies.}
	
\end{document}